%% file: arXivV3.tex
\documentclass[reprint,titlepage,amsmath,amssymb,aps,prd,floatfix,nofootinbib,eqsecnum]{revtex4-2}

\usepackage{silence} 
\usepackage{amsthm} 
    \theoremstyle{definition}
    \newtheorem{definition}{Definition}
\usepackage{graphicx}
\usepackage[utf8]{inputenc}
\usepackage[T1]{fontenc}
\usepackage{newtxtext}
\usepackage{newtxmath}
\usepackage{microtype} 
\usepackage{physics} 
\usepackage{tensor} 
\usepackage{pict2e} 
\usepackage[svgnames]{xcolor} 
    \definecolor{accent}{HTML}{df2d16}
\usepackage{pgfplots} 
    \pgfplotsset{compat=1.18}
\usepackage{tikz}
\usepackage[compat=1.1.0]{tikz-feynman} 
\usepackage{enumitem} 
\usepackage[allcolors=accent,colorlinks,pdfusetitle,pdfauthor={Níckolas de Aguiar Alves and André G. S. Landulfo}]{hyperref} 
\usepackage{orcidlink}

\input{texnical} 

\begin{document}
\title{Null infinity as a Killing horizon}

\author{\firstname{Níckolas} de \surname{Aguiar Alves}\,\orcidlink{0000-0002-0309-735X}}
\email{alves.nickolas@ufabc.edu.br}

\author{André G. S. Landulfo\,\orcidlink{0000-0002-3717-4966}}
\email{andre.landulfo@ufabc.edu.br}

\affiliation{Center for Natural and Human Sciences, \href{https://ror.org/028kg9j04}{Federal University of ABC}, Avenida dos Estados 5001, Bangú, Santo André, São Paulo 09280-560, Brazil}
\date{August 20, 2025}

\begin{abstract}
   Symmetries are ubiquitous in modern physics. They not only allow for a more simplified description of physical systems but also, from a more fundamental perspective, can be seen as determining a theory itself. In the present paper, we propose a new definition of asymptotic symmetries that unifies and generalizes the usual notions of symmetry considered in asymptotically flat spacetimes and expanding universes with cosmological horizons. This is done by considering BMS-like symmetries for ``asymptotic (conformal) Killing horizons'', or A(C)KHs, here defined as null hypersurfaces that are tangent to a vector field satisfying the (conformal) Killing equation in a limiting sense. The construction is theory-agnostic and extremely general, for it makes no use of the Einstein equations and can be applied to a wide range of scenarios with different dimensions or hypersurface cross sections. While we reproduce the results by Dappiaggi, Moretti, and Pinamonti in the case of asymptotic Killing horizons, the conformal generalization does not yield only the BMS group, but a larger group. The enlargement is due to the presence of ``superdilations''.  We speculate on many implications and possible continuations of this work, including the exploration of gravitational memory effects beyond general relativity, understanding antipodal matching conditions at spatial infinity in terms of bifurcate horizons, and the absence of superrotations in de Sitter spacetime and Killing horizons.
\end{abstract}

\maketitle

\section{Introduction}\label{sec: intro}
     Symmetry is the backbone of physics. From a practical perspective, the occurrence of symmetries in a physical system makes it easier to understand. From a fundamental perspective, the symmetries of a physical theory can be understood as determining a theory itself. In a sense, theoretical physics is a continuous pursuit for the symmetries of the universe.

    Unfortunately, real-world scenarios often are less symmetric than one would wish for. While Minkowski spacetime enjoys Poincaré symmetry, this is only possible due to its vacuum nature, and realistic spacetimes explicitly break many of the symmetries one could be interested in exploiting. You remember reading the previous phrase, and thus time-translation symmetry is broken. This document is in front of you, at rest, and at a fixed distance, and thus Lorentz transformations and spatial translations are not symmetries either. How can symmetry be useful in such an irregular world? 

    One possibility is to move away from any matter fields with the hope of regaining control of the symmetries of the spacetime. For instance, Earth violates all sorts of Poincaré symmetries. Could they be restored if the Earth was suddenly very far away? If we choose to neglect all content of the universe except for  Earth---or some other compact object---could we recover the (proper, orthochronous) Poincaré group, given by the semidirect product
        \begin{equation}\label{eq: Poincare-group}
            \mathrm{ISO}^+(3,1) = \mathrm{SO}^+(3,1) \ltimes \mathbb{R}^4,
        \end{equation}
    as the symmetry group when standing infinitely far from the said object?  
    
    Although intuitive, this was shown not to be the case by Bondi, Metzner, and Sachs (BMS) in their seminal papers~\cite{bondi1962GravitationalWavesGeneral,sachs1962AsymptoticSymmetriesGravitational,sachs1962GravitationalWavesGeneral}. They rather found a much larger symmetry group. Namely, they discovered that the symmetry group of (future) null infinity should be given by
        \begin{equation}\label{eq: BMS-group}
            \mathrm{BMS}_4 = \mathrm{SO}^+(3,1) \ltimes \mathcal{C}^{\infty}(\mathbb{S}^2),
        \end{equation}
    where \(\mathcal{C}^{\infty}(\mathbb{S}^2)\) is the additive group of smooth functions on a sphere. The BMS group is thus an extension of the Poincaré group by the new ``supertranslations''---the elements of \(\mathcal{C}^{\infty}(\mathbb{S}^2)\) generalizing the ordinary spacetime translations \(\mathbb{R}^4\).

    Analyzing the properties of infinity in asymptotically flat spacetimes is useful to make predictions about observables measured far away from a compact object. However, we know the universe we live in is not asymptotically flat. Instead, it is dominated at late times by a positive cosmological constant, making it ``asymptotically de Sitter''. Furthermore, inflation requires a de Sitter-like behavior at very early times, and it is a successful approach to understanding the formation of primordial perturbations---see, \eg, Refs.~\cite{baumann2022Cosmology,weinberg2008Cosmology,planckcollaboration2020AX}. Hence, it would be interesting to understand whether or not it is possible to define asymptotic symmetries in contexts that do not necessarily correspond to null infinity. 

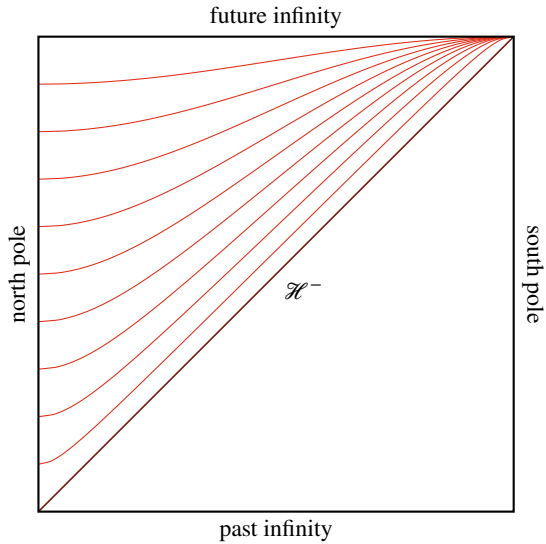
\begin{figure}[tbp]
        \centering
        \begin{tikzpicture}[scale=2]
            \foreach \a in {9,18,...,81}{
                \draw[domain=2*rad(\a)-pi/2:pi/2, samples=200, smooth, variable=\x, accent] plot ({rad(acos(tan(\a)*cos(deg(\x))-sin(deg(\x))))},{\x+pi/2});
            };
            \draw[semithick, accent!50!black] (0,0) -- (pi,pi);
            \draw[thick] (0,0) -- ++(pi,0) -- ++(0,pi) -- ++(-pi,0) -- cycle;
            
            \node[anchor=south] at (0,pi/2) [rotate=90] {north pole};
            \node[anchor=south] at (pi,pi/2) [rotate=270] {south pole};
            \node[anchor=south] at (pi/2,pi) {future infinity};
            \node[anchor=north] at (pi/2,0) {past infinity};
            \node[anchor=north west] at (pi/2,pi/2) {\(\mathscr{H}^-\)};
        \end{tikzpicture}
        \caption{Carter--Penrose diagram for de Sitter spacetime. The cosmological patch of de Sitter spacetime is the region above and to the left of \(\mathscr{H}^-\), which is the past cosmological horizon. The thin lines represent spatial planes foliating the cosmological patch. Notice their past limit is the cosmological horizon.}
        \label{fig: desitter}
    \end{figure}

    One approach to asymptotic symmetries in ``de Sitter-like'' spacetimes was given by Dappiaggi, Moretti, and Pinamonti (DMP)~\cite{dappiaggi2009CosmologicalHorizonsReconstruction,dappiaggi2009DistinguishedQuantumStates,dappiaggi2017HadamardStatesLightlike}. They considered a class of spacetimes with past particle horizons similar to \(\mathscr{H}^-\simeq \mathbb{R}\times \mathbb{S}^2\) in de Sitter (see Fig.~\ref{fig: desitter}). These are now known as expanding universes with cosmological horizons, which need not be homogeneous nor isotropic. Within this class, they established the horizon enjoys a BMS-like symmetry group, which we will refer to as the DMP group. It is given by
    \begin{equation}\label{eq: DMP-group}
        \mathrm{DMP}_4 = \mathrm{SO}(3) \ltimes (\mathcal{C}^{\infty}(\mathbb{S}^2) \ltimes \mathcal{C}^{\infty}(\mathbb{S}^2)).
    \end{equation}
    We can see that \(\mathrm{DMP}_4\) contains an \(\mathrm{SO}(3)\) factor, which corresponds to the isometry group of the sphere (the transverse space in \(\mathscr{H}^-\)), as well as two \(\mathcal{C}^{\infty}(\mathbb{S}^2)\) factors. One of these factors describes ``supertranslations'' while the other corresponds to ``superdilations'', \ie, direction-dependent generalizations of dilations. 

    While the BMS group extends the Poincaré group, the DMP group does not extend the de Sitter isometry group \(\mathrm{SO}^+(4,1)\), because the latter does not preserve the horizon. Nonetheless, \(\mathrm{DMP}_4\) does extend the ``cosmological de Sitter group''
    \begin{equation}\label{eq: cosmological-de-sitter-group}
        \mathrm{SO}(3) \ltimes (\mathbb{R} \ltimes \mathbb{R}^3),
    \end{equation}
    which is comprised of the de Sitter isometries that preserve the cosmological horizon. The generators included in \(\mathrm{SO}^+(4,1)\), but absent in the cosmological de Sitter group, correspond to de Sitter boosts.

    \begin{figure*}[tbp]
        \centering
        \begin{equation*}
            \vcenter{\hbox{\begin{tikzpicture}
                \begin{feynman}
                    \node[blob, pattern color=white] (c) at (0,0);
                    \vertex (a1) at (180:2);
                    \vertex (a2) at (135:2);
                    \vertex (a3) at (225:2);
                    \vertex (a4) at (225:1);
                    \vertex (a5) at (300:1.5);
                    \vertex (b1) at (30:2);
                    \vertex (b2) at (330:2);
                    \vertex (b3) at (330:1);
                    \vertex (b4) at (0:2) {\footnotesize\(\mu\nu\)};
                    \diagram*{
                        (a1) -- [fermion] (c),
                        (a2) -- [fermion] (c),
                        (a3) -- [fermion] (c),
                        (c) -- [fermion] (b1),
                        (c) -- [fermion] (b3),
                        (b3) -- [fermion,momentum'=\(p\)] (b2),
                        (b3) -- [graviton] (b4),
                        (b3) -- [draw=none,momentum=\(q\)] (b4),
                    };
                \end{feynman}
            \end{tikzpicture}}} = \qty(\frac{\sqrt{8\pi G} \tensor{p}{_\mu}\tensor{p}{_\nu}}{p \cdot q - i \epsilon}) \times \qty(\vcenter{\hbox{\begin{tikzpicture}
                \begin{feynman}
                    \node[blob, pattern color=white] (c) at (0,0);
                    \vertex (a1) at (180:2);
                    \vertex (a2) at (135:2);
                    \vertex (a3) at (225:2);
                    \vertex (b1) at (30:2);
                    \vertex (b2) at (330:2);
                    \diagram*{
                        (a1) -- [fermion] (c),
                        (a2) -- [fermion] (c),
                        (a3) -- [fermion] (c),
                        (c) -- [fermion] (b1),
                        (c) -- [fermion,momentum'=\(p\)] (b2),
                    };
                \end{feynman}
            \end{tikzpicture}}}) + o\qty(\frac{1}{\tensor{q}{^0}}).
        \end{equation*}
        \caption{Illustration of the Weinberg soft graviton theorem. To leading order, adding a soft graviton to an external leg of a Feynman diagram amounts to multiplying the original diagram by a ``soft factor''. This soft factor depends on the graviton's momentum and on the momentum of the leg to which it is attached. This picture considers the case in which the soft graviton attaches to an outgoing particle, but a similar result holds for incoming particles.}
        \label{fig: soft-graviton-theorem-out}
    \end{figure*}

    Although very appealing, the idea of asymptotic symmetries will be useful only if it can be used to make physical predictions. In the asymptotically flat case, for instance, it has been noticed by Strominger and collaborators~\cite{strominger2014BMSInvarianceGravitational, strominger2016GravitationalMemoryBMS, he2015BMSSupertranslationsWeinbergs} that three separate results from the 1960s and 1970s---the BMS group, the Weinberg soft graviton theorem~\cite{weinberg1965InfraredPhotonsGravitons}, and the gravitational wave displacement memory effect~\cite{zeldovich1974RadiationGravitationalWaves, braginsky1987GravitationalwaveBurstsMemory, christodoulou1991NonlinearNatureGravitation}---are corners of the same ``infrared triangle''. The three phenomena are all expressions of the same underlying physics, in spite of having originated in different research programs and being formulated in different theoretical frameworks. Chronologically, the first corner of the infrared triangle was the BMS group itself. It was soon followed by the second corner, Weinberg's soft graviton theorem~\cite{weinberg1965InfraredPhotonsGravitons} (see Refs.~\cite{weinberg1995Foundations,aguiaralves2025LecturesBondiMetzner,mcloughlin2022SAGEXReviewScattering,mohanty2023GravitationalWavesQuantum} for reviews). This is a relation between scattering amplitudes in quantum field theory (QFT) which states that the leading-order effect of adding a very low energy graviton (a ``soft graviton'') to an external leg of any Feynman diagram amounts to merely multiplying the diagram by an overall ``soft factor''. This is pictured in Fig. \ref{fig: soft-graviton-theorem-out}. Later,  Strominger and collaborators \cite{strominger2014BMSInvarianceGravitational,he2015BMSSupertranslationsWeinbergs} showed that the Weinberg soft graviton theorem can be recast as the Ward identity for invariance under supertranslations at past and future null infinity that are antipodally related near spatial infinity---see Refs.~\cite{aguiaralves2025LecturesBondiMetzner,strominger2018LecturesInfraredStructure,strominger2014BMSInvarianceGravitational,he2015BMSSupertranslationsWeinbergs} for details. The third corner of the infrared triangle is the gravitational wave displacement memory effect, discovered by \textcite{zeldovich1974RadiationGravitationalWaves} and reviewed, for example, in Refs.~\cite{bieri2015GravitationalWavesTheir,bieri2024GravitationalWaveDisplacement,aguiaralves2025LecturesBondiMetzner}. The memory effect predicts that the passage of a gravitational wave will often cause a permanent displacement of the relative positions of a pair of nearby inertial detectors located far away from its source. This effect turns out to be mathematically equivalent to a supertranslation---see, for example, Refs.~\cite{aguiaralves2025LecturesBondiMetzner,strominger2016GravitationalMemoryBMS}. Furthermore, it was shown by \textcite{strominger2016GravitationalMemoryBMS} that the memory in the gravitational field is given by a Fourier transform of Weinberg's soft factor. Hence, all three phenomena are related to each other. Since the memory effect can in principle be detected by future gravitational wave observatories  \cite{favata2010GravitationalwaveMemoryEffect,grant2023OutlookDetectingGravitationalwave,*grant2023ErratumOutlookDetecting}, there are reasonable prospects that infrared effects in gravity will soon be measured.

    Supertranslations find physical meaning through the memory effect and the Weinberg soft theorem. Hence, these results are explicit examples of how the knowledge of symmetries is useful in understanding the behavior of a theory or of spacetime itself---even if the symmetries are only exact at infinity. 

    The BMS group was merely a first step in understanding the symmetries of null infinity. The Lorentz group \(\mathrm{SO}^+(3,1)\) is the group of global conformal transformations on the sphere. As a consequence, it is possible to consider an extension to the local conformal algebra on the sphere, given by two copies of the so-called Witt algebra. This leads to the ``extended BMS algebra'' \cite{banks2003CritiquePureString,barnich2010AspectsBMSCFT,barnich2010SymmetriesAsymptoticallyFlat}, with the new transformations being known as superrotations. While this extension may seem arbitrary, it is one of the cornerstones of modern two-dimensional conformal field theory (CFT) \cite{belavin1984InfiniteConformalSymmetry,difrancesco1997ConformalFieldTheory,blumenhagen2009IntroductionConformalField,schottenloher2008MathematicalIntroductionConformal,[{}][{, Chap. 2.}]{polchinski1998IntroductionBosonicString}}. Upon quantization, the Witt algebra leads to the Virasoro algebra, and these Virasoro-like superrotations suggest a correspondence between gravity in asymptotically flat spacetimes and two-dimensional CFTs. In fact, one of the main outcomes of soft physics has been the development of the celestial holography program~\cite{pasterski2021LecturesCelestialAmplitudes, pasterski2025ChapterCelestialHolography, pasterski2021LecturesCelestialAmplitudes, raclariu2021LecturesCelestialHolography, donnay2024CelestialHolographyAsymptotic}, which aims at constructing a holographic correspondence between quantum gravity in four-dimensional asymptotically flat spacetimes and a two-dimensional CFT living on the celestial sphere. 

    Another example of the use of asymptotic symmetries is in the context of quantum field theory in curved spacetimes. DMP \cite{dappiaggi2006RigorousStepsHolography,moretti2006UniquenessTheoremBMSInvariant,moretti2008QuantumOutstatesHolographically,dappiaggi2011RigorousConstructionHadamard} noticed that one may relate a quantum field theory in the bulk of an asymptotically flat spacetime to a QFT at null infinity. At null infinity, one may exploit the infinite-dimensional BMS group to define a boundary quantum state. Once the state is defined at the boundary, it is pulled back to the quantum field theory in the bulk. It turns out that the resulting bulk quantum state is of Hadamard form, and thus of distinguished physical interest\footnote{There are some technical restrictions for the construction---for example the necessity that the field be conformally coupled---which are discussed in detail in Ref.~\cite{dappiaggi2017HadamardStatesLightlike}.}. These ideas reach beyond the asymptotically flat case. As shown by DMP~\cite{dappiaggi2009CosmologicalHorizonsReconstruction,dappiaggi2009DistinguishedQuantumStates}, it is possible to adapt the above construction to de Sitter-like cosmological spacetimes by using the DMP group in the cosmological horizon  \(\mathscr{H}^-\). Details can be found in Ref.~\cite{dappiaggi2017HadamardStatesLightlike}. Notice, in particular, that this construction gives meaning to \(\mathrm{DMP}_4\) and establishes it has physical interest.
    
    There are many more lines of investigation. For example, Prabhu, Satishchandran, and Wald~\cite{prabhu2022InfraredFiniteScattering, prabhu2024InfraredFiniteScatteringStatesRepresentations, prabhu2024InfraredFiniteScatteringAmplitudesSoftTheorems} have employed the algebraic approach to QFT to develop an infrared-finite approach to scattering theory in quantum gravity. Danielson, Satishchandran, and Wald~\cite{danielson2022BlackHolesDecohere,danielson2022KillingHorizonsDecohere,danielson2025LocalDescriptionDecoherence} have also shown how soft physics can be interesting even at finite distances, by noticing that radiation in the form of soft gravitons (or photons) leads to the decoherence of quantum superpositions in the presence of Killing horizons. These examples showcase how there is considerable interest in infrared physics throughout many different subareas of fundamental physics, and even when considering hypersurfaces that do not lie at infinity.

    It is curious, though, that the BMS and DMP groups are so similar, yet so different. This raises a few natural questions.
    \begin{enumerate}
        \item Why does the DMP group has an extra factor of \(\mathcal{C}^{\infty}(\mathbb{S}^2)\)? This factor corresponds to ``superdilations'' and it is indispensable if one wants the DMP group to extend the cosmological de Sitter group \eqref{eq: cosmological-de-sitter-group}. However, it does not occur in the BMS group.
        \item As one can see from Eqs. \eqref{eq: BMS-group} and \eqref{eq: DMP-group}, the DMP group contains an \(\mathrm{SO}(3)\) factor (the isometry group on the sphere) as opposed to BMS's \(\mathrm{SO}^+(3,1)\) (the conformal group on the sphere). This suggests it is not possible to enhance the DMP group to include Virasoro-like superrotation symmetries, which would be interesting to pursue a CFT dual to quantum gravity on an expanding universe with cosmological horizon. If so, this could be relevant for the construction of a dS/CFT duality~\cite{strominger2001DSCFTCorrespondence} on the cosmological horizon. 
    \end{enumerate}

    To address these questions, we propose that the definitions of the DMP and BMS groups should be ``unified'' so that one can see where their similarities start and end. While Refs.~\cite{dappiaggi2009CosmologicalHorizonsReconstruction, dappiaggi2009DistinguishedQuantumStates,dappiaggi2017HadamardStatesLightlike, koga2001AsymptoticSymmetriesKilling,compere2019LBMS4GroupdS4,ruzziconi2020VariousExtensionsBMS,duval2014CarrollNewtonGalilei, duval2014ConformalCarrollGroups, duval2014ConformalCarrollGroupsa,ciambelli2019CarrollStructuresNull,chandrasekaran2018SymmetriesChargesGeneral} have attempted such constructions before, we are unaware of any work in which the notion of a Killing horizon is the prominent feature of a null hypersurface (including null infinity) on which asymptotic symmetries are defined. Since a Killing vector field is the gold standard in general relativity to define a symmetry, it seems natural that asymptotic symmetries should always be considered on Killing-like horizons. 
    
    There is an important caveat. When discussing symmetries at, \eg, timelike or spacelike infinities, no Killing-like horizon is obvious. In this paper we ignore these cases and focus instead on asymptotic symmetries on null hypersurfaces. Notice, though, that if we relax the definition of a ``Killing-like horizon'' to include all surfaces invariant under some symmetry (\ie, nonnecessarily null surfaces), then symmetries at spacelike and timelike infinity may also fit this description \cite{ashtekar1992SpatialInfinityBoundary}. For instance, the structure of timelike and spacelike infinities can be understood in terms of hyperbolic slicings of Minkowski spacetime, with each hyperbola being invariant under boosts \cite{strominger2018LecturesInfraredStructure}. 

    To avoid a definition that relies on the notion of an exact Killing horizon (which is too stringent), we propose instead working with asymptotic (conformal) Killing horizons\footnote{Asymptotic Killing horizons have been considered by Koga~\cite{koga2001AsymptoticSymmetriesKilling}, but the conformal case (which would be interesting for asymptotically flat spacetimes) was not. Furthermore, the group obtained by Koga was larger than the DMP group and resembled the Newman–Unti group~\cite{newman1962BehaviorAsymptoticallyFlat} more than the BMS group.}. This invites one to consider ``the sky as a Killing horizon''. One may then use the Carrollian methods developed in Refs.~\cite{duval2014CarrollNewtonGalilei, duval2014ConformalCarrollGroups,duval2014ConformalCarrollGroupsa,ciambelli2019CarrollStructuresNull} to discuss asymptotic symmetries.

    Identifying the asymptotic symmetries of spacetime could lead, via the infrared triangle, to the prediction of other aspects of soft physics. For instance, if a certain asymptotic symmetry occurs naturally in a geometric, theory-agnostic analysis, it may be related to a memory effect that is not predicted by general relativity. Metric theories of gravity may have up to six propagating modes \cite{eardley1973GravitationalWaveObservationsPRD,eardley1973GravitationalWaveObservationsPRL,will2014ConfrontationGeneralRelativity}, as opposed to only two graviton polarizations in Einstein gravity, and thus these modes could carry memory in modified gravity. By identifying such memories, one could ``bootstrap'' a theory of gravity by considering the infrared symmetries it should possess.

    The paper is structured in the following manner. Section \ref{sec: carrollian-manifolds} discusses the Carrollian structures introduced by \cite{duval2014CarrollNewtonGalilei,duval2014ConformalCarrollGroups,duval2014ConformalCarrollGroupsa}, which will be needed for our construction of asymptotic (conformal) Killing horizons. Our original contributions are presented in Secs.~\ref{sec: generalized-killing-horizons} and \ref{sec: generalized-BMS-DMP}. The former introduces the generalized notions of Killing horizons needed for the latter, which discusses the natural conformal Carroll groups of these geometric constructions and thus establishes the core results of the paper. The remaining sections explore possible applications and physical consequences. In Sec.~\ref{sec:superdilations}, we discuss the occurrence of superdilations in the symmetry group for the asymptotic conformal Killing horizons previously discussed. In Sec.~\ref{sec: outlook} we discuss some other future perspectives and we conclude in Sec.~\ref{sec:conclusion}.

    We use abstract index notation~\cite{wald1984GeneralRelativity} and  \({-}{+}{+}{+}\) metric signature convention. Lowercase Latin letters indicate abstract indices, Greek letters indicate spacetime coordinate indices, and uppercase Latin letters indicate indices in a codimension-two submanifold (with no distinction between abstract or coordinate indices in this case). The round metric on the two-sphere is denoted by \(\tensor{\gamma}{_A_B}\). We use geometrical units with \(G = c = 1\) unless otherwise specified.

\section{Carrollian Structures}\label{sec: carrollian-manifolds}
    Let us begin by discussing some properties of null submanifolds of Lorentzian spacetimes following the work of \textcite{duval2014CarrollNewtonGalilei,duval2014ConformalCarrollGroups,duval2014ConformalCarrollGroupsa}. This will later be useful to characterize the symmetry groups we are looking for.

    Firstly, we note that the induced metric $h_{ab}$ on a null submanifold \(\mathscr{N}\) of a spacetime \((M,\tensor{g}{_a_b})\) is degenerate, \ie, it always has a nontrivial kernel on the tangent space of the manifold. A way of obtaining this behavior is by considering a Lorentzian metric such as
    \begin{equation}
        \dd{s}^2 = - c^2 \dd{t}^2 + \dd{x}^2 + \dd{y}^2 + \dd{z}^2
    \end{equation}
    and taking the ``Carollian limit'' \(c \to 0\) \cite{levy-leblond1965NouvelleLimiteNonrelativiste,sengupta1966AnalogueGalileiGroup}, where \(c\) is the speed of light.

    The basic structures we would like to ask of a ``Carrollian manifold'' are the differentiable manifold itself and a metric. However, since the metric is degenerate, we will also need to introduce in the structure a vector field along the one-dimensional kernel of the metric. We thus get the following definition.
    \begin{definition}[Carrollian Structure]
        A \emph{Carrollian structure} is a triple \((\mathscr{N}, \tensor{h}{_a_b}, \tensor{n}{^a})\) formed by a three-dimensional manifold \(\mathscr{N}\), a symmetric and positive-semidefinite rank-\(2\) tensor field \(\tensor{h}{_a_b}\) with one-dimensional kernel, and a nonvanishing vector field \(\tensor{n}{^a}\) with \(\tensor{h}{_a_b}\tensor{n}{^b} = 0\) at all points.
    \end{definition}
    Due to the metric being degenerate, it is not possible to single out a preferred Levi-Civita connection. Hence, it can be useful to add a connection as part of the definition of a Carrollian structure. Nevertheless, this definition turns out to be inappropriate when discussing asymptotic symmetry groups, as it rules out some Poincaré transformations from the symmetry group at null infinity---see, for example, Ref.~\cite{aguiaralves2025LecturesBondiMetzner} for a discussion.

    The choice of including \(\tensor{n}{^a}\) in the definition of a Carrollian structure may seem arbitrary. Since it is a key aspect of our analysis (and BMS-like analyses in general), we take a moment to discuss this in more depth.

    Consider a null surface \(\mathscr{N}\) having the form \(\mathbb{R} \times \Sigma\) for some two-dimensional manifold \(\Sigma\). \(\mathscr{N}\)  is naturally endowed with a projection onto \(\Sigma\), \(\pi \colon \mathscr{N} \to \Sigma\). This projection has the important property that \(\pi^{-1}(\qty{x}) \cong F = \mathbb{R}\) for all points \(x \in \Sigma\), where \(\pi^{-1}\) denotes the preimage. Notice \(F\) could have been a more general manifold instead of \(\mathbb{R}\). The quadruple \((\mathscr{N},\pi,\Sigma,F)\) is an example of a fiber bundle with base space $\Sigma$ and fiber $F$~\cite{hamilton2017MathematicalGaugeTheory,kobayashi1963FoundationsDifferentialGeometry,kolar1993NaturalOperationsDifferential,tu2017DifferentialGeometryConnections}. At this point, we have not used a metric. However, we already have enough structure to discuss the ``vertical direction'' of \(\mathscr{N}\). We say a vector \(\tensor{\xi}{^a} \in \mathrm{T}_p\mathscr{N}\) is vertical if its pushforward to \(\mathrm{T}_{\pi(p)}\Sigma\) through \(\pi\) vanishes. Hence, we define the vertical spaces \(\mathrm{V}_p \mathscr{N}\) at each point of \(\mathscr{N}\) as \(\mathrm{V}_p \mathscr{N} = \mathrm{Ker}(\pi^*|_p)\), where \(\pi^*|_p\) is the pushforward of \(\pi\) evaluated at \(p\). Since the vertical spaces are one-dimensional by construction, they naturally yield a vector field up to normalization. Once we embed this structure in a Lorentzian spacetime, the vertical direction will be interpreted as the ``null direction''. Notice that \(\pi^{-1}(\qty{x})\) will become (the image of) a null geodesic on \(\mathscr{N}\) for each possible \(x \in \Sigma\). In this sense, \(\Sigma\) will be the space of geodesic generators of \(\mathscr{N}\).

    The existence of the vector field \(\tensor{n}{^a}\) ensures we can do this process backward. Given the three-dimensional manifold \(\mathscr{N}\), we can use the integral lines of $n^a$ to rebuild the structure of a fiber bundle. In this way, we ensure the desired structure of something that locally looks like the product of the real line with a spatial manifold. If we did not have a vector field, this could have been impossible.
        
    From a different perspective, recall that a manifold admits a Lorentzian metric if, and only if, it admits a continuous, nonvanishing line element field \cite{hawking1973LargeScaleStructure}. A line element field is an attribution of pairs of opposite vectors \((\tensor{t}{^a},-\tensor{t}{^a})\). Once the Lorentzian metric is given, there is always such a line element field that is everywhere timelike. The existence of the line element field ensures there is no topological obstruction to the existence of a Lorentzian metric\footnote{An example of a manifold that does not have such a field, and thus does not admit a Lorentzian metric, is a two-dimensional sphere (this follows from the ``hairy ball theorem'').}. The stronger requirement that there exists an everywhere nonvanishing (timelike) vector field ensures the spacetime to be time-orientable. Since Carrollian manifolds can be understood as the \(c \to 0\) limit of Lorentzian manifolds, the existence of a nonvanishing vector field can be seen as enforcing the appropriate topological behavior of the underlying manifold. 

    The definitions of asymptotically flat spacetimes and of expanding universes with cosmological horizons naturally equip (future) null infinity \(\mathscr{I}^+\) or the cosmological horizon \(\mathscr{H}^-\) with a Carrollian structure, but not in a unique way. This ambiguity is precisely the origin of the asymptotic symmetries in this formalism. To illustrate the properties and structure of Carrollian structures, let us look at how they emerge both in the asymptotically flat and cosmological cases. 

    \subsection{Asymptotically Flat Spacetimes}    
        We first consider the case of an asymptotically flat spacetime. To keep the construction of both asymptotically flat and cosmological cases similar, we work with asymptotically flat spacetimes defined in terms of the conformal compactification originally given by \textcite{penrose1963AsymptoticPropertiesFields,penrose1965ZeroRestmassFields}. We note, however, that there are more general classes of asymptotically flat spacetimes, such as those considered by \textcite{christodoulou1993GlobalNonlinearStability} and other authors \cite{bieri2007ExtensionStabilityTheorem,bieri2009ExtensionStabilityTheorem,bieri2009ExtensionsStabilityTheorem,bieri2024GravitationalWaveDisplacement}.
        
        We want an asymptotically flat spacetime to be a spacetime that ``looks Minkowskian'' at infinity. Hence, we use the following definition---see Refs. \cite{wald1984GeneralRelativity,dappiaggi2017HadamardStatesLightlike,aguiaralves2025LecturesBondiMetzner} for further discussion.
        \begin{definition}[Asymptotic Flatness]
            Let \((M,\tensor{g}{_a_b})\) be a four-dimensional Lorentzian spacetime. We assume there to be an ``unphysical spacetime'' \((\tilde{M},\tensor{\tilde{g}}{_a_b})\), a smooth embedding \(\psi\colon M \to \tilde{M}\) with open range, and a smooth function \(\Omega \colon \psi(M) \to \mathbb{R}\) such that \(\Omega > 0\) and 
            \begin{equation}
                \tensor{\tilde{g}}{_a_b}|_{\psi(M)} = \Omega^2 \psi^* \tensor{g}{_a_b},
            \end{equation}
            where \(\psi^*\) denotes the push-forward. \((M,\tensor{g}{_a_b})\) is said to be \emph{asymptotically flat at future null infinity} with conformal extension \((\tilde{M},\tensor{\tilde{g}}{_a_b})\) and conformal factor \(\Omega\) if the following conditions are met.
            \begin{enumerate}
                \item \(\psi(M)\) is the interior of a manifold with boundary \(\mathscr{I}^+\), the latter being a three-dimensional submanifold of \(\tilde{M}\) such that \(\mathscr{I}^+ \cap J^-(\psi(M)) = \varnothing\) (where the causal past is meant in \(\tilde{M}\)).
                \item \((\tilde{M},\tensor{\tilde{g}}{_a_b})\) is strongly causal in a neighborhood of \(\mathscr{I}^+\).
                \item \(\Omega\) can be smoothly extended to a function defined on \(\tilde{M}\) (the extension being denoted by \(\Omega\) as well) such that \(\Omega|_{\mathscr{I}^+} = 0\) and \(\dd{\Omega}|_{\mathscr{I}^+} \neq 0\).
                \item Given \(\tensor{n}{^a} = \tensor{\tilde{g}}{^a^b}\tensor{\tilde{\nabla}}{_b}\Omega\), there is a smooth function \(\omega \colon \tilde{M} \to \mathbb{R}\) such that \(\omega > 0\), \(\tensor{\tilde{\nabla}}{_a}\pqty{\omega^4 \tensor{n}{^a}}|_{\mathscr{I}^+} = 0\), and with the integral lines of \(\omega^{-1}\tensor{n}{^a}\) being complete.
                \item The vacuum Einstein equations hold for \((M,\tensor{g}{_a_b})\) on a neighborhood of infinity, or at least asymptotically as one approaches infinity (see Ref. \cite{wald1984GeneralRelativity} for details).
            \end{enumerate}
        \end{definition}

        These conditions are meant to ensure the hypersurface \(\mathscr{I}^+\) is similar to future null infinity in Minkowski spacetime. In particular, it holds that it has the manifold structure \(\mathbb{R} \times \mathbb{S}^2\) and it is always possible to choose coordinates so that the induced metric on \(\mathscr{I}^+\) [as a submanifold of \((\tilde{M},\tensor{\tilde{g}}{_a_b})\)] is 
        \begin{equation}\label{eq: induced-metric-null-infinity}
            \dd{\tilde{\sigma}}^2 = - 0 \dd{u}^2 + \tensor{\gamma}{_A_B}\dd{\tensor{x}{^A}}\dd{\tensor{x}{^B}},
        \end{equation}
        while
        \begin{equation}\label{eq: na-null-infinity}
            \tensor{n}{^a} = \tensor{\tilde{g}}{^a^b}\tensor{\tilde{\nabla}}{_b}\Omega|_{\mathscr{I}^+} = \tensor{\qty(\pdv{u})}{^a}.
        \end{equation}
        It is important to mention that this is a possible choice of coordinates, but the construction and definition of asymptotic flatness are coordinate-independent. There is, however, an unphysical dependence on the conformal parameter \(\Omega\).  We will typically refer to a spacetime which is ``asymptotically flat at future null infinity'' merely as an ``asymptotically flat spacetime'', for simplicity.

        For these spacetimes, the Carrollian structure at the boundary is given by \(\mathscr{I}^+\), by the induced metric on \(\mathscr{I}^+\) as a submanifold of \((\tilde{M},\tensor{\tilde{g}}{_a_b})\), and by the vector \(\tensor{n}{^a} = \tensor{\tilde{g}}{^a^b}\tensor{\tilde{\nabla}}{_b}\Omega\), where \(\Omega\) is the conformal factor. The key observation in this case is that we can still perform other conformal transformations: \(\Omega\) is not uniquely determined. If \((M,\tensor{g}{_a_b})\) is asymptotically flat with conformal extension \((\tilde{M},\tensor{\tilde{g}}{_a_b})\) and conformal parameter \(\Omega\), then the conformal parameter \(\omega\Omega\) with smooth \(\omega > 0\) on \(\tilde{M}\) would also be a valid option. This ambiguity is not capable of changing the differential structure of \(\mathscr{I}^+\) (which is always diffeomorphic to \(\mathbb{R} \times \mathbb{S}^2\)), but it can alter the induced metric \(\tensor{h}{_a_b}\) and the vector \(\tensor{n}{^a}\) according to
        \begin{subequations}
            \begin{align}
                \tensor{h}{_a_b} &\to \omega^2 \tensor{h}{_a_b}, \\
                \intertext{and}
                \tensor{n}{^a} &\to \omega^{-1} \tensor{n}{^a}.
            \end{align}
        \end{subequations}
        Hence, we must establish an equivalence relation 
        \begin{equation}\label{eq: equivalence-relation-BMS}
            (\mathscr{I}^+, \tensor{h}{_a_b}, \tensor{n}{^a}) \sim (\mathscr{I}^+, \omega^2\tensor{h}{_a_b}, \omega^{-1}\tensor{n}{^a}).
        \end{equation}
        Any transformation between Carrollian structures that preserves this equivalence relation will need to be considered a symmetry for null infinity.

        Since we can always choose coordinates so that Eqs. (\ref{eq: induced-metric-null-infinity}) and (\ref{eq: na-null-infinity}) hold, we can establish the symmetries of null infinity by studying the transformations of the form 
        \begin{equation}\label{eq: extended-BMS-transformations}
            \qty(\mathbb{R} \times \mathbb{S}^2, \tensor{\gamma}{_A_B}, \pdv{u}) \to \qty(\mathbb{R} \times \mathbb{S}^2, \omega^2 \tensor{\gamma}{_A_B}, \frac{1}{\omega}\pdv{u}).
        \end{equation}
        We will explore this in Sec.~\ref{sec: carroll-groups}. 

    \subsection{Expanding Universes with Cosmological Horizons}
        Next we discuss the case of expanding universes with cosmological horizons. As discussed in the introduction, this notion was introduced by DMP~\cite{dappiaggi2009CosmologicalHorizonsReconstruction,dappiaggi2009DistinguishedQuantumStates,dappiaggi2017HadamardStatesLightlike} to characterize a class of spacetimes that resembles the cosmological patch of de Sitter spacetime. With this paradigmatic example in mind, we may pursue a general definition of ``expanding universe with cosmological horizon''.

        \begin{definition}[Expanding Universe with Cosmological Horizon]
            We consider a four-dimensional Lorentzian spacetime \((M,\tensor{g}{_a_b})\) where \(M\) has the differential structure \(M = \mathbb{R} \times \Sigma\). \(\Sigma\) is to be interpreted as diffeomorphic to the spatial sections of \(M\). We assume there to be an unphysical spacetime \((\tilde{M},\tensor{\tilde{g}}{_a_b})\), a smooth isometric embedding \(\psi \colon M \to \tilde{M}\) with open range, a function\footnote{This function is analogous to the scale factor in FLRW cosmologies.} \(\Omega \colon M \to \mathbb{R}\) with \(\Omega > 0\), a future-directed timelike vector field \(\tensor{X}{^a}\) in \(M\), and a positive constant \(H > 0\). We will say \((M,\tensor{g}{_a_b},\Omega,\tensor{X}{^a},H)\) is an \emph{expanding universe with geodesically complete cosmological particle horizon} with extension \((\tilde{M}, \tensor{\tilde{g}}{_a_b})\) if the following conditions are met.
            \begin{enumerate}
                \item \(\psi(M)\) is the interior of a manifold with boundary \(\mathscr{H}^-\), the latter being a three-dimensional submanifold of \(\tilde{M}\) such that \(\mathscr{H}^- \cap J^+\qty(\psi(M)) = \varnothing\).
                \item The function \(\Omega\) can be extended to a smooth function on \(\tilde{M}\) (denoted by the same symbol) such that \(\Omega|_{\mathscr{H}^-} = 0\) and \(\dd{\Omega}|_{\mathscr{H}^-} \neq 0\).
                \item The vector field \(\tensor{X}{^a}\) is a conformal Killing vector field for \(\tensor{g}{_a_b}\) in a neighborhood of \(\mathscr{H}^-\) with 
                \begin{equation}\label{eq: conformal-Killing-expanding-universe}
                    \pounds_{X}\tensor{\tilde{g}}{_a_b} = -2 \tensor{X}{^c}\tensor{\tilde{\nabla}}{_c}\qty(\log \Omega) \tensor{\tilde{g}}{_a_b}.
                \end{equation}
                Furthermore, \(\tensor{X}{^c}\tensor{\tilde{\nabla}}{_c}(\log \Omega)\) tends to zero as one approaches \(\mathscr{H}^-\) without \(\tensor{X}{^a}\) tending identically to zero.
                \item \(\mathscr{H}^-\) is diffeomorphic to \(\mathbb{R} \times \mathbb{S}^2\) and the metric \(\tensor{\tilde{g}}{_a_b}\) admits the almost-Bondi form 
                \begin{equation}
                    \dd{s}^2|_{\mathscr{H}^-} = H^{-2}\qty[-2 \dd{\Omega} \dd{v} + \tensor{\gamma}{_A_B}\dd{\tensor{x}{^A}}\dd{\tensor{x}{^B}}]
                \end{equation}
                where \(v\) is an affine parameter along the geodesics spanning \(\mathscr{H}^-\). In particular, the geodesics \(v \mapsto (v,\tensor{x}{^A}) \in \mathscr{H}^-\) are complete. 
            \end{enumerate}
            We say \(\mathscr{H}^-\) is the (cosmological) particle horizon of \((M,\tensor{g}{_a_b})\), while the integral parameter of \(\tensor{X}{^a}\) is the conformal (cosmological) time. 
        \end{definition}

        This definition mimics the definition of asymptotically flat spacetimes by defining a distinguished null hypersurface (in this case, the particle horizon) by considering an extension of the spacetime. Notice, however, that a key distinction between this definition and the definition of asymptotically flat spacetimes is that we are now considering an isometric extension, not a conformal extension. This distinction will later change the symmetry group of the distinguished null hypersurface, \ie, it is the origin of the difference between the BMS and DMP groups. We notice that, at the particle horizon,
        \begin{equation}\label{eq: na-particle-horizon}
            \tensor{\tilde{g}}{^a^b}\tensor{\tilde{\nabla}}{_b} \Omega|_{\mathscr{H}^-} = - H^2 \tensor{\qty(\pdv{v})}{^a}.
        \end{equation}
        Furthermore, the induced metric at the particle horizon can be seen to be 
        \begin{equation}\label{eq: induced-metric-particle-horizon}
            \dd{\tilde{\sigma}}^2 = - 0 \dd{v}^2 + H^{-2}\tensor{\gamma}{_A_B}\dd{\tensor{x}{^A}}\dd{\tensor{x}{^B}}.
        \end{equation}
        It can also be proven that the unique smooth extension of \(\tensor{X}{^a}\) to \(\mathscr{H}^-\) has the form \cite{dappiaggi2009CosmologicalHorizonsReconstruction}
        \begin{equation}\label{eq: Xa-particle-horizon}
            \tensor{X}{^a}|_{\mathscr{H}^-} = f(\tensor{x}{^A}) \tensor{\qty(\pdv{v})}{^a}
        \end{equation}
        for some function \(f \in \mathcal{C}^{\infty}(\mathbb{S}^2)\) vanishing at most in a set with an empty interior.
    
        When dealing with expanding universes with cosmological horizons, the extended spacetime \((\tilde{M},\tensor{\tilde{g}}{_a_b})\) enforces a Carrollian structure on the cosmological horizon \(\mathscr{H}^-\), but it still includes an ambiguity. The differentiable manifold is \(\mathscr{H}^-\) (which is always diffeomorphic to \(\mathbb{R} \times \mathbb{S}^2\) by construction), the metric is the induced metric on \(\mathscr{H}^-\) as a submanifold of \(\tilde{M}\), while the vector \(\tensor{n}{^a}\) is given by \(\tensor{\tilde{\nabla}}{^a}\Omega\). 
        
        The ambiguity lies again in the function \(\Omega\), which could be replaced by \(\omega \Omega\) for a smooth function \(\omega > 0\) defined on \(\tilde{M}\). Such a transformation now preserves the induced metric, but still modifies \(\tensor{n}{^a}\) according to
        \begin{equation}
            \tensor{n}{^a} \to \omega \tensor{n}{^a}.
        \end{equation}
        We note, however, that imposing condition (\ref{eq: conformal-Killing-expanding-universe}) to $\omega \Omega$ will yield that $n^a\tilde{\nabla}_a \omega =0$ on \(\mathscr{H}^-\)~\cite{aguiaralves2025LecturesBondiMetzner}. 
        
        Hence, this time the equivalence relation is 
        \begin{equation}\label{eq: equivalence-relation-DMP}
            (\mathscr{H}^-, \tensor{h}{_a_b}, \tensor{n}{^a}) \sim (\mathscr{H}^-, \tensor{h}{_a_b}, \omega\tensor{n}{^a}),
        \end{equation}
        where $\omega$ is a smooth function on $\mathbb{S}^2$. Transformations preserving the equivalence classes of this equivalence relation are to be understood as symmetries of the cosmological horizon. 

        By picking (\ref{eq: na-particle-horizon}) and (\ref{eq: induced-metric-particle-horizon}) as a canonical choice, we see that the symmetries of the cosmological horizon should be the transformations of the form
        \begin{equation}\label{eq: DMP-transformations}
            \qty(\mathbb{R} \times \mathbb{S}^2, \frac{\tensor{\gamma}{_A_B}}{H^2}, -H^2\pdv{v}) \to \qty(\mathbb{R} \times \mathbb{S}^2, \frac{\tensor{\gamma}{_A_B}}{H^2}, -\omega H^2\pdv{v}).
        \end{equation}
        
    \subsection{BMS and DMP as Conformal Carroll Groups}\label{sec: carroll-groups}
        We are now ready to discuss the BMS and DMP groups as conformal Carroll groups. The basic approach is to study the transformations that preserve the equivalence relations \eqref{eq: equivalence-relation-BMS} and \eqref{eq: equivalence-relation-DMP}. This can be done at either the level of groups (by considering the global transformations) or at the level of Lie algebras (by working with the vector fields generating these transformations). We will choose the latter option. In Ref.~\cite{aguiaralves2025LecturesBondiMetzner}, and references therein, the global approach in both the asymptotically flat and cosmological scenarios is discussed. 
    
        \subsubsection{Asymptotically Flat Spacetimes}
            The symmetries of null infinity should be of the form given in Eq.~(\ref{eq: extended-BMS-transformations}). Since they preserve the differential structure, we take them to be locally generated by vector fields. When globally defined, they correspond to diffeomorphisms\footnote{It is important to notice that not all of the transformations we may consider are globally defined. For example, Virasoro-like superrotations are not, and thus they do not exponentiate to genuine diffeomorphisms. In this sense, Eq.~(\ref{eq: extended-BMS-transformations}) can be violated at isolated points.}.
    
            The equivalence relation on Eq.~(\ref{eq: extended-BMS-transformations}) requires the vector fields generating the transformation (let us denote them by \(\tensor{\xi}{^a}\)) to satisfy the conditions
            \begin{subequations}
                \begin{align}
                    \pounds_{\xi} \tensor{\gamma}{_a_b} &= \lambda \tensor{\gamma}{_a_b} \label{eq:lie_gamma} \\
                    \intertext{and}
                    \pounds_{\xi} \tensor{n}{^a} &= - \frac{\lambda}{2} \tensor{n}{^a}, \label{eq:lie_n}
                \end{align}
            \end{subequations}
            where \(\tensor{\gamma}{_a_b}\) is the round metric on the sphere understood as a tensor on the ambient spacetime and \(\tensor{n}{^a} = \tensor{\qty(\pdv*{u})}{^a}\). 
            
            These conditions establish differential equations for \(\tensor{\xi}{^a}\). To solve them, let us note that Lie derivatives can be evaluated with any choice of covariant derivative \cite{wald1984GeneralRelativity}. It will be convenient for us to pick the one induced by the sphere: all Christoffel symbols with an \(u\) index are taken to vanish, while the remaining ones are taken to be the same as the Levi-Civita connection on the two-sphere. With such a connection, it will be useful to  write \(\tensor{\xi}{^a}\) as
            \begin{equation}\label{eq: decomposition-xi-BMS}
                \tensor{\xi}{^a} = F \tensor{n}{^a} + \tensor{Y}{^a},
            \end{equation}
            where \(\tensor{Y}{^a}\) is a vector on the sphere and \(F\) is a function on \(\mathscr{I}^+\).
            
            Using this covariant derivative together with Eq.~(\ref{eq: decomposition-xi-BMS}), we find that Eqs.~(\ref{eq:lie_gamma}) and~(\ref{eq:lie_n}) yield
            \begin{subequations}
                \begin{gather}
                    \tensor{D}{_a}\tensor{Y}{_b} + \tensor{D}{_b}\tensor{Y}{_a} = \tensor{D}{_c}\tensor{Y}{^c} \tensor{\gamma}{_a_b} \\
                    \intertext{and}
                    F(u,\tensor{x}{^A}) = \frac{\tensor{D}{_c}\tensor{Y}{^c}}{2} u + f(\tensor{x}{^A}). 
                \end{gather}
            \end{subequations}
            Hence, \(\tensor{Y}{^a}\) is a conformal Killing vector field on the sphere, while \(f \in \mathcal{C}^{\infty}(\mathbb{S}^2)\)\footnote{We restrict attention to smooth functions because we are working with smooth manifolds. At this point, we are considering globally defined functions.}.
    
            At this point, one may compute the commutators of all Killing vector fields of this form and find that they establish a Lie algebra, as expected. There is one important caveat: which vector fields \(\tensor{Y}{^a}\) are admissible. If one restricts attention to globally defined vector fields, then they span the Lorentz algebra \(\mathfrak{so}^+(3,1)\). This is the Lie algebra for \(\mathrm{SO}^+(3,1)\), which is the Lorentz group and the (global) conformal group on the sphere. If this is the choice made for \(\tensor{Y}{^a}\), then the Killing vector fields obtained give the Lie algebra 
            \begin{equation}
                \mathfrak{bms}_4 = \mathfrak{so}^+(3,1) \loplus \mathcal{Y},
            \end{equation}
            where \(\mathcal{Y}\) is the algebra of supertranslations [the Lie algebra of \(\mathcal{C}^{\infty}(\mathbb{S}^2)\)] and \(\loplus\) denotes the semidirect sum of Lie algebras. This is the Lie algebra of the BMS group (\ref{eq: BMS-group}).
    
            A second possibility is to consider the vector fields \(\tensor{Y}{^a}\) as elements of the local conformal algebra, as one does in two-dimensional CFT \cite{belavin1984InfiniteConformalSymmetry,difrancesco1997ConformalFieldTheory,blumenhagen2009IntroductionConformalField,schottenloher2008MathematicalIntroductionConformal,[{}][{, Chap. 2.}]{polchinski1998IntroductionBosonicString},banks2003CritiquePureString}. In this case, we get the ``extended BMS algebra'' \cite{barnich2010AspectsBMSCFT,barnich2010SymmetriesAsymptoticallyFlat}
            \begin{equation}
                \mathfrak{ebms}_4 = (\mathcal{W} \oplus \mathcal{W}) \loplus \mathcal{Y},
            \end{equation}
            where \(\mathcal{W}\) is the Witt algebra and \(\oplus\) denotes the direct sum of Lie algebras. This time, \(\mathcal{Y}\) is larger than the algebra of smooth functions on the sphere, and instead corresponds to locally defined objects. Namely, to meromorphic functions on the Riemann sphere.

            As mentioned in the introduction, the new transformations obtained in going from \(\mathfrak{so}^+(3,1)\) to \(\mathcal{W} \oplus \mathcal{W}\) are known as superrotations. They are a possible way toward a holographic duality between quantum gravity in four-dimensional asymptotically flat spacetimes and two-dimensional CFTs, which justifies their interest. They also enjoy an infrared triangle \cite{cachazo2014EvidenceNewSoft,pasterski2016NewGravitationalMemories,pasterski2019ImplicationsSuperrotations}, which gives them physical meaning.
    
        \subsubsection{Expanding Universes with Cosmological Horizon}
            To obtain the DMP group, we proceed analogously. The key difference is that we now consider transformations of the form given by Eq.~(\ref{eq: DMP-transformations}). This means the generating vector fields \(\tensor{\xi}{^a}\) now satisfy
            \begin{subequations}
                \begin{align}
                    \pounds_{\xi} \tensor{\gamma}{_a_b} &= 0 \\
                    \intertext{and}
                    \pounds_{\xi} \tensor{n}{^a} &= \lambda \tensor{n}{^a},
                \end{align}
            \end{subequations}
            where \(\tensor{n}{^a} = \tensor{\qty(\pdv*{v})}{^a}\) and $\lambda$ is a smooth function on $\mathbb{S}^2$. Notice the constant \(H\) does not play any relevant role regarding the symmetry structure. 
    
            By performing a decomposition analogous to Eq.~(\ref{eq: decomposition-xi-BMS}) and once again using the connection \(\tensor{D}{_a}\) induced by the sphere we find 
            \begin{subequations}
                \begin{gather}
                    \tensor{D}{_a}\tensor{Y}{_b} + \tensor{D}{_b}\tensor{Y}{_a} = 0 \\
                    \intertext{and}
                    F(v,\tensor{x}{^A}) = g(\tensor{x}{^A}) v + f(\tensor{x}{^A}),
                \end{gather}
            \end{subequations}
            with \(f, g \in \mathcal{C}^{\infty}(\mathbb{S}^2)\). 
    
            This time, there is no ambiguity with our choice of \(\tensor{Y}{^a}\) vector fields, because there is no such thing as a ``local isometry algebra''. The \(\tensor{Y}{^a}\) vector fields must span the isometry algebra of the sphere, which is \(\mathfrak{so}(3)\). Hence, the algebra we obtain is the DMP algebra 
            \begin{equation}
                \mathfrak{dmp}_4 = \mathfrak{so}(3) \loplus (\mathcal{Y} \loplus \mathcal{Y}),
            \end{equation}
            with \(\mathcal{Y}\) standing for the algebra of smooth functions on the sphere. \(\mathfrak{dmp}_4\) can be exponentiated to the DMP group (\ref{eq: DMP-group}).

            There is no obvious way to extend \(\mathfrak{dmp}_4\) to an algebra that more closely resembles what is found in two-dimensional CFTs. Such an extension by superrotations would be of interest to the construction of a dS/CFT correspondence. Notice that manually substituting \(\mathfrak{so}(3)\) with a larger algebra would be an \emph{ad hoc} approach, and thus the physical meaning of doing so would be dubious at best.

\section{Asymptotic (Conformal) Killing Horizons}\label{sec: generalized-killing-horizons}
    Many authors have considered how to generalize the BMS group to other null surfaces, some of which we notice in the following (in chronological order).
    \begin{enumerate}
        \item Koga \cite{koga2001AsymptoticSymmetriesKilling} considered spherically symmetric Killing horizons and obtained a subgroup of \(\mathrm{O}(3) \ltimes \mathcal{C}^{\infty}(\mathbb{R} \times \mathbb{S}^2)\) which relates to the DMP group in a similar way to how the Newman--Unti group \cite{newman1962BehaviorAsymptoticallyFlat} relates to the BMS group.
        \item Duval \emph{et al.}~\cite{duval2014CarrollNewtonGalilei,duval2014ConformalCarrollGroups,duval2014ConformalCarrollGroupsa} considered the notion of ``conformal Carroll groups'', which are generalizations of the notion of conformal group of a pseudo-Riemannian manifold to the case of a Carrollian manifold (this being a particular notion of null manifold introduced by them). 
        \item \textcite{chandrasekaran2018SymmetriesChargesGeneral} studied null hypersurfaces of the form \(\mathbb{R}\times\Sigma\) embedded in a Lorentzian spacetime and argued the symmetry group should be \(\mathrm{Diff}(\Sigma) \ltimes (\mathcal{C}^{\infty}(\Sigma) \ltimes \mathcal{C}^{\infty}(\Sigma))\).
        \item \textcite{ciambelli2019CarrollStructuresNull} reformulated the work of Duval \emph{et al.}~\cite{duval2014CarrollNewtonGalilei,duval2014ConformalCarrollGroups,duval2014ConformalCarrollGroupsa} in the language of fiber bundles.
    \end{enumerate}
    There are many other works considering the behavior of symmetries and degrees of freedom at boundaries and horizons~\cite{donnay2016ExtendedSymmetriesBlack,donnay2016SupertranslationsSuperrotationsBlack,compere2019LBMS4GroupdS4,ruzziconi2020VariousExtensionsBMS,grumiller2020Horizons2020,grumiller2020SpacetimeStructureGeneric,adami2021NullBoundaryPhase}, and our list is surely not comprehensive.
    
    We do not find these constructions completely satisfactory. We believe a good definition of the notion of asymptotic symmetry groups should have the following properties:
    \begin{enumerate}
        \item The definition should be sufficiently general to discuss null infinity in asymptotically flat spacetimes and cosmological horizons in expanding universes. These are the two fiducial cases on which others should be based.
        \item In the case of asymptotically flat spacetimes, superrotations should come in the form of two copies of the Witt algebra (or an extension of this). Not only does the superrotation infrared triangle suggests these symmetries should be taken seriously, but they are also at the foundation of two-dimensional CFT.
        \item In the case of expanding universes with cosmological horizons, the DMP symmetries should be reproduced or extended. This is motivated by their success in constructing Hadamard states through holographic techniques.
    \end{enumerate}
    
    In fact, we take one step further and add another condition to these desiderata. While general relativity is the most attractive theory of gravitation so far, it is certainly incomplete. Classical general relativity and quantum mechanics have serious compatibility issues and, thus, they ought to be modified even in the most conservative scenario. Therefore, we choose to focus on spacetime symmetries depending as least as possible on the specific theory describing the spacetime. This is also inspired by the geometrical notion of (conformal) Killing vector fields not depending on the underlying theory of gravity\footnote{One could argue that defining symmetries with Killing vector fields assumes spacetime to be described by a Lorentzian manifold, as opposed to a more general geometric space. We do not challenge this point of view, but will not focus on it in this work. In an attempt to work with a general class of theories, but still keep the calculations manageable, we consider metric theories of gravity.}.

    This last requirement is particularly strong. The standard perspective on asymptotic symmetries is that one should consider a physical theory (say, general relativity), impose suitable falloff conditions on the fields, and then study the resulting symmetries. These falloff conditions should be strong enough to allow interesting quantities (such as conserved charges) to be defined, but loose enough to allow the consideration of physically interesting cases. For example, a typical derivation of the BMS group in this approach would take the following steps.
    \begin{enumerate}
        \item Write the most general metric in Bondi gauge \cite{bondi1960GravitationalWavesGeneral} in an \(1/r\) expansion, assuming the leading contribution to the metric to be of Minkowski form.
        \item Impose the Einstein equations to relate the different metric coefficients (see, \eg, Ref. \cite{barnich2010AspectsBMSCFT}).
        \item Compute which vector fields \(\tensor{\xi}{^a}\) preserve the falloff conditions previously found, \ie, which vector fields are such that \(\tensor{g}{_a_b} + \pounds_\xi \tensor{g}{_a_b}\) has the same form as the original metric \(\tensor{g}{_a_b}\) found in the previous item.
        \item Find the commutators between the vector fields obtained in the previous item to establish what is the Lie algebra of the asymptotic symmetry group.
    \end{enumerate}
    We are overlooking many subtleties in this derivation. For instance, one may choose whether to consider only vector fields that are globally defined, or allow more general meromorphic vector fields\footnote{Superrotations in the original sense of \textcite{barnich2010AspectsBMSCFT,barnich2010SymmetriesAsymptoticallyFlat} require meromorphic vector fields, which are similar to the generators of the Virasoro algebra in CFT. Other authors, such as \textcite{campiglia2014AsymptoticSymmetriesSubleading}, prefer to consider globally defined vector fields.}. 

    Notably, the construction above makes use of the Einstein equations to establish the falloffs of interest. This is justified from a field theoretic perspective, since one is studying the symmetries of a particular field theory. In a different language, the symmetries are a property of a (pre)symplectic form on some phase space, and are thus intrinsically linked to a choice of theory---a perspective particularly explored in Refs.  \cite{wald2000GeneralDefinitionConserved,gieres2023CovariantCanonicalFormulations,ruzziconi2020AsymptoticSymmetriesGauge}, for instance. Nevertheless, at least some spacetime symmetries can also be viewed as a geometrical feature. In this sense, they do not depend on the underlying theory of gravity. For example, the isometries of Minkowski spacetime do not depend on gravity being described by general relativity or a modification thereof. More importantly, Weinberg's soft graviton theorem is a universal feature of all gravitational theories consistent in the infrared. Therefore, supertranslations are theory-agnostic symmetries. To search for other extensions of general relativity, it is interesting to consider asymptotic symmetries by following an approach that depends as least as possible on the underlying theory of gravity.
 
    In order to do so, let us first note that the notion of symmetry for a cosmological horizon and for null infinity are different because they were defined in different ways. One of them admits a conformal freedom, which yields the opportunity to introduce both Lorentz transformations (as opposed to only rotations) and superrotations. The other admits only an isometry freedom, which restricts the result to the DMP group. While it is possible to extend some DMP transformations to consider their local versions (for instance, by replacing smooth functions on the sphere with meromorphic functions on the Riemann sphere), the isometries do not admit a local version.
 
    To extend this notion to other sorts of null surfaces, we will adapt the main properties of null infinity and cosmological horizons while still preserving one key aspect: they are not arbitrary null surfaces, but rather ``asymptotic (conformal) Killing horizons''. In more detail, these null surfaces have the distinguished property of being generated by a null vector field satisfying the Killing equation at the surface, although it may not be a Killing vector field elsewhere. 

    First note that, in Bondi gauge,  we can write the metric near null infinity as
    \begin{equation}\label{eq: conformal-metric-bondi-gauge-scri}
        \Omega^2 \dd{s}^2 = - \Omega^2 \dd{u}^2 - 2 \dd{u} \dd{\Omega} + \tensor{\gamma}{_A_B}\dd{\tensor{x}{^A}}\dd{\tensor{x}{^B}} + \cdots,
    \end{equation}
    where the dots denote terms subleading in \(\Omega \sim 1/r\). In this coordinate system, one possible BMS transformation is \(\tensor{\xi}{^a} = \tensor{\qty(\pdv*{u})}{^a} + \order{\Omega}\), corresponding to a time translation. Equation  (\ref{eq: conformal-metric-bondi-gauge-scri}) tells us \(\tensor{\xi}{^a}\) is causal on a neighborhood of \(\mathscr{I}^+\), but is only null at \(\mathscr{I}^+\). In this sense, \(\mathscr{I}^+\) is a horizon for \(\tensor{\xi}{^a}\). With this in mind, we propose the following definition.
    
    \begin{definition}[Asymptotic Conformal Killing Horizon]\label{def: asymptotic-conformal-killing-horizon}
        Let \((M,\tensor{g}{_a_b})\) be a \(d\)-dimensional Lorentzian spacetime. Suppose there is a manifold \(\tilde{M}\) and an embedding \(\psi \colon M \to \tilde{M}\). We say a null surface \(\mathscr{N}\) contained in the closure of \(M\) in \(\tilde{M}\) is an \emph{asymptotic conformal Killing horizon} (or ACKH) for \((M,\tensor{g}{_a_b})\) if there is a vector field \(\tensor{X}{^a}\) (which we will call asymptotic conformal Killing vector field) with the following properties.
        \begin{enumerate}
            \item \(\tensor{X}{^a}\) can be extended to \(\mathscr{N}\) with complete integral lines thereon.
            \item \(\tensor{X}{^a}\tensor{X}{_a}\) can be extended to \(\mathscr{N}\) and vanishes thereon.
            \item There is a function \(\Omega \in \mathcal{C}^{\infty}(M)\) satisfying
            \begin{equation}
                \pounds_X(\Omega^2) = -\frac{2}{d} \Omega^2 \tensor{\nabla}{_a}\tensor{X}{^a} 
            \end{equation}
            and such that \(\pounds_X (\Omega^2 \tensor{g}{_a_b})\) can be extended to \(\mathscr{N}\) vanishing thereon.
        \end{enumerate}
    \end{definition}
    
    Note that
    \begin{equation}
        \pounds_X (\Omega^2 \tensor{g}{_a_b}) = \Omega^2 \pounds_X (\tensor{g}{_a_b}) + \pounds_X (\Omega^2) \tensor{g}{_a_b}.
    \end{equation}
    This observation motivates the last requirement on the definition of an asymptotic conformal Killing horizon. Namely, \(\pounds_X (\Omega^2 \tensor{g}{_a_b}) = 0\) implies the conformal Killing equation for \(\tensor{X}{^a}\) and the metric \(\tensor{g}{_a_b}\).

    By construction, it holds that \(\mathscr{I}^+\) is an ACKH. To see this, let us first note a convenient characterization of BMS transformations given by \textcite{wald1984GeneralRelativity}. One has that the one-parameter generators of the BMS group are characterized by vector fields \(\tensor{\xi}{^a} \) with the following properties. 
        \begin{enumerate}
            \item \(\tensor{\xi}{^a}\) can be smoothly extended to future null infinity.
            \item \(\Omega^2 \pounds_{\xi} \tensor{g}{_a_b}\) can be smoothly extended to future null infinity and vanishes thereon.
            \item Two vector fields \(\tensor*{\xi}{^a_1}, \tensor*{\xi}{^a_2} \) are associated to the same generator of \(\mathrm{BMS}_4\) if, and only if, their extensions to \(\mathscr{I}^+\) coincide. 
        \end{enumerate}
     Above, \(\Omega\) is (as usual) the conformal factor, which must be chosen so that \(\tensor{\tilde{\nabla}}{_a}\tensor{\tilde{\nabla}}{_b}\Omega = 0\) at \(\mathscr{I}^+\)---a condition which is always possible \cite{wald1984GeneralRelativity}. This condition ensures \(\Omega^{-1} \tensor{\tilde{g}}{^a^b} \tensor{\tilde{\nabla}}{_a}\Omega \tensor{\tilde{\nabla}}{_b}\Omega = 0\) at \(\mathscr{I}^+\) which, in turn, can be shown to imply \(\pounds_\xi(\Omega^2)\tensor{g}{_a_b} = 0\) for \(\tensor{\xi}{^a} =\tensor{\tilde{\nabla}}{^a}\Omega\)~\cite{wald1984GeneralRelativity}.

    Notice, however, that the conformal factor in an ACKH does not play a central role. Instead, the main role is played by the vector field \(\tensor{X}{^a}\), which bears physical meaning because it is associated with an asymptotic symmetry.

    Similar results can be established for the DMP group. Namely, by definition, the cosmological horizon is defined as a conformal Killing horizon which is asymptotically Killing. Inspired by this behavior, we define an asymptotic Killing horizon.
    \begin{definition}[Asymptotic Killing Horizon]\label{def: asymptotic-killing-horizon}
        Let \((M,\tensor{g}{_a_b})\) be a Lorentzian spacetime. Consider an isometric extension \((\tilde{M},\tensor{g}{_a_b})\) of \((M,\tensor{g}{_a_b})\). We say a null surface \(\mathscr{N}\) contained in the closure of \(M\) in \(\tilde{M}\) is an \emph{asymptotic Killing horizon} (or AKH) for \((M,\tensor{g}{_a_b})\) if there is a vector field \(\tensor{X}{^a} \) (which we will call asymptotic Killing vector field) with the following properties.
        \begin{enumerate}
            \item \(\tensor{X}{^a}\) can be extended to \(\mathscr{N}\) with complete integral lines thereon.
            \item \(\tensor{X}{^a}\tensor{X}{_a}\) can be extended to \(\mathscr{N}\) and vanishes thereon.
            \item \(\pounds_X \tensor{g}{_a_b}\) can be extended to \(\mathscr{N}\) and vanishes thereon.
        \end{enumerate}
    \end{definition}

    Definitions \ref{def: asymptotic-conformal-killing-horizon} and \ref{def: asymptotic-killing-horizon} generalize the original notion of Killing horizon \cite{carter1966CompleteAnalyticExtension,carter1969KillingHorizonsOrthogonally} by considering vector fields that only satisfy the (conformal) Killing equation asymptotically. Definition \ref{def: asymptotic-killing-horizon} is a more technical version of the one employed by \textcite{koga2001AsymptoticSymmetriesKilling}, while Def. \ref{def: asymptotic-conformal-killing-horizon} concerns a different scenario in which the horizon may be located an infinite distance away, or is related to a confomorphism as opposed to an isometry. Notice the notion of conformal Killing horizon has been considered before, especially when considering the behavior of some dynamical black hole horizons \cite{dyer1979ConformalKillingHorizons,sultana2004ConformalKillingHorizons,nielsen2018ConformalKillingHorizons}.

\section{Asymptotic (Conformal) Killing Horizon Group}\label{sec: generalized-BMS-DMP}
    We may now consider the symmetry groups on top of each of these structures.

    \subsection{Asymptotic Killing Horizons}
        First we let \((M,\tensor{g}{_a_b})\) be a spacetime with an asymptotic Killing horizon \(\mathscr{N}\), the vector field associated to it being \(\tensor{X}{^a}\). The ambient spacetime naturally induces a metric \(\tensor{h}{_a_b}\) on \(\mathscr{N}\), and thus we have a pair \((\mathscr{N},\tensor{h}{_a_b})\). We still need a normal null vector. 

        In principle, we may feel tempted to work with the asymptotic Killing vector field \(\tensor{X}{^a}\). Notice, however, that our current structure \((\mathscr{N},\tensor{h}{_a_b})\) only remembers the metric structure of spacetime along spacelike directions, not along the null directions. Using \(\tensor{X}{^a}\) would give information about the Killing behavior, but the resulting Carrollian manifold would have no knowledge of what are the geodesics of the spacetime, for example. In order to preserve the geodesic generators of the horizon, we define the vector field \(\tensor{n}{^a} \) given by
        \begin{equation}
            \tensor{n}{^a} = \sigma \tensor{X}{^a},
        \end{equation}
        with \(\sigma\) satisfying
        \begin{equation}\label{eq: sigma-eq}
            \pounds_X \sigma = \kappa,
        \end{equation}
        where \(\kappa\) is the inaffinity for \(\tensor{X}{^a}\): \(\tensor{X}{^b}\tensor{\nabla}{_b}\tensor{X}{^a} = \kappa \tensor{X}{^a}\). It follows that
        \begin{equation}
            \tensor{n}{^a}\tensor{\nabla}{_a}\tensor{n}{^b} = 0.
        \end{equation}
        The integral lines of \(\tensor{n}{^a}\) are then not only geodesics in the ambient spacetime but affinely parameterized geodesics. In this sense, \(\tensor{n}{^a}\) is chosen in a manner that is compatible with the ambient metric of the spacetime. 

        We have thus constructed a triple \((\mathscr{N},\tensor{h}{_a_b},\tensor{n}{^a})\), which is a Carrollian structure associated with the asymptotic Killing horizon. We will assume, from now on, that $\mathscr{N}$ has no compact vertical directions. While \((\mathscr{N},\tensor{h}{_a_b})\) is fixed (the induced metric is unambiguous, and so is the manifold \(\mathscr{N}\)), \(\tensor{n}{^a}\) is defined up to a ``gauge redundancy''. Equation (\ref{eq: sigma-eq}) admits a gauge-like freedom in the initial conditions, which can depend on spacelike variables. We thus get the equivalence relation
        \begin{equation}
            (\mathscr{N},\tensor{h}{_a_b},\tensor{n}{^a}) \sim (\mathscr{N},\tensor{h}{_a_b}, \sigma\tensor{n}{^a}),
        \end{equation}
        where \(\pounds_v \sigma = 0\) for any vertical vector \(\tensor{v}{^a}\).

        In a local description, this equivalence relation is translated into the Killing-like equations
        \begin{subequations}
            \begin{align}
                \pounds_{\xi} \tensor{h}{_a_b} &= 0, \\
                \pounds_{\xi} \tensor{n}{^a} &= \lambda \tensor{n}{^a},
            \end{align}
        \end{subequations}
        where \(\lambda \in \mathcal{C}^{\infty}(\Sigma)\) (with \(\Sigma\) the space of integral lines of \(\tensor{n}{^a}\), \ie, the transverse space of \(\mathscr{N}\)). The general vector field \(\tensor{\xi}{^a}\) satisfying these equations is given by
        \begin{equation}\label{eq: killing-like-AKH}
            \tensor{\xi}{^a} = \tensor{Y}{^a} + (g(\tensor{x}{^A}) u + f(\tensor{x}{^A}))\tensor{n}{^a},
        \end{equation}
        where \(\tensor{n}{^a}\tensor{\nabla}{_a}u = 1\), \(\{\tensor{x}{^A}\}\) are coordinates for \(\Sigma\), and \(\tensor{Y}{^a}\) is a Killing vector field for \((\Sigma,\tensor{h}{_a_b})\). 

       It is important to notice that the Riemannian manifold \((\Sigma,\tensor{h}{_a_b})\) is well-defined up to isometries. Consider two arbitrary smooth functions  \(s, s' \colon \Sigma \to \mathscr{N}\) such that, for each \(x \in \Sigma\), \(s(x)\) and \(s'(x)\) are points in the generator \(x\) of \(\mathscr{N}\).  Then the surfaces \((s(\Sigma),\tensor{h}{_a_b})\) and  \((s'(\Sigma),\tensor{h}{_a_b})\) must be isometric. This follows from the Killing equation on the hypersurface \(\mathscr{N}\). On the spacetime \(M\), we have
        \begin{equation}
            \pounds_{\zeta X} \tensor{g}{_a_b} = \tensor{\nabla}{_(_a}\zeta\tensor{X}{_b_)} + \zeta\tensor{\nabla}{_(_a}\tensor{X}{_b_)}.
        \end{equation}
        Once this is pulled back to \(\mathscr{N}\), we find \(\pounds_{\zeta X} \tensor{h}{_a_b} = 0\) for any function \(\zeta\), because the pullback of \(\tensor{X}{_a}\) vanishes on \(\mathscr{N}\) (as it is given by $h_{ab}X^b$) and \(\tensor{\nabla}{_(_a}\tensor{X}{_b_)}=\pounds_X \tensor{g}{_a_b}\) vanishes on \(\mathscr{N}\) due to the (asymptotic) Killing equation. Hence, the flow along the integral lines of any vertical vector field is isometric. It follows that all surfaces \((s(\Sigma),\tensor{h}{_a_b})\) are isometric, irrespective of the particular choice of \(s\). In particular, the statement that ``\(\tensor{Y}{^a}\) is a Killing vector field for \((\Sigma,\tensor{h}{_a_b})\)'' is well-defined.
        
        With this in mind, we see the group of asymptotic symmetries at the horizon is
        \begin{equation}\label{eq: AKH-group}
            G_{\text{AKH}} = \mathrm{Isom}(\Sigma) \ltimes \qty(\mathcal{C}^{\infty}(\Sigma) \ltimes \mathcal{C}^{\infty}(\Sigma)), 
        \end{equation}
        where \(\mathrm{Isom}(\Sigma)\) is the isometry group of \((\Sigma, \tensor{h}{_a_b})\). This generalizes the DMP group for other possible choices of cross-sections. As an example, one may consider the Rindler horizon in Minkowski spacetime, in which case \(\Sigma = \mathbb{R}^2\). 

        This result should be contrasted to the group obtained by \textcite{koga2001AsymptoticSymmetriesKilling}, which resembles \(\mathrm{O}(3) \ltimes \mathcal{C}^{\infty}(\mathbb{R} \times \mathbb{S}^2)\). Koga considered spherically symmetric Killing horizons, which means we should take \(\Sigma = \mathbb{S}^2\) on Eq. \eqref{eq: AKH-group}. The difference between \(\mathrm{Isom}(\mathbb{S}^2) = \mathrm{SO}(3)\) [as in Eq. \eqref{eq: AKH-group}] or \(\mathrm{O}(3)\) (as given by \textcite{koga2001AsymptoticSymmetriesKilling}) merely reflects our choice of neglecting the discrete parity transformations. The difference between Koga's \(\mathcal{C}^{\infty}(\mathbb{R} \times \mathbb{S}^2)\) factor and the \(\mathcal{C}^{\infty}(\mathbb{S}^2) \ltimes \mathcal{C}^{\infty}(\mathbb{S}^2)\) factor on Eq. \eqref{eq: AKH-group} is due to our choice of preserving the vertical direction in the sense of imposing \(\pounds_\xi \tensor{n}{^a} = \lambda \tensor{n}{^a}\). Koga does not make this imposition, and as a consequence Eq. \eqref{eq: killing-like-AKH} becomes
        \begin{equation}
            \tensor{\xi}{^a} = \tensor{Y}{^a} + F(u,\tensor{x}{^A}) \tensor{n}{^a},
        \end{equation}
        with \(F \in \mathcal{C}^{\infty}(\mathbb{R} \times \mathbb{S}^2)\)\footnote{More precisely, Koga considers \(F \in \mathcal{C}^{\infty}(\mathbb{R} \times \mathbb{S}^2)\) such that \(u \mapsto F(u,\tensor{x}{^A})\) is a diffeomorphism with strictly positive derivative.}. This choice resembles the Newman--Unti group \cite{newman1962BehaviorAsymptoticallyFlat}, while our approach reproduces the results by DMP \cite{dappiaggi2009CosmologicalHorizonsReconstruction,dappiaggi2009DistinguishedQuantumStates,dappiaggi2017HadamardStatesLightlike}. Since the ``cosmological de Sitter group'' has the structure given on Eq. \eqref{eq: cosmological-de-sitter-group}, we believe this choice to be more natural. 

    \subsection{Asymptotic Conformal Killing Horizons}\label{subsec: ACKH}
        Suppose now that the spacetime \((M,\tensor{g}{_a_b})\) has an asymptotic conformal Killing horizon \(\mathscr{N}\) with vector field \(\tensor{X}{^a}\). There is an (ambiguous) choice of conformal factor \(\Omega\) such that, for \((\tilde{M},\Omega^2\tensor{g}{_a_b})\), \(\mathscr{N}\) is an asymptotic Killing horizon. If we proceed as in the last case, we get a Carrollian structure. Namely, the Carrollian structure is \((\mathscr{N},\tensor{\tilde{h}}{_a_b},\tensor{\tilde{n}}{^a})\), where \(\tensor{\tilde{h}}{_a_b}\) is the metric induced by \(\Omega^2 \tensor{g}{_a_b}\) on \(\mathscr{N}\) and \(\tensor{\tilde{n}}{^a}\) is a geodesic generator obtained from the asymptotic conformal Killing vector field \(\tensor{X}{^a}\) by considering the metric \(\Omega^2 \tensor{g}{_a_b}\). Notice, in particular, that
        \begin{equation}
            \tensor{\tilde{n}}{^a}\tensor{\tilde{\nabla}}{_a}\tensor{\tilde{n}}{^b} = 0,
        \end{equation}
        where the covariant derivative with a tilde refers to the metric \(\Omega^2 \tensor{g}{_a_b}\), not to the original metric \(\tensor{g}{_a_b}\).

        Any other choice of conformal factor \(\Omega\) with \(\pounds_X \Omega = 0\) would yield equally acceptable results. For this reason, we get the equivalence relation
        \begin{equation}\label{eq: equivalence-relation-ACKH}
            (\mathscr{N},\tensor{\tilde{h}}{_a_b},\tensor{\tilde{n}}{^a}) \sim (\mathscr{N},\omega^2 \tensor{\tilde{h}}{_a_b},\sigma \tensor{\tilde{n}}{^a})
        \end{equation}
        for \(\omega, \sigma \in \mathcal{C}^{\infty}(\Sigma)\). As before, \(\Sigma\) is the cross section of \(\mathscr{N}\). Notice there is a key difference between Eq. \eqref{eq: equivalence-relation-ACKH} and the analogous expression for asymptotically flat spacetimes, Eq. \eqref{eq: equivalence-relation-BMS}. Namely, when working with asymptotically flat spacetimes it is common to impose \(\sigma = \omega^{-1}\) (equivalently, to pick \(\tensor{\tilde{n}}{^a} = \tensor{\tilde{g}}{^a^b}\tensor{\tilde{\nabla}}{_b}\Omega\)). We see no clear reason to make this imposition, and hence allow \(\omega\) and \(\sigma\) to be determined independently\footnote{Notice that other authors, such as \textcite{ciambelli2019CarrollStructuresNull}, preferred to impose a relation between the conformal freedom on the metric and on the normal vector to preserve Weyl invariance.}.

        At the level of vector fields, we get the equations
        \begin{subequations}\label{eq: killing-equations-ACKH}
            \begin{align}
                \pounds_{\xi} \tensor{\tilde{h}}{_a_b} &= \mu \tensor{\tilde{h}}{_a_b}, \\
                \pounds_{\xi} \tensor{\tilde{n}}{^a} &= \lambda \tensor{\tilde{n}}{^a},
            \end{align}
        \end{subequations}
        from which we get the vector field 
        \begin{equation}\label{eq: killing-like-ACKH}
            \tensor{\xi}{^a} = \tensor{Y}{^a} + (g(\tensor{x}{^A}) u + f(\tensor{x}{^A}))\tensor{n}{^a},
        \end{equation}
        where \(\tensor{Y}{^a}\) is now a \emph{conformal} Killing vector field for \(\Sigma\). 
        
        Using the same argument we used in the isometric case, we find the the different slices \((\Sigma,\tensor{\tilde{h}}{_a_b})\) are confomorphic. The associated symmetry group is
        \begin{equation}\label{eq: ACKH-group}
            G_{\text{ACKH}} = \mathrm{Conf}(\Sigma) \ltimes \qty(\mathcal{C}^{\infty}(\Sigma) \ltimes \mathcal{C}^{\infty}(\Sigma)), 
        \end{equation}
        where \(\mathrm{Conf}(\Sigma)\) is the conformal group of \((\Sigma,\tensor{h}{_a_b})\). This generalizes the BMS group and includes new transformations in the form of superdilations (the first \(\mathcal{C}^{\infty}(\mathbb{S}^2)\) factor in \(G_{\text{ACKH}}\)).

    \subsection{Non-Triviality of the Notion of Asymptotic (Conformal) Killing Horizon}
        Our definition of asymptotic (conformal) Killing horizons is meant to be as general as possible. It is thus important to verify it is not trivial. In other words, are all null hypersurfaces A(C)KHs (thus rendering the notion of A(C)KH superfluous), or does the definition actually restrict the class of hypersurfaces we are studying?

        Let us start with the isometric case. Not all null hypersurfaces are AKHs. Indeed, consider a null hypersurface \((\mathscr{N},\tensor{h}{_a_b},\tensor{n}{^a})\) embedded in the Lorentzian spacetime \((M,\tensor{g}{_a_b})\). Consider the extrinsic curvature
        \begin{equation}
            \tensor{K}{_a_b} = - \frac{1}{2}\pounds_n \tensor{h}{_a_b} = - \tensor{h}{_a^c} \tensor{h}{_b^d} \tensor{\nabla}{_(_c}\tensor{n}{_d_)}.
        \end{equation}
        If we are working with an asymptotic Killing horizon, we know \(\tensor{n}{^a} = \sigma \tensor{X}{^a}\) for some asymptotic Killing vector field \(\tensor{X}{^a}\). Hence, 
        \begin{equation}
            \tensor{K}{_a_b} = -\tensor{h}{_a^c} \tensor{h}{_b^d} \tensor{X}{_(_c}\tensor{\nabla}{_d_)}\sigma=0.
        \end{equation}
         Geometrically, this is to be interpreted as the cross sections \(\Sigma\) being isometric when transported along the flow of the vector \(\tensor{n}{^a}\).
        
        For an ACKH, the same argument shows the extrinsic curvature is proportional to the metric on the null hypersurface. This means the geodesics spanning the null hypersurface are shearless when considered in the ambient spacetime, although they are allowed to expand. The cross sections \(\Sigma\) are thus conformally related when transported along a vertical field, as previously claimed. \textcite{ciambelli2019CarrollStructuresNull} impose a shearless condition in their derivation of the BMS group when formulating Carrollian structures in the language of fiber bundles. In their analysis, this assumption enters at a technical level to allow the solution of a system of differential equations. Notice our construction naturally interprets this condition in terms of the existence of an asymptotic (conformal) Killing vector field.

        Notice that, in four dimensions, the shearless condition is stronger than requiring different slices to be confomorphic. If a pair of two-dimensional Riemannian manifolds is diffeomorphic, then they are confomorphic. This follows from all Riemannian metrics in two dimensions being locally conformal to the plane. Hence, a null hypersurface with diffeomorphic cross sections will have confomorphic cross sections in \(d=4\), even if it is not shearless.

\section{Superdilations}
\label{sec:superdilations}
    Perhaps the most striking result of our analysis is the occurrence of superdilations in the symmetry group for asymptotic conformal Killing horizons. We now discuss this result and its consequences in more detail. 

    \subsection{Where do they come from?}
        The origin of superdilations in \(G_{\text{ACKH}}\) is the conformal freedom in \(\tensor{n}{^a}\) as present in Eqs. (\ref{eq: equivalence-relation-ACKH}) and (\ref{eq: killing-equations-ACKH}). This freedom naturally occurs in expanding universes with cosmological horizons, but is prohibited in the traditional analysis of asymptotically flat spacetimes because \(\tensor{n}{^a}\) is taken to be \(\tensor{\tilde{\nabla}}{^a}\Omega\). This constrains the conformal freedom in \(\tensor{n}{^a}\) and forces it to be dictated by the conformal transformations of the metric. Since \(\Omega\) is an unphysical parameter from the start, it is unclear why it should be used to dictate the normal vector \(\tensor{n}{^a}\) when other options are available---for instance, our approach in which the normal vector is any geodesic generator normal to the surface. 

        These sorts of transformations were hinted at before, but---to our knowledge---never seriously considered to have a similar status to the BMS transformations. Let us briefly discuss some of these occurrences and how our proposal differs from them.
        
        The first occurrence of transformations resembling superdilations would be in the Newman--Unti group \cite{newman1962BehaviorAsymptoticallyFlat}, which allows transformations with a more general form for the expression \eqref{eq: killing-like-ACKH}. One would have 
        \begin{equation}
            \tensor{\xi}{^a} = \tensor{Y}{^a} + F(u,\tensor{x}{^A}) \tensor{n}{^a},
        \end{equation}
        where for each fixed \(\tensor{x}{^A}\) the function \(u \mapsto F(u,\tensor{x}{^A})\) is a diffeomorphism with strictly positive derivative. Our construction restricts \(F\) to have the form 
        \begin{equation}
            F(u,\tensor{x}{^A}) = g(\tensor{x}{^A}) u + f(\tensor{x}{^A}),
        \end{equation}
        which naturally arises when enforcing the appropriate scaling behavior of the null vector \(\tensor{n}{^a}\). Notice, thus, that equipping a null manifold with a null vector is an important restriction. 

        In their seminal papers introducing the concept of superrotations, \textcite{barnich2010AspectsBMSCFT,barnich2010SymmetriesAsymptoticallyFlat} also described the Newman--Unti transformations. Their construction worked in a generalized form of Bondi gauge and ultimately discarded all Newman--Unti transformations lying outside of the BMS group through the use of a modified Lie bracket. 
        
        In more detail, they considered a metric with the form 
        \begin{multline}\label{eq: Barnich-Troessaert-Ansatz}
            \dd{s}^2 = - (2 r \dot{\varphi} + e^{-2\varphi} - e^{-2\varphi}\tensor{D}{_A}\tensor{D}{^A}\varphi)\dd{u}^2 \\ - 2 \dd{u} \dd{r} + r^2 e^{2 \varphi} \tensor{\gamma}{_A_B}\dd{\tensor{x}{^A}}\dd{\tensor{x}{^B}} + \cdots,
        \end{multline}
        where the trailing dots stand for subleading terms to ensure asymptotic flatness in a suitable sense and \(\dot{\varphi}\) is the derivative of \(\varphi\) with respect to \(u\). \(\tensor{D}{_A}\) is the Levi-Civita connection on the round sphere. It will be convenient to define \(\tensor{\bar{\gamma}}{_A_B} = e^{2\varphi}\tensor{\gamma}{_A_B}\) and denote its Levi-Civita connection by \(\tensor{\bar{D}}{_A}\). 
        
       By analyzing transformations in which the Ansatz \eqref{eq: Barnich-Troessaert-Ansatz} is preserved up to conformal rescalings of \(\varphi\) by a function \(\omega(u,\tensor{x}{^A})\), Barnich and Troessaert find that the (bulk) vector fields generating them are given by
        \begin{subequations}\label{eq: NU-vector-field-bondi-gauge}
        \begin{multline}
            \tensor*{\xi}{^a} = f \tensor{\qty(\pdv{u})}{^a} +\qty[\tensor{Y}{^A}  -\frac{\tensor{\bar{\gamma}}{^A^B}\tensor{\bar{D}}{_B}f}{r}]\tensor{\qty(\pdv{\tensor{x}{^A}})}{^a} \\ + \qty[- r \qty(f \dot{\varphi} + \frac{1}{2}\psi - \omega) + \frac{1}{2} \tensor{\bar{D}}{^A}\tensor{\bar{D}}{_A}f]\tensor{\qty(\pdv{r})}{^a},
        \end{multline}
        where 
          \begin{gather}
            \psi(\tensor{x}{^A}) = \tensor{\bar{D}}{_A}\tensor{Y}{^A}, \\
            f(u,\tensor{x}{^A}) = e^{\varphi}T + \frac{1}{2}e^{\varphi}\int_0^u e^{-\varphi} \psi \dd{u'} - e^{\varphi}\int_0^u e^{-\varphi} \omega \dd{u'},
        \end{gather}
        \(T \in \mathcal{C}^{\infty}(\mathbb{S}^2)\), \(\tensor{Y}{^A}\) is a conformal Killing vector field on the sphere, and \(\omega \in \mathcal{C}^{\infty}(\mathbb{R} \times \mathbb{S}^2)\).
        \end{subequations}
        Approaching null infinity, these vector fields asymptote to
        \begin{equation}\label{eq: NU-vector-field-bondi-gauge-scri}
            \tensor*{\xi}{^a}|_{\mathscr{I}^+} = f \tensor{\qty(\pdv{u})}{^a} +\tensor{Y}{^A} \tensor{\qty(\pdv{\tensor{x}{^A}})}{^a}.
        \end{equation}
        Then, by considering the three vector fields 
        \begin{align}
            \tensor*{\xi}{^a_T} &= e^{\varphi}T \tensor{\qty(\pdv{u})}{^a}, \\
            \tensor*{\xi}{^a_Y} &= \frac{1}{2}e^{\varphi}\int_0^u e^{-\varphi} \tensor{\bar{D}}{_A}\tensor{Y}{^A} \dd{u'} \tensor{\qty(\pdv{u})}{^a} +\tensor{Y}{^A} \tensor{\qty(\pdv{\tensor{x}{^A}})}{^a} , \\
            \tensor*{\xi}{^a_\omega} &= - e^{\varphi}\int_0^u e^{-\varphi} \omega \dd{u'} \tensor{\qty(\pdv{u})}{^a}.
        \end{align}
     they found that \(\tensor*{\xi}{^a_T}\) and \(\tensor*{\xi}{^a_Y}\) yield the BMS algebra. One can now compute the commutators between these ``old''  vector fields,  \(\tensor*{\xi}{^a_T}\) and \(\tensor*{\xi}{^a_Y}\), and the new vector fields \(\tensor*{\xi}{^a_\omega}\). For our purposes, the most relevant one is
        \begin{multline}\label{eq: commutator-lorentz-superdilation}
            \tensor{\comm{\tensor{\xi}{_Y}}{\tensor{\xi}{_\omega}}}{^a} = - \tensor{Y}{^A}\tensor{\partial}{_A}\varphi e^{\varphi}\int_0^u e^{-\varphi} \omega \dd{u'} \tensor{\qty(\pdv{u})}{^a} \\ + e^{\varphi}\int_0^u \tensor{Y}{^A}\tensor{\partial}{_A}\varphi e^{-\varphi} \omega \dd{u'} \tensor{\qty(\pdv{u})}{^a} \\ - e^{\varphi}\int_0^u e^{-\varphi} \tensor{Y}{^A}\tensor{\partial}{_A}\omega \dd{u'} \tensor{\qty(\pdv{u})}{^a}.
        \end{multline}
        Notice the commutators depend on the background value of \(\varphi\), and in this sense fail to be universal. \textcite{barnich2010AspectsBMSCFT} addressed this by working with a modified Lie bracket, which takes into consideration the fact that the change in background due to a diffeomorphism also changes the vector fields themselves. 
        
        In our proposal, there is a key difference. As one can tell by comparing Eqs. \eqref{eq: killing-like-ACKH}, \eqref{eq: NU-vector-field-bondi-gauge}, and \eqref{eq: NU-vector-field-bondi-gauge-scri}, the analysis based on ACKHs restricts attention to the case with \(\dot{\varphi} = 0\) (and, consequently, \(\dot{\omega} = 0\) as well). Hence, Eq. \eqref{eq: NU-vector-field-bondi-gauge} becomes
        \begin{subequations}\label{eq: ACKH-vector-field-bondi-gauge}
        \begin{multline}
            \tensor*{\xi}{^a} = f \tensor{\qty(\pdv{u})}{^a} +\qty[\tensor{Y}{^A}  -\frac{\tensor{\bar{\gamma}}{^A^B}\tensor{\bar{D}}{_B}f}{r}]\tensor{\qty(\pdv{\tensor{x}{^A}})}{^a} \\ + \qty[- r \qty(\frac{1}{2}\tensor{\bar{D}}{_A}\tensor{Y}{^A} - \omega) + \frac{1}{2} \tensor{\bar{D}}{^A}\tensor{\bar{D}}{_A}f]\tensor{\qty(\pdv{r})}{^a},
        \end{multline}
        where \(T, \omega \in \mathcal{C}^{\infty}(\mathbb{S}^2)\) and \(\tensor{Y}{^A}\) is a conformal Killing vector field on the sphere. This time we have the definition
        \begin{equation}
            f(u,\tensor{x}{^A}) = e^{\varphi}T + \frac{u}{2} \tensor{\bar{D}}{_A}\tensor{Y}{^A} - \omega u.
        \end{equation}
        \end{subequations}
        The vector field obtained at \(\mathscr{I}^+\) is then of the form \eqref{eq: killing-like-ACKH}. Finally, the (previously problematic) commutator \eqref{eq: commutator-lorentz-superdilation} is now simply
        \begin{equation}
            \tensor{\comm{\tensor{\xi}{_Y}}{\tensor{\xi}{_\omega}}}{^a} = \tensor*{\xi}{^a_{\tensor{Y}{^A}\tensor{\partial}{_A}\omega}},
        \end{equation}
        with no specific references to the background. There is no need to modify the Lie bracket.
        
        A third occurrence of superdilations in the literature is in the work of \textcite{ciambelli2019CarrollStructuresNull}. As mentioned in Sec. \ref{subsec: ACKH}, \textcite{ciambelli2019CarrollStructuresNull} chose to eliminate the freedom between the two equations \eqref{eq: killing-equations-ACKH} by imposing a relation between \(\mu\) and \(\lambda\) to preserve Weyl invariance. This removes the superdilations and recovers the standard BMS group. 

        Finally, \textcite{freidel2021WeylBMSGroup} considered very similar symmetries in a slightly different setting. Namely, they consider a group (named the Weyl BMS group) given by 
        \begin{equation}
            \mathrm{BMSW}_4 = (\mathrm{Diff}(\mathbb{S}^2) \ltimes \mathcal{C}^{\infty}(\mathbb{S}^2)) \ltimes \mathcal{C}^{\infty}(\mathbb{S}^2).
        \end{equation}
        Notice the change in the parentheses when writing the iterated semidirect product. Note also that, in this case, one is considering \(\mathrm{Diff}(\mathbb{S}^2)\) as opposed to \(\mathrm{Conf}(\mathbb{S}^2)\). In Ref. \cite{freidel2021WeylBMSGroup}, the extra factor of \(\mathcal{C}^{\infty}(\mathbb{S}^2)\) can be used to recover the local conformal transformations. Hence, Virasoro-like superrotations come at the expense of superdilations, not alongside them.

    \subsection{Why were they ignored?}
        BMS transformations are typically defined (in a conformal extension formalism) by requiring that \(\Omega^2 \pounds_\xi \tensor{g}{_a_b}\) vanishes as one approaches infinity, where \(\Omega\) is the conformal factor. Our definition of ACKH used the fact that
        \begin{equation}
            \pounds_\xi (\Omega^2 \tensor{g}{_a_b}) = \Omega^2 \pounds_\xi \tensor{g}{_a_b} + \tensor{g}{_a_b} \pounds_{\xi} \Omega^2
        \end{equation}
        to provide a definition based on requiring that \(\pounds_\xi (\Omega^2 \tensor{g}{_a_b})\) vanishes at the horizon. The two definitions must coincide if  \(\tensor{g}{_a_b} \pounds_{\xi} \Omega^2\) vanishes at the horizon, but this need not be the case. 

        In fact, superdilations violate this condition. Using the construction employed by \textcite{barnich2010AspectsBMSCFT}, we see a generic superdilation of the form 
        \begin{multline}
            \tensor*{\xi}{^a} = - \omega u \tensor{\qty(\pdv{u})}{^a} + \frac{u \tensor{\bar{\gamma}}{^A^B}\tensor{\bar{D}}{_B}\omega}{r} \tensor{\qty(\pdv{\tensor{x}{^A}})}{^a} \\ + \qty[\omega r - \frac{u}{2} \tensor{\bar{D}}{^A}\tensor{\bar{D}}{_A}\omega]\tensor{\qty(\pdv{r})}{^a}
        \end{multline}
        will act on, say, the Minkowski metric such that 
        \begin{equation}
            \lim_{r \to +\infty} \frac{1}{r^2} \pounds_{\xi} \tensor{\eta}{_A_B} = 2 \omega \tensor{\gamma}{_A_B},
        \end{equation}
        with the action on all other components vanishing. Using \(\Omega = 1/r\), we thus see \(\Omega^2 \pounds_\xi \tensor{g}{_a_b}\) does not vanish, but rather is proportional to the induced metric. Hence, superdilations are asymptotic confomorphisms, not asymptotic isometries. Similar ideas were explored by \textcite{haco2017ConformalBMSGroup}, but the new transformations they introduced do not resemble superdilations.

        While one may argue isometries should hold a higher standard than confomorphisms---which is certainly true for the bulk---this seems artificial at the conformal boundary. Null infinity is, by principle, constructed through conformal methods, and hence it seems natural to allow it to enjoy conformal symmetries. A further argument for considering these transformations seriously is their role in expanding universes with cosmological horizons. Considering the freedom in scaling \(\tensor{n}{^a} \to \lambda \tensor{n}{^a}\) in horizons defined isometrically is valid and useful, as attested by the DMP construction of Hadamard states \cite{dappiaggi2009CosmologicalHorizonsReconstruction,dappiaggi2009DistinguishedQuantumStates,dappiaggi2017HadamardStatesLightlike}. Therefore, it seems arbitrary to restrict this freedom in the (in principle less stringent) conformal case. 
        
        One may then wonder why focus on superdilations instead of the larger Newman--Unti group, which has similar transformations properties \cite{barnich2010AspectsBMSCFT}. The first reason is that superdilations are the particular case of Newman--Unti transformations in which one does not need to modify the Lie bracket at infinity, and thus they are ``special'' relative to other reparameterizations of the coordinate at infinity. The reason for this can be traced back to the way we defined A(C)KHs in section \ref{sec: generalized-killing-horizons}. We specifically chose to define the normal null vector \(\tensor{n}{^a}\) of an A(C)KH so that it was affinely parameterized on the ambient spacetime. In this sense, the hypersurface ``remembered'' the metric along the null direction. This structure turns out to be fruitful because it naturally restricted the Newman--Unti transformations to geodesic-preserving transformations. The same argument had been used before by DMP in the cosmological case \cite{dappiaggi2009CosmologicalHorizonsReconstruction,dappiaggi2009DistinguishedQuantumStates}.

        A second reason is that the Newman--Unti group is particularly involved from the point of view of Lie theory, and does not admit a Lie group structure \cite{prinz2022LieTheoryAsymptoticNU}. The ACKH group, on the other hand, is a simple generalization of the BMS group by means of an additional semidirect product. The BMS group does have a well-behaved Lie group structure \cite{prinz2022LieTheoryAsymptoticBMS}, and we thus may expect the ACKH group to be an improvement of the Newman--Unti group in this regard.

    \subsection{Why should they be considered?}
        In spite of these arguments, definitions in mathematics (and often in theoretical physics) are subjective. The true question is not whether we can define the group of asymptotic symmetries so that it includes superdilations (we can), but rather whether this perspective is useful. 
        
\begin{figure}
            \centering
            \begin{tikzpicture}[>=Stealth,scale=0.5]
                \begin{scope}
                    \draw[thick,accent!50!white] (0,0) ellipse (1 and 2);
                    \draw[thick,accent] (0,0) ellipse (2 and 1);
                    \draw[thick,->] (0,1.9) -- (0,1.1);
                    \draw[thick,->] (0,-1.9) -- (0,-1.1);
                    \draw[thick,<-] (1.9,0) -- (1.1,0);
                    \draw[thick,<-] (-1.9,0) -- (-1.1,0);
                \end{scope}
    
                \begin{scope}[shift={(7,0)},rotate=45]
                    \draw[thick,accent!50!white] (0,0) ellipse (1 and 2);
                    \draw[thick,accent] (0,0) ellipse (2 and 1);
                    \draw[thick,->] (0,1.9) -- (0,1.1);
                    \draw[thick,->] (0,-1.9) -- (0,-1.1);
                    \draw[thick,<-] (1.9,0) -- (1.1,0);
                    \draw[thick,<-] (-1.9,0) -- (-1.1,0);
                \end{scope}
    
                \begin{scope}[shift={(3.5,-5)},rotate=45]
                    \draw[thick,accent!50!white] (0,0) circle (1);
                    \draw[thick,accent] (0,0) circle (2);
                    \draw[thick,<-] (0,1.9) -- (0,1.1);
                    \draw[thick,<-] (0,-1.9) -- (0,-1.1);
                    \draw[thick,<-] (1.9,0) -- (1.1,0);
                    \draw[thick,<-] (-1.9,0) -- (-1.1,0);
                \end{scope}
            \end{tikzpicture}
            \caption{Gravitational waves in general relativity modify the shape of an array of inertial detectors by causing shears, as illustrated on the top for the \(+\) and \(\times\) polarizations of the graviton. A breathing mode in modified gravity can cause expansions and contractions of the array, as depicted on the bottom.}
            \label{fig: GW-shear-or-not-shear}
        \end{figure}
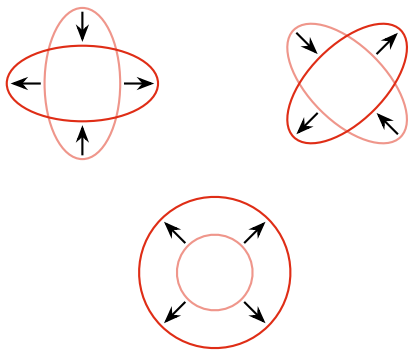

        Superdilations appear to pass this test. Recalling the infrared triangle, it seems natural to speculate that superdilations are related to a memory effect and to a soft theorem. Since superdilations generalize standard dilations, we may intuit that they should be related to a permanent expansion or contraction effect on an array of inertial detectors upon the passage of a gravitational wave. In general relativity, this is not possible. General-relativistic gravitational waves have transverse and traceless polarizations, which mean they are able to cause shears on congruences of geodesics, but not expansions. From a different perspective, general-relativistic gravitational waves are codified in the Weyl tensor, which does not cause expansion or contraction of congruences of geodesics at leading order. Nevertheless, other theories of gravity can codify radiation in the Ricci tensor. In such theories, one possible gravitational wave polarization is the ``transverse breathing mode'', which is a transverse excitation with the effect of expanding and contracting an array of inertial particles---see Fig. \ref{fig: GW-shear-or-not-shear}.

        Scalar-tensor theories with a massless mode often excite the transverse breathing mode. For instance, Brans--Dicke gravity~\cite{brans1961MachsPrincipleRelativistic} with a massless scalar field (often named dilaton) encodes radiative information on its additional scalar mode, whose effects can be transmitted to the Ricci tensor and cause expansion and contraction of geodesic congruences away from matter sources. It is known that Brans--Dicke gravity predicts a memory effect beyond general relativity \cite{hou2021GravitationalMemoryEffects,seraj2021GravitationalBreathingMemory}, and it is possible to relate it to soft theorems and asymptotic symmetries \cite{campiglia2018CanScalarsHave,campiglia2019ScalarAsymptoticCharges}. While the superdilations we described above seem ``too large'' in an \(1/r\) expansion to generate this memory effect, we expect them to be associated to subleading effects, similarly to how superrotations lead to subleading memory effects in general relativity.

        Notice that a theory-agnostic definition of asymptotic symmetries suggested the existence of a memory effect in a wider class of gravitational theories. Spacetime symmetries, even when asymptotic, should be considered as a property of spacetime, regardless of the underlying theory ruling the behavior of spacetime. The symmetries of Minkowski spacetime would still be the same if gravity was described by modified gravity, for they are a statement about the geometric invariance properties of spacetime. It is thus compelling to speculate that the symmetries available at infinity can---or should---be used to constrain which degrees of freedom should be in a full theory of gravity. Just like the equivalence between inertial and gravitational masses suggests the equivalence principle or gauge and conformal invariance constrain the behavior of quantum field theories, the spacetime isometries and confomorphisms at infinity could be a clue to what are the fundamental components of reality.

        To obtain a candidate theory of gravity based on the symmetries at infinity, a possible procedure is to exploit covariant phase space methods \cite{gieres2023CovariantCanonicalFormulations,wald2000GeneralDefinitionConserved}. For instance, \textcite{nagy2023RadiativePhaseSpace} consider how to extend the covariant phase space of a class of theories by exploiting the Stückelberg trick \cite{stuckelberg1938WechselwirkungskrafteElektrodynamikUnd,hinterbichler2012TheoreticalAspectsMassive}. We expect similar techniques can be applied in this case to extend general relativity in a natural way.
    
    \section{Outlook}\label{sec: outlook}
        Let us now briefly discuss some future perspectives emerging from our present work. 
    
        \subsection{Bifurcate Asymptotic (Conformal) Killing Horizons}
            Bifurcate Killing horizons \cite{boyer1969GeodesicKillingOrbits} are of center importance to quantum field theory in curved spacetimes, since they are the basic geometric structure behind the Unruh effect \cite{kay1991TheoremsUniquenessThermal,wald1994QuantumFieldTheory}. By mimicking the definition of bifurcate Killing horizons, we can define the notion of bifurcate asymptotic (conformal) Killing horizons. 

\begin{figure}[tbp]
                \centering
                \begin{tikzpicture}
                    \begin{axis}[
                        view={70}{25},
                        axis line style={draw=none},
                        xtick=\empty,
                        ytick=\empty,
                        ztick=\empty,
                        width = 8cm,
                        height = 8cm,
                    ]
                        \addplot3[surf, color=accent, faceted color=accent!75!white, opacity=0.5, mesh/rows=41] table [col sep=comma,x=x,y=y,z=z] {lightcone.csv};
                    \end{axis}
                \end{tikzpicture}
                \caption{While the two future and past lightcones of the origin in Minkowski spacetime are conformal Killing horizons, their intersection is a single point, not a bifurcation surface. This is related to the fact that conformal Killing horizons admit nonvanishing geodesic expansion.}
                \label{fig: lightcone}
            \end{figure}
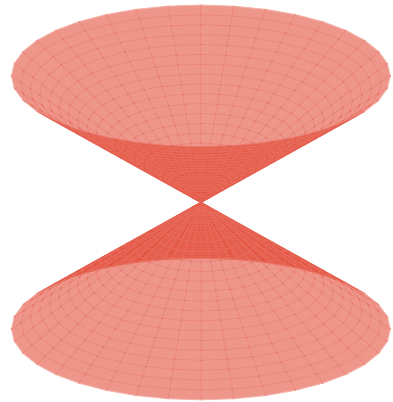

            There are some subtleties in pursuing such a construction. Typical Killing horizons have isometric cross sections, and their bifurcations always include a bifurcation surface. However, conformal horizons may have a nonvanishing geodesic expansion. This makes them more complicated, because the horizons may collapse to a single point. This is illustrated by lightcones in Minkowski spacetime, for example, as pictured on Fig. \ref{fig: lightcone}. They are a conformal Killing horizon for the vector field associated with dilations. A proper definition of a bifurcate ACKH should be able to describe these sorts of cases.
            
            A driving motivation for pursuing these notions is that null infinity could then be understood as a bifurcate ACKH. If a spacetime is asymptotically flat at null and spatial infinity, then \(\mathscr{I}^+\) and \(\mathscr{I}^-\) are the future and past lightcones for spatial infinity \(i^0\). Based on the work of \textcite{strominger2014BMSInvarianceGravitational}, it is now natural to speculate that identifying other spacetimes with bifurcate conformal Killing horizons could lead to analogs of soft theorems in curved spacetime. Notably, we mention the notion of conformal Killing horizon has had particular attention due to the study of dynamical black holes \cite{dyer1979ConformalKillingHorizons,sultana2004ConformalKillingHorizons,nielsen2018ConformalKillingHorizons}, which would be an interesting application, if possible.

    \subsection{Superrotations in de Sitter Spacetime}
        One of the main motivations for our introduction of asymptotic (conformal) Killing horizons was investigating whether the DMP group admits a natural geometric extension that accommodates conformal transformations on its cross sections. If it was possible to extend \(\mathrm{SO}(3)\) to \(\mathrm{SO}^+(3,1)\), it would be natural to expect a second extension to the local conformal algebra on the sphere. These would be analogs of Virasoro-like superrotations in de Sitter spacetime. These symmetries would then prompt the identification of a conformal field theory on the cross sections of the cosmological horizon, which would be an interesting step toward a dS/CFT correspondence. 

        Our results disfavor this possibility. The natural structure of an asymptotic Killing horizon---such as the cosmological horizons in expanding universes---encompasses only the isometry group of the horizon's cross-section. This suggests there are not sufficiently many gravitational degrees of freedom near the horizon to justify a two-dimensional conformal field theory. This result is in agreement with the analysis by \textcite{koga2001AsymptoticSymmetriesKilling}. 

        One may ask whether it is possible to bypass the AKH structure and consider the cosmological horizon as an asymptotic conformal Killing horizon, which does include conformal transformations in its symmetry group. This, however, does not seem to be the case. AKHs signal symmetries of the pseudo-Riemannian manifold itself and have isometric cross sections, while ACKHs are weaker in the sense they signal symmetries present in the conformal equivalence class of the manifold and have confomorphic cross sections. Given a spacetime and an asymptotic (conformal) Killing horizon, we can tell whether the horizon is conformal or not by investigating whether the cross sections are isometric or merely confomorphic. This reflects whether we have a symmetry of the metric or merely of the conformal structure. 
        
        Finally, we briefly comment on the possibility of a dS/CFT correspondence. \textcite{strominger2001DSCFTCorrespondence} originally conjectured that such a duality could happen between the bulk and the future (or past) infinity of de Sitter spacetime (the top line on Fig. \ref{fig: desitter}) or between the bulk and the cosmological horizon. The holographic analyses by \textcite{bousso1999HolographyGeneralSpacetimes,bousso2002HolographicPrinciple} favor a duality with the cosmological horizon. Nevertheless, the absence of a notion of superrotations in de Sitter spacetime---and the subsequent difficulty in obtaining two copies of the Virasoro algebra---make it seem difficult to formulate a dS/CFT correspondence by exploiting asymptotic symmetries. 

    \subsection{Construction of Hadamard States}
        A major motivation for the work of DMP was the construction of Hadamard states in various curved spacetimes---see Ref. \cite{dappiaggi2017HadamardStatesLightlike} for a review. Notably, their work in expanding universes with cosmological horizons \cite{dappiaggi2009CosmologicalHorizonsReconstruction,dappiaggi2009DistinguishedQuantumStates,dappiaggi2017HadamardStatesLightlike} allowed the construction of Hadamard states in spacetimes similar to de Sitter, as mentioned in the introduction. For exact de Sitter spacetime, the construction singles out the Bunch--Davies vacuum \cite{bunch1978QuantumFieldTheory,chernikov1968QuantumTheoryScalar,hartle1976PathintegralDerivationBlackhole}. In this sense, the DMP group is used to bootstrap a state for a quantum field theory in a curved spacetime. Notably, the ``DMP bootstrap'' selects exactly the state of major interest in de Sitter spacetime. It seems possible that the DMP techniques can be extended to other spacetimes by exploiting A(C)KH groups. As mentioned in the introduction, DMP themselves had already considered the case of asymptotically flat spacetimes \cite{dappiaggi2006RigorousStepsHolography,moretti2006UniquenessTheoremBMSInvariant,moretti2008QuantumOutstatesHolographically,dappiaggi2011RigorousConstructionHadamard,dappiaggi2017HadamardStatesLightlike}.

\section{Conclusions}
\label{sec:conclusion}
    We have explored a definition of asymptotic symmetries in asymptotic (conformal) Killing horizons. These sorts of hypersurfaces generalize null infinity in asymptotically flat spacetimes and cosmological horizons in expanding universes. They also admit natural structures to define groups similar to the BMS and DMP groups. For the case of an asymptotic Killing horizon, we recover a natural generalization of the DMP group, while for asymptotic conformal Killing horizons we obtain an extension of the BMS group with ``superdilations''. 

    Focusing on hypersurfaces with a Killing horizon-like structure is natural because these hypersurfaces encode symmetry properties of the ambient spacetime, and admit a definition of the isometry or conformal group of their cross sections. Such a definition may not be possible with less structure, for different cross-sections of the null hypersurface may not be isometric or confomorphic. 

    Since this construction is agnostic about the underlying theory of gravity, we expect it to inspire modifications of gravity by performing suitable phase space extensions to general relativity with the goal of ``fitting'' superdilations as a field-theoretic symmetry. Other prospective outlooks are the investigation of bifurcate horizons and their application to spacetimes with a spatial infinity, and eventual applications to the construction of distinguished states in quantum field theory in curved spacetimes. Finally, our analysis disfavors the possibility of considering Virasoro-like superrotations on the de Sitter cosmological horizon.

\begin{acknowledgments}
    Part of the calculations in this work were carried out with the aid of \texttt{Mathematica 14.0} \cite{wolframresearch2024Mathematica140} (and in particular the \texttt{OGRe} package \cite{shoshany2021OGReObjectOrientedGeneral}) in a license to the University of São Paulo (N. A. A.'s \emph{alma mater}). Feynman diagrams were drawn with \texttt{tikz-feynman} \cite{ellis2017TikZFeynmanFeynmanDiagrams}.
    
    This study was inspired by discussions that took place during and after the \href{https://www.ictp-saifr.org/wqag2024/}{Witnessing Quantum Aspects of Gravity in a Lab} conference, and thus we thank the organizers and participants. In particular, N. A. A. thanks Gautam Satishchandran, Daine L. Danielson, Maitá C. Micol, and Gabriel S. Menezes for conversations on many different aspects of infrared physics.
    
    The work of N. A. A. was supported by the São Paulo Research Foundation (FAPESP) under grant 2025/05161-0 and by the Coordenação de Aperfeiçoamento de Pessoal de Nível Superior---Brasil (CAPES)---Finance Code 001. 
\end{acknowledgments}

\bibliography{shortbib}
\end{document}

%% file: texnical.tex
\usetikzlibrary{arrows, arrows.meta, calc, bending, positioning, decorations.markings, decorations.pathmorphing, patterns, shapes, scopes}
    \tikzfeynmanset{
      warn luatex=false,
      with arrow/.style = {
       decoration={
         markings,
         mark=at position 0.5
              with {\arrow[xshift=2.5pt]{Triangle[width=3pt,length=5pt]}}
         },
       postaction=decorate},
       with reversed arrow/.style = {
       decoration={
         markings,
         mark=at position 0.5
              with {\arrowreversed[xshift=-2.5pt]{Triangle[width=3pt,length=5pt]}}
         },
       postaction=decorate},
      every blob={/tikz/fill=gray!30},
      graviton/.style={accent,decorate, decoration={coil,aspect=0.75,mirror,segment length=1.5mm}, double},
      photon/.append style={accent},
      scalar/.append style={accent, thick}
}

\makeatletter
\DeclareRobustCommand{\loplus}{\mathbin{\mathpalette\dog@lsemi{+}}}
\DeclareRobustCommand{\lotimes}{\mathbin{\mathpalette\dog@lsemi{\times}}}
\DeclareRobustCommand{\roplus}{\mathbin{\mathpalette\dog@rsemi{+}}}
\DeclareRobustCommand{\rotimes}{\mathbin{\mathpalette\dog@rsemi{\times}}}
\let\oplus\relax
\DeclareRobustCommand{\oplus}{\mathbin{\mathpalette\dog@fullsemi{+}}}
\let\otimes\relax
\DeclareRobustCommand{\otimes}{\mathbin{\mathpalette\dog@fullsemi{\times}}}

\newcommand{\dog@rsemi}[2]{\dog@semi{#1}{#2}{-90,90}}
\newcommand{\dog@lsemi}[2]{\dog@semi{#1}{#2}{270,90}}
\newcommand{\dog@fullsemi}[2]{\dog@semi{#1}{#2}{0,360}}
\newcommand{\dog@semi}[3]{%
  \begingroup
  \sbox\z@{$\m@th#1#2$}%
  \setlength{\unitlength}{\dimexpr\ht\z@+\dp\z@\relax}%
  \makebox[\wd\z@]{\raisebox{-\dp\z@}{%
    \begin{picture}(1,1)
    \linethickness{\variable@rule{#1}}
    \roundcap
    \put(0.5,0.5){\makebox(0,0){\raisebox{\dp\z@}{$\m@th#1#2$}}}
    \put(0.5,0.5){\arc[#3]{0.5}}
    \end{picture}%
  }}%
  \endgroup
}
\newcommand{\variable@rule}[1]{%
  \fontdimen8  
  \ifx#1\displaystyle\textfont3\else
    \ifx#1\textstyle\textfont3\else
      \ifx#1\scriptstyle\scriptfont3\else
        \scriptscriptfont3\relax
  \fi\fi\fi
}
\makeatother

\newcommand{\ie}{\emph{i.e.}} 
\newcommand{\eg}{\emph{e.g.}} 

\renewcommand{\pounds}{\text{\textsterling}}

%% file: arXivV3.bbl
\begin{thebibliography}{117}%
\makeatletter
\providecommand \@ifxundefined [1]{%
 \@ifx{#1\undefined}
}%
\providecommand \@ifnum [1]{%
 \ifnum #1\expandafter \@firstoftwo
 \else \expandafter \@secondoftwo
 \fi
}%
\providecommand \@ifx [1]{%
 \ifx #1\expandafter \@firstoftwo
 \else \expandafter \@secondoftwo
 \fi
}%
\providecommand \natexlab [1]{#1}%
\providecommand \enquote  [1]{``#1''}%
\providecommand \bibnamefont  [1]{#1}%
\providecommand \bibfnamefont [1]{#1}%
\providecommand \citenamefont [1]{#1}%
\providecommand \href@noop [0]{\@secondoftwo}%
\providecommand \href [0]{\begingroup \@sanitize@url \@href}%
\providecommand \@href[1]{\@@startlink{#1}\@@href}%
\providecommand \@@href[1]{\endgroup#1\@@endlink}%
\providecommand \@sanitize@url [0]{\catcode `\\12\catcode `\$12\catcode `\&12\catcode `\#12\catcode `\^12\catcode `\_12\catcode `\%12\relax}%
\providecommand \@@startlink[1]{}%
\providecommand \@@endlink[0]{}%
\providecommand \url  [0]{\begingroup\@sanitize@url \@url }%
\providecommand \@url [1]{\endgroup\@href {#1}{\urlprefix }}%
\providecommand \urlprefix  [0]{URL }%
\providecommand \Eprint [0]{\href }%
\providecommand \doibase [0]{https://doi.org/}%
\providecommand \selectlanguage [0]{\@gobble}%
\providecommand \bibinfo  [0]{\@secondoftwo}%
\providecommand \bibfield  [0]{\@secondoftwo}%
\providecommand \translation [1]{[#1]}%
\providecommand \BibitemOpen [0]{}%
\providecommand \bibitemStop [0]{}%
\providecommand \bibitemNoStop [0]{.\EOS\space}%
\providecommand \EOS [0]{\spacefactor3000\relax}%
\providecommand \BibitemShut  [1]{\csname bibitem#1\endcsname}%
\let\auto@bib@innerbib\@empty
\bibitem [{\citenamefont {Bondi}\ \emph {et~al.}(1962)\citenamefont {Bondi}, \citenamefont {{Van der Burg}},\ and\ \citenamefont {Metzner}}]{bondi1962GravitationalWavesGeneral}%
  \BibitemOpen
  \bibfield  {author} {\bibinfo {author} {\bibfnamefont {H.}~\bibnamefont {Bondi}}, \bibinfo {author} {\bibfnamefont {M.~G.~J.}\ \bibnamefont {{Van der Burg}}},\ and\ \bibinfo {author} {\bibfnamefont {A.~W.~K.}\ \bibnamefont {Metzner}},\ }\bibfield  {title} {\bibinfo {title} {Gravitational waves in general relativity, {{VII}}. {{Waves}} from axi-symmetric isolated system},\ }\href {https://doi.org/10.1098/rspa.1962.0161} {\bibfield  {journal} {\bibinfo  {journal} {Proceedings of the Royal Society of London. Series A. Mathematical and Physical Sciences}\ }\textbf {\bibinfo {volume} {269}},\ \bibinfo {pages} {21} (\bibinfo {year} {1962})}\BibitemShut {NoStop}%
\bibitem [{\citenamefont {Sachs}(1962{\natexlab{a}})}]{sachs1962AsymptoticSymmetriesGravitational}%
  \BibitemOpen
  \bibfield  {author} {\bibinfo {author} {\bibfnamefont {R.~K.}\ \bibnamefont {Sachs}},\ }\bibfield  {title} {\bibinfo {title} {Asymptotic symmetries in gravitational theory},\ }\href {https://doi.org/10.1103/PhysRev.128.2851} {\bibfield  {journal} {\bibinfo  {journal} {Physical Review}\ }\textbf {\bibinfo {volume} {128}},\ \bibinfo {pages} {2851} (\bibinfo {year} {1962}{\natexlab{a}})}\BibitemShut {NoStop}%
\bibitem [{\citenamefont {Sachs}(1962{\natexlab{b}})}]{sachs1962GravitationalWavesGeneral}%
  \BibitemOpen
  \bibfield  {author} {\bibinfo {author} {\bibfnamefont {R.~K.}\ \bibnamefont {Sachs}},\ }\bibfield  {title} {\bibinfo {title} {Gravitational waves in general relativity {{VIII}}. {{Waves}} in asymptotically flat space-time},\ }\href {https://doi.org/10.1098/rspa.1962.0206} {\bibfield  {journal} {\bibinfo  {journal} {Proceedings of the Royal Society of London. Series A. Mathematical and Physical Sciences}\ }\textbf {\bibinfo {volume} {270}},\ \bibinfo {pages} {103} (\bibinfo {year} {1962}{\natexlab{b}})}\BibitemShut {NoStop}%
\bibitem [{\citenamefont {Baumann}(2022)}]{baumann2022Cosmology}%
  \BibitemOpen
  \bibfield  {author} {\bibinfo {author} {\bibfnamefont {D.}~\bibnamefont {Baumann}},\ }\href {https://doi.org/10.1017/9781108937092} {\emph {\bibinfo {title} {Cosmology}}}\ (\bibinfo  {publisher} {Cambridge University Press},\ \bibinfo {address} {Cambridge},\ \bibinfo {year} {2022})\BibitemShut {NoStop}%
\bibitem [{\citenamefont {Weinberg}(2008)}]{weinberg2008Cosmology}%
  \BibitemOpen
  \bibfield  {author} {\bibinfo {author} {\bibfnamefont {S.}~\bibnamefont {Weinberg}},\ }\href {https://doi.org/10.1093/oso/9780198526827.001.0001} {\emph {\bibinfo {title} {Cosmology}}}\ (\bibinfo  {publisher} {Oxford University Press},\ \bibinfo {address} {Oxford},\ \bibinfo {year} {2008})\BibitemShut {NoStop}%
\bibitem [{\citenamefont {Akrami}\ \emph {et~al.}(2020)\citenamefont {Akrami} \emph {et~al.}}]{planckcollaboration2020AX}%
  \BibitemOpen
  \bibfield  {author} {\bibinfo {author} {\bibfnamefont {Y.}~\bibnamefont {Akrami}} \emph {et~al.} (\bibinfo {collaboration} {Planck Collaboration}),\ }\bibfield  {title} {\bibinfo {title} {{\emph{Planck}} 2018 results: {{X}}. {{Constraints}} on inflation},\ }\href {https://doi.org/10.1051/0004-6361/201833887} {\bibfield  {journal} {\bibinfo  {journal} {Astronomy \& Astrophysics}\ }\textbf {\bibinfo {volume} {641}},\ \bibinfo {eid} {A10} (\bibinfo {year} {2020})},\ \Eprint {https://arxiv.org/abs/1807.06211} {arXiv:1807.06211 [astro-ph.CO]} \BibitemShut {NoStop}%
\bibitem [{\citenamefont {Dappiaggi}\ \emph {et~al.}(2009{\natexlab{a}})\citenamefont {Dappiaggi}, \citenamefont {Moretti},\ and\ \citenamefont {Pinamonti}}]{dappiaggi2009CosmologicalHorizonsReconstruction}%
  \BibitemOpen
  \bibfield  {author} {\bibinfo {author} {\bibfnamefont {C.}~\bibnamefont {Dappiaggi}}, \bibinfo {author} {\bibfnamefont {V.}~\bibnamefont {Moretti}},\ and\ \bibinfo {author} {\bibfnamefont {N.}~\bibnamefont {Pinamonti}},\ }\bibfield  {title} {\bibinfo {title} {Cosmological horizons and reconstruction of quantum field theories},\ }\href {https://doi.org/10.1007/s00220-008-0653-8} {\bibfield  {journal} {\bibinfo  {journal} {Communications in Mathematical Physics}\ }\textbf {\bibinfo {volume} {285}},\ \bibinfo {pages} {1129} (\bibinfo {year} {2009}{\natexlab{a}})},\ \Eprint {https://arxiv.org/abs/0712.1770} {arXiv:0712.1770 [gr-qc]} \BibitemShut {NoStop}%
\bibitem [{\citenamefont {Dappiaggi}\ \emph {et~al.}(2009{\natexlab{b}})\citenamefont {Dappiaggi}, \citenamefont {Moretti},\ and\ \citenamefont {Pinamonti}}]{dappiaggi2009DistinguishedQuantumStates}%
  \BibitemOpen
  \bibfield  {author} {\bibinfo {author} {\bibfnamefont {C.}~\bibnamefont {Dappiaggi}}, \bibinfo {author} {\bibfnamefont {V.}~\bibnamefont {Moretti}},\ and\ \bibinfo {author} {\bibfnamefont {N.}~\bibnamefont {Pinamonti}},\ }\bibfield  {title} {\bibinfo {title} {Distinguished quantum states in a class of cosmological spacetimes and their {{Hadamard}} property},\ }\href {https://doi.org/10.1063/1.3122770} {\bibfield  {journal} {\bibinfo  {journal} {Journal of Mathematical Physics}\ }\textbf {\bibinfo {volume} {50}},\ \bibinfo {eid} {062304} (\bibinfo {year} {2009}{\natexlab{b}})},\ \Eprint {https://arxiv.org/abs/0812.4033} {arXiv:0812.4033 [gr-qc]} \BibitemShut {NoStop}%
\bibitem [{\citenamefont {Dappiaggi}\ \emph {et~al.}(2017)\citenamefont {Dappiaggi}, \citenamefont {Moretti},\ and\ \citenamefont {Pinamonti}}]{dappiaggi2017HadamardStatesLightlike}%
  \BibitemOpen
  \bibfield  {author} {\bibinfo {author} {\bibfnamefont {C.}~\bibnamefont {Dappiaggi}}, \bibinfo {author} {\bibfnamefont {V.}~\bibnamefont {Moretti}},\ and\ \bibinfo {author} {\bibfnamefont {N.}~\bibnamefont {Pinamonti}},\ }\href {https://doi.org/10.1007/978-3-319-64343-4} {\emph {\bibinfo {title} {Hadamard States From Light-like Hypersurfaces}}},\ \bibinfo {series} {{{SpringerBriefs}} in Mathematical Physics}\ No.~\bibinfo {number} {25}\ (\bibinfo  {publisher} {Springer},\ \bibinfo {address} {Cham},\ \bibinfo {year} {2017})\ \Eprint {https://arxiv.org/abs/1706.09666} {arXiv:1706.09666 [math-ph]} \BibitemShut {NoStop}%
\bibitem [{\citenamefont {Strominger}(2014)}]{strominger2014BMSInvarianceGravitational}%
  \BibitemOpen
  \bibfield  {author} {\bibinfo {author} {\bibfnamefont {A.}~\bibnamefont {Strominger}},\ }\bibfield  {title} {\bibinfo {title} {On {{BMS}} invariance of gravitational scattering},\ }\href {https://doi.org/10.1007/JHEP07(2014)152} {\bibfield  {journal} {\bibinfo  {journal} {Journal of High Energy Physics}\ }\textbf {\bibinfo {volume} {2014}},\ \bibinfo {eid} {152} (\bibinfo {year} {2014})},\ \Eprint {https://arxiv.org/abs/1312.2229} {arXiv:1312.2229 [hep-th]} \BibitemShut {NoStop}%
\bibitem [{\citenamefont {Strominger}\ and\ \citenamefont {Zhiboedov}(2016)}]{strominger2016GravitationalMemoryBMS}%
  \BibitemOpen
  \bibfield  {author} {\bibinfo {author} {\bibfnamefont {A.}~\bibnamefont {Strominger}}\ and\ \bibinfo {author} {\bibfnamefont {A.}~\bibnamefont {Zhiboedov}},\ }\bibfield  {title} {\bibinfo {title} {Gravitational memory, {{BMS}} supertranslations and soft theorems},\ }\href {https://doi.org/10.1007/JHEP01(2016)086} {\bibfield  {journal} {\bibinfo  {journal} {Journal of High Energy Physics}\ }\textbf {\bibinfo {volume} {2016}},\ \bibinfo {eid} {86} (\bibinfo {year} {2016})},\ \Eprint {https://arxiv.org/abs/1411.5745} {arXiv:1411.5745 [hep-th]} \BibitemShut {NoStop}%
\bibitem [{\citenamefont {He}\ \emph {et~al.}(2015)\citenamefont {He}, \citenamefont {Lysov}, \citenamefont {Mitra},\ and\ \citenamefont {Strominger}}]{he2015BMSSupertranslationsWeinbergs}%
  \BibitemOpen
  \bibfield  {author} {\bibinfo {author} {\bibfnamefont {T.}~\bibnamefont {He}}, \bibinfo {author} {\bibfnamefont {V.}~\bibnamefont {Lysov}}, \bibinfo {author} {\bibfnamefont {P.}~\bibnamefont {Mitra}},\ and\ \bibinfo {author} {\bibfnamefont {A.}~\bibnamefont {Strominger}},\ }\bibfield  {title} {\bibinfo {title} {{{BMS}} supertranslations and {{Weinberg}}'s soft graviton theorem},\ }\href {https://doi.org/10.1007/JHEP05(2015)151} {\bibfield  {journal} {\bibinfo  {journal} {Journal of High Energy Physics}\ }\textbf {\bibinfo {volume} {2015}},\ \bibinfo {eid} {151} (\bibinfo {year} {2015})},\ \Eprint {https://arxiv.org/abs/1401.7026} {arXiv:1401.7026 [hep-th]} \BibitemShut {NoStop}%
\bibitem [{\citenamefont {Weinberg}(1965)}]{weinberg1965InfraredPhotonsGravitons}%
  \BibitemOpen
  \bibfield  {author} {\bibinfo {author} {\bibfnamefont {S.}~\bibnamefont {Weinberg}},\ }\bibfield  {title} {\bibinfo {title} {Infrared photons and gravitons},\ }\href {https://doi.org/10.1103/PhysRev.140.B516} {\bibfield  {journal} {\bibinfo  {journal} {Physical Review}\ }\textbf {\bibinfo {volume} {140}},\ \bibinfo {pages} {B516} (\bibinfo {year} {1965})}\BibitemShut {NoStop}%
\bibitem [{\citenamefont {Zel'dovich}\ and\ \citenamefont {Polnarev}(1974)}]{zeldovich1974RadiationGravitationalWaves}%
  \BibitemOpen
  \bibfield  {author} {\bibinfo {author} {\bibfnamefont {{\relax Ya}.~B.}\ \bibnamefont {Zel'dovich}}\ and\ \bibinfo {author} {\bibfnamefont {A.~G.}\ \bibnamefont {Polnarev}},\ }\bibfield  {title} {\bibinfo {title} {Radiation of gravitational waves by a cluster of superdense stars},\ }\href {https://ui.adsabs.harvard.edu/abs/1974SvA....18...17Z} {\bibfield  {journal} {\bibinfo  {journal} {Soviet Astronomy}\ }\textbf {\bibinfo {volume} {18}},\ \bibinfo {pages} {17} (\bibinfo {year} {1974})}\BibitemShut {NoStop}%
\bibitem [{\citenamefont {Braginsky}\ and\ \citenamefont {Thorne}(1987)}]{braginsky1987GravitationalwaveBurstsMemory}%
  \BibitemOpen
  \bibfield  {author} {\bibinfo {author} {\bibfnamefont {V.~B.}\ \bibnamefont {Braginsky}}\ and\ \bibinfo {author} {\bibfnamefont {K.~S.}\ \bibnamefont {Thorne}},\ }\bibfield  {title} {\bibinfo {title} {Gravitational-wave bursts with memory and experimental prospects},\ }\href {https://doi.org/10.1038/327123a0} {\bibfield  {journal} {\bibinfo  {journal} {Nature}\ }\textbf {\bibinfo {volume} {327}},\ \bibinfo {pages} {123} (\bibinfo {year} {1987})}\BibitemShut {NoStop}%
\bibitem [{\citenamefont {Christodoulou}(1991)}]{christodoulou1991NonlinearNatureGravitation}%
  \BibitemOpen
  \bibfield  {author} {\bibinfo {author} {\bibfnamefont {D.}~\bibnamefont {Christodoulou}},\ }\bibfield  {title} {\bibinfo {title} {Nonlinear nature of gravitation and gravitational-wave experiments},\ }\href {https://doi.org/10.1103/PhysRevLett.67.1486} {\bibfield  {journal} {\bibinfo  {journal} {Physical Review Letters}\ }\textbf {\bibinfo {volume} {67}},\ \bibinfo {pages} {1486} (\bibinfo {year} {1991})}\BibitemShut {NoStop}%
\bibitem [{\citenamefont {Weinberg}(1995)}]{weinberg1995Foundations}%
  \BibitemOpen
  \bibfield  {author} {\bibinfo {author} {\bibfnamefont {S.}~\bibnamefont {Weinberg}},\ }\href {https://doi.org/10.1017/CBO9781139644167} {\emph {\bibinfo {title} {The Quantum Theory of Fields, Vol I: Foundations}}}\ (\bibinfo  {publisher} {Cambridge University Press},\ \bibinfo {address} {Cambridge},\ \bibinfo {year} {1995})\BibitemShut {NoStop}%
\bibitem [{\citenamefont {de~Aguiar~Alves}(2025)}]{aguiaralves2025LecturesBondiMetzner}%
  \BibitemOpen
  \bibfield  {author} {\bibinfo {author} {\bibfnamefont {N.}~\bibnamefont {de~Aguiar~Alves}},\ }\href@noop {} {\bibinfo {title} {Lectures on the {{Bondi}}–{{Metzner}}–{{Sachs}} group and related topics in infrared physics}} (\bibinfo {year} {2025}),\ \Eprint {https://arxiv.org/abs/2504.12521} {arXiv:2504.12521 [gr-qc]} \BibitemShut {NoStop}%
\bibitem [{\citenamefont {McLoughlin}\ \emph {et~al.}(2022)\citenamefont {McLoughlin}, \citenamefont {Puhm},\ and\ \citenamefont {Raclariu}}]{mcloughlin2022SAGEXReviewScattering}%
  \BibitemOpen
  \bibfield  {author} {\bibinfo {author} {\bibfnamefont {T.}~\bibnamefont {McLoughlin}}, \bibinfo {author} {\bibfnamefont {A.}~\bibnamefont {Puhm}},\ and\ \bibinfo {author} {\bibfnamefont {A.-M.}\ \bibnamefont {Raclariu}},\ }\bibfield  {title} {\bibinfo {title} {The {{SAGEX}} review on scattering amplitudes, chapter 11: {{Soft}} theorems and celestial amplitudes},\ }\href {https://doi.org/10.1088/1751-8121/ac9a40} {\bibfield  {journal} {\bibinfo  {journal} {Journal of Physics A: Mathematical and Theoretical}\ }\textbf {\bibinfo {volume} {55}},\ \bibinfo {eid} {443012} (\bibinfo {year} {2022})},\ \Eprint {https://arxiv.org/abs/2203.13022} {arXiv:2203.13022 [hep-th]} \BibitemShut {NoStop}%
\bibitem [{\citenamefont {Mohanty}(2023)}]{mohanty2023GravitationalWavesQuantum}%
  \BibitemOpen
  \bibfield  {author} {\bibinfo {author} {\bibfnamefont {S.}~\bibnamefont {Mohanty}},\ }\href {https://doi.org/10.1007/978-3-031-23770-6} {\emph {\bibinfo {title} {Gravitational Waves from a Quantum Field Theory Perspective}}},\ \bibinfo {series} {Lecture Notes in Physics}\ No.\ \bibinfo {number} {1013}\ (\bibinfo  {publisher} {Springer},\ \bibinfo {address} {Cham},\ \bibinfo {year} {2023})\BibitemShut {NoStop}%
\bibitem [{\citenamefont {Strominger}(2018)}]{strominger2018LecturesInfraredStructure}%
  \BibitemOpen
  \bibfield  {author} {\bibinfo {author} {\bibfnamefont {A.}~\bibnamefont {Strominger}},\ }\href {https://doi.org/10.23943/9781400889853} {\emph {\bibinfo {title} {Lectures on the Infrared Structure of Gravity and Gauge Theory}}}\ (\bibinfo  {publisher} {Princeton University Press},\ \bibinfo {address} {Princeton, NJ},\ \bibinfo {year} {2018})\ \Eprint {https://arxiv.org/abs/1703.05448} {arXiv:1703.05448 [hep-th]} \BibitemShut {NoStop}%
\bibitem [{\citenamefont {Bieri}\ \emph {et~al.}(2015)\citenamefont {Bieri}, \citenamefont {Garfinkle},\ and\ \citenamefont {Yau}}]{bieri2015GravitationalWavesTheir}%
  \BibitemOpen
  \bibfield  {author} {\bibinfo {author} {\bibfnamefont {L.}~\bibnamefont {Bieri}}, \bibinfo {author} {\bibfnamefont {D.}~\bibnamefont {Garfinkle}},\ and\ \bibinfo {author} {\bibfnamefont {S.-T.}\ \bibnamefont {Yau}},\ }\bibfield  {title} {\bibinfo {title} {Gravitational waves and their memory in general relativity},\ }in\ \href {https://doi.org/10.4310/SDG.2015.v20.n1.a4} {\emph {\bibinfo {booktitle} {One Hundred Years of General Relativity: {{A}} Jubilee Volume on General Relativity and Mathematics}}},\ \bibinfo {series and number} {\bibinfo {series} {Surveys in Differential Geometry}\ No.~\bibinfo {number} {20}},\ \bibinfo {editor} {edited by\ \bibinfo {editor} {\bibfnamefont {L.}~\bibnamefont {Bieri}}\ and\ \bibinfo {editor} {\bibfnamefont {S.-T.}\ \bibnamefont {Yau}}}\ (\bibinfo  {publisher} {International Press of Boston},\ \bibinfo {address} {Boston, MA},\ \bibinfo {year} {2015})\ pp.\ \bibinfo {pages} {75--97},\ \Eprint {https://arxiv.org/abs/1505.05213} {arXiv:1505.05213 [gr-qc]} \BibitemShut {NoStop}%
\bibitem [{\citenamefont {Bieri}\ and\ \citenamefont {Polnarev}(2024)}]{bieri2024GravitationalWaveDisplacement}%
  \BibitemOpen
  \bibfield  {author} {\bibinfo {author} {\bibfnamefont {L.}~\bibnamefont {Bieri}}\ and\ \bibinfo {author} {\bibfnamefont {A.}~\bibnamefont {Polnarev}},\ }\bibfield  {title} {\bibinfo {title} {Gravitational wave displacement and velocity memory effects},\ }\href {https://doi.org/10.1088/1361-6382/ad4dfe} {\bibfield  {journal} {\bibinfo  {journal} {Classical and Quantum Gravity}\ }\textbf {\bibinfo {volume} {41}},\ \bibinfo {eid} {135012} (\bibinfo {year} {2024})},\ \Eprint {https://arxiv.org/abs/2402.02594} {arXiv:2402.02594 [gr-qc]} \BibitemShut {NoStop}%
\bibitem [{\citenamefont {Favata}(2010)}]{favata2010GravitationalwaveMemoryEffect}%
  \BibitemOpen
  \bibfield  {author} {\bibinfo {author} {\bibfnamefont {M.}~\bibnamefont {Favata}},\ }\bibfield  {title} {\bibinfo {title} {The gravitational-wave memory effect},\ }\href {https://doi.org/10.1088/0264-9381/27/8/084036} {\bibfield  {journal} {\bibinfo  {journal} {Classical and Quantum Gravity}\ }\textbf {\bibinfo {volume} {27}},\ \bibinfo {eid} {084036} (\bibinfo {year} {2010})},\ \Eprint {https://arxiv.org/abs/1003.3486} {arXiv:1003.3486 [gr-qc]} \BibitemShut {NoStop}%
\bibitem [{\citenamefont {Grant}\ and\ \citenamefont {Nichols}(2023{\natexlab{a}})}]{grant2023OutlookDetectingGravitationalwave}%
  \BibitemOpen
  \bibfield  {author} {\bibinfo {author} {\bibfnamefont {A.~M.}\ \bibnamefont {Grant}}\ and\ \bibinfo {author} {\bibfnamefont {D.~A.}\ \bibnamefont {Nichols}},\ }\bibfield  {title} {\bibinfo {title} {Outlook for detecting the gravitational-wave displacement and spin memory effects with current and future gravitational-wave detectors},\ }\href {https://doi.org/10.1103/PhysRevD.107.064056} {\bibfield  {journal} {\bibinfo  {journal} {Physical Review D}\ }\textbf {\bibinfo {volume} {107}},\ \bibinfo {eid} {064056} (\bibinfo {year} {2023}{\natexlab{a}})},\ \Eprint {https://arxiv.org/abs/2210.16266} {arXiv:2210.16266 [gr-qc]} \BibitemShut {NoStop}%
\bibitem [{\citenamefont {Grant}\ and\ \citenamefont {Nichols}(2023{\natexlab{b}})}]{grant2023ErratumOutlookDetecting}%
  \BibitemOpen
  \bibfield  {author} {\bibinfo {author} {\bibfnamefont {A.~M.}\ \bibnamefont {Grant}}\ and\ \bibinfo {author} {\bibfnamefont {D.~A.}\ \bibnamefont {Nichols}},\ }\bibfield  {title} {\bibinfo {title} {Erratum: {{Outlook}} for detecting the gravitational-wave displacement and spin memory effects with current and future gravitational-wave detectors [{{Phys}}. {{Rev}}. {{D}} {\textbf{107}} , 064056 (2023)]},\ }\href {https://doi.org/10.1103/PhysRevD.108.029901} {\bibfield  {journal} {\bibinfo  {journal} {Physical Review D}\ }\textbf {\bibinfo {volume} {108}},\ \bibinfo {eid} {029901} (\bibinfo {year} {2023}{\natexlab{b}})}\BibitemShut {NoStop}%
\bibitem [{\citenamefont {Banks}(2003)}]{banks2003CritiquePureString}%
  \BibitemOpen
  \bibfield  {author} {\bibinfo {author} {\bibfnamefont {T.}~\bibnamefont {Banks}},\ }\href@noop {} {\bibinfo {title} {A critique of pure string theory: {{Heterodox}} opinions of diverse dimensions}} (\bibinfo {year} {2003}),\ \Eprint {https://arxiv.org/abs/hep-th/0306074} {arXiv:hep-th/0306074} \BibitemShut {NoStop}%
\bibitem [{\citenamefont {Barnich}\ and\ \citenamefont {Troessaert}(2010{\natexlab{a}})}]{barnich2010AspectsBMSCFT}%
  \BibitemOpen
  \bibfield  {author} {\bibinfo {author} {\bibfnamefont {G.}~\bibnamefont {Barnich}}\ and\ \bibinfo {author} {\bibfnamefont {C.}~\bibnamefont {Troessaert}},\ }\bibfield  {title} {\bibinfo {title} {Aspects of the {{BMS}}/{{CFT}} correspondence},\ }\href {https://doi.org/10.1007/JHEP05(2010)062} {\bibfield  {journal} {\bibinfo  {journal} {Journal of High Energy Physics}\ }\textbf {\bibinfo {volume} {2010}},\ \bibinfo {eid} {62} (\bibinfo {year} {2010}{\natexlab{a}})},\ \Eprint {https://arxiv.org/abs/1001.1541} {arXiv:1001.1541 [hep-th]} \BibitemShut {NoStop}%
\bibitem [{\citenamefont {Barnich}\ and\ \citenamefont {Troessaert}(2010{\natexlab{b}})}]{barnich2010SymmetriesAsymptoticallyFlat}%
  \BibitemOpen
  \bibfield  {author} {\bibinfo {author} {\bibfnamefont {G.}~\bibnamefont {Barnich}}\ and\ \bibinfo {author} {\bibfnamefont {C.}~\bibnamefont {Troessaert}},\ }\bibfield  {title} {\bibinfo {title} {Symmetries of asymptotically flat 4 dimensional spacetimes at null infinity revisited},\ }\href {https://doi.org/10.1103/PhysRevLett.105.111103} {\bibfield  {journal} {\bibinfo  {journal} {Physical Review Letters}\ }\textbf {\bibinfo {volume} {105}},\ \bibinfo {eid} {111103} (\bibinfo {year} {2010}{\natexlab{b}})},\ \Eprint {https://arxiv.org/abs/0909.2617} {arXiv:0909.2617 [gr-qc]} \BibitemShut {NoStop}%
\bibitem [{\citenamefont {Belavin}\ \emph {et~al.}(1984)\citenamefont {Belavin}, \citenamefont {Polyakov},\ and\ \citenamefont {Zamolodchikov}}]{belavin1984InfiniteConformalSymmetry}%
  \BibitemOpen
  \bibfield  {author} {\bibinfo {author} {\bibfnamefont {A.~A.}\ \bibnamefont {Belavin}}, \bibinfo {author} {\bibfnamefont {A.~M.}\ \bibnamefont {Polyakov}},\ and\ \bibinfo {author} {\bibfnamefont {A.~B.}\ \bibnamefont {Zamolodchikov}},\ }\bibfield  {title} {\bibinfo {title} {Infinite conformal symmetry in two-dimensional quantum field theory},\ }\href {https://doi.org/10.1016/0550-3213(84)90052-X} {\bibfield  {journal} {\bibinfo  {journal} {Nuclear Physics B}\ }\textbf {\bibinfo {volume} {241}},\ \bibinfo {pages} {333} (\bibinfo {year} {1984})}\BibitemShut {NoStop}%
\bibitem [{\citenamefont {Di~Francesco}\ \emph {et~al.}(1997)\citenamefont {Di~Francesco}, \citenamefont {Mathieu},\ and\ \citenamefont {Sénéchal}}]{difrancesco1997ConformalFieldTheory}%
  \BibitemOpen
  \bibfield  {author} {\bibinfo {author} {\bibfnamefont {P.}~\bibnamefont {Di~Francesco}}, \bibinfo {author} {\bibfnamefont {P.}~\bibnamefont {Mathieu}},\ and\ \bibinfo {author} {\bibfnamefont {D.}~\bibnamefont {Sénéchal}},\ }\href {https://doi.org/10.1007/978-1-4612-2256-9} {\emph {\bibinfo {title} {Conformal Field Theory}}},\ Graduate Texts in Contemporary Physics\ (\bibinfo  {publisher} {Springer},\ \bibinfo {address} {New York, NY},\ \bibinfo {year} {1997})\BibitemShut {NoStop}%
\bibitem [{\citenamefont {Blumenhagen}\ and\ \citenamefont {Plauschinn}(2009)}]{blumenhagen2009IntroductionConformalField}%
  \BibitemOpen
  \bibfield  {author} {\bibinfo {author} {\bibfnamefont {R.}~\bibnamefont {Blumenhagen}}\ and\ \bibinfo {author} {\bibfnamefont {E.}~\bibnamefont {Plauschinn}},\ }\href {https://doi.org/10.1007/978-3-642-00450-6} {\emph {\bibinfo {title} {Introduction to Conformal Field Theory: {{With}} Applications to String Theory}}},\ \bibinfo {series} {Lecture Notes in Physics}\ No.\ \bibinfo {number} {779}\ (\bibinfo  {publisher} {Springer},\ \bibinfo {address} {Berlin},\ \bibinfo {year} {2009})\BibitemShut {NoStop}%
\bibitem [{\citenamefont {Schottenloher}(2008)}]{schottenloher2008MathematicalIntroductionConformal}%
  \BibitemOpen
  \bibfield  {author} {\bibinfo {author} {\bibfnamefont {M.}~\bibnamefont {Schottenloher}},\ }\href {https://doi.org/10.1007/978-3-540-68628-6} {\emph {\bibinfo {title} {A Mathematical Introduction to Conformal Field Theory}}},\ \bibinfo {series} {Lecture Notes in Physics}\ No.\ \bibinfo {number} {759}\ (\bibinfo  {publisher} {Springer},\ \bibinfo {address} {Berlin},\ \bibinfo {year} {2008})\BibitemShut {NoStop}%
\bibitem [{\citenamefont {Polchinski}(1998)}]{polchinski1998IntroductionBosonicString}%
  \BibitemOpen
  \bibfield  {author} {\bibinfo {author} {\bibfnamefont {J.}~\bibnamefont {Polchinski}},\ }\href {https://doi.org/10.1017/CBO9780511816079} {\emph {\bibinfo {title} {String Theory, Vol 1: An Introduction to the Bosonic String}}},\ Cambridge Monographs on Mathematical Physics\ (\bibinfo  {publisher} {Cambridge University Press},\ \bibinfo {address} {Cambridge},\ \bibinfo {year} {1998})\BibitemShut {NoStop}%
\bibitem [{\citenamefont {Pasterski}(2021)}]{pasterski2021LecturesCelestialAmplitudes}%
  \BibitemOpen
  \bibfield  {author} {\bibinfo {author} {\bibfnamefont {S.}~\bibnamefont {Pasterski}},\ }\bibfield  {title} {\bibinfo {title} {Lectures on celestial amplitudes},\ }\href {https://doi.org/10.1140/epjc/s10052-021-09846-7} {\bibfield  {journal} {\bibinfo  {journal} {The European Physical Journal C}\ }\textbf {\bibinfo {volume} {81}},\ \bibinfo {eid} {1062} (\bibinfo {year} {2021})},\ \Eprint {https://arxiv.org/abs/2108.04801} {arXiv:2108.04801 [hep-th]} \BibitemShut {NoStop}%
\bibitem [{\citenamefont {Pasterski}(2025)}]{pasterski2025ChapterCelestialHolography}%
  \BibitemOpen
  \bibfield  {author} {\bibinfo {author} {\bibfnamefont {S.}~\bibnamefont {Pasterski}},\ }\bibfield  {title} {\bibinfo {title} {A chapter on celestial holography},\ }in\ \href {https://doi.org/10.1016/B978-0-323-95703-8.00108-7} {\emph {\bibinfo {booktitle} {Encyclopedia of Mathematical Physics}}},\ Vol.~\bibinfo {volume} {1},\ \bibinfo {editor} {edited by\ \bibinfo {editor} {\bibfnamefont {R.}~\bibnamefont {Szabo}}\ and\ \bibinfo {editor} {\bibfnamefont {M.}~\bibnamefont {Bojowald}}}\ (\bibinfo  {publisher} {Academic Press},\ \bibinfo {address} {Oxford},\ \bibinfo {year} {2025})\ \bibinfo {edition} {2nd}\ ed.,\ pp.\ \bibinfo {pages} {471--479},\ \Eprint {https://arxiv.org/abs/2310.04932} {arXiv:2310.04932 [hep-th]} \BibitemShut {NoStop}%
\bibitem [{\citenamefont {Raclariu}(2021)}]{raclariu2021LecturesCelestialHolography}%
  \BibitemOpen
  \bibfield  {author} {\bibinfo {author} {\bibfnamefont {A.-M.}\ \bibnamefont {Raclariu}},\ }\href@noop {} {\bibinfo {title} {Lectures on celestial holography}} (\bibinfo {year} {2021}),\ \Eprint {https://arxiv.org/abs/2107.02075} {arXiv:2107.02075 [hep-th]} \BibitemShut {NoStop}%
\bibitem [{\citenamefont {Donnay}(2024)}]{donnay2024CelestialHolographyAsymptotic}%
  \BibitemOpen
  \bibfield  {author} {\bibinfo {author} {\bibfnamefont {L.}~\bibnamefont {Donnay}},\ }\bibfield  {title} {\bibinfo {title} {Celestial holography: {{An}} asymptotic symmetry perspective},\ }\href {https://doi.org/10.1016/j.physrep.2024.04.003} {\bibfield  {journal} {\bibinfo  {journal} {Physics Reports}\ }\textbf {\bibinfo {volume} {1073}},\ \bibinfo {pages} {1} (\bibinfo {year} {2024})},\ \Eprint {https://arxiv.org/abs/2310.12922} {arXiv:2310.12922 [hep-th]} \BibitemShut {NoStop}%
\bibitem [{\citenamefont {Dappiaggi}\ \emph {et~al.}(2006)\citenamefont {Dappiaggi}, \citenamefont {Moretti},\ and\ \citenamefont {Pinamonti}}]{dappiaggi2006RigorousStepsHolography}%
  \BibitemOpen
  \bibfield  {author} {\bibinfo {author} {\bibfnamefont {C.}~\bibnamefont {Dappiaggi}}, \bibinfo {author} {\bibfnamefont {V.}~\bibnamefont {Moretti}},\ and\ \bibinfo {author} {\bibfnamefont {N.}~\bibnamefont {Pinamonti}},\ }\bibfield  {title} {\bibinfo {title} {Rigorous steps towards holography in asymptotically flat spacetimes},\ }\href {https://doi.org/10.1142/S0129055X0600270X} {\bibfield  {journal} {\bibinfo  {journal} {Reviews in Mathematical Physics}\ }\textbf {\bibinfo {volume} {18}},\ \bibinfo {pages} {349} (\bibinfo {year} {2006})},\ \Eprint {https://arxiv.org/abs/gr-qc/0506069} {arXiv:gr-qc/0506069} \BibitemShut {NoStop}%
\bibitem [{\citenamefont {Moretti}(2006)}]{moretti2006UniquenessTheoremBMSInvariant}%
  \BibitemOpen
  \bibfield  {author} {\bibinfo {author} {\bibfnamefont {V.}~\bibnamefont {Moretti}},\ }\bibfield  {title} {\bibinfo {title} {Uniqueness theorem for {{BMS}}-invariant states of scalar {{QFT}} on the null boundary of asymptotically flat spacetimes and bulk-boundary observable algebra correspondence},\ }\href {https://doi.org/10.1007/s00220-006-0107-0} {\bibfield  {journal} {\bibinfo  {journal} {Communications in Mathematical Physics}\ }\textbf {\bibinfo {volume} {268}},\ \bibinfo {pages} {727} (\bibinfo {year} {2006})},\ \Eprint {https://arxiv.org/abs/gr-qc/0512049} {arXiv:gr-qc/0512049} \BibitemShut {NoStop}%
\bibitem [{\citenamefont {Moretti}(2008)}]{moretti2008QuantumOutstatesHolographically}%
  \BibitemOpen
  \bibfield  {author} {\bibinfo {author} {\bibfnamefont {V.}~\bibnamefont {Moretti}},\ }\bibfield  {title} {\bibinfo {title} {Quantum out-states holographically induced by asymptotic flatness: {{Invariance}} under spacetime symmetries, energy positivity and {{Hadamard}} property},\ }\href {https://doi.org/10.1007/s00220-008-0415-7} {\bibfield  {journal} {\bibinfo  {journal} {Communications in Mathematical Physics}\ }\textbf {\bibinfo {volume} {279}},\ \bibinfo {pages} {31} (\bibinfo {year} {2008})},\ \Eprint {https://arxiv.org/abs/gr-qc/0610143} {arXiv:gr-qc/0610143} \BibitemShut {NoStop}%
\bibitem [{\citenamefont {Dappiaggi}\ \emph {et~al.}(2011)\citenamefont {Dappiaggi}, \citenamefont {Moretti},\ and\ \citenamefont {Pinamonti}}]{dappiaggi2011RigorousConstructionHadamard}%
  \BibitemOpen
  \bibfield  {author} {\bibinfo {author} {\bibfnamefont {C.}~\bibnamefont {Dappiaggi}}, \bibinfo {author} {\bibfnamefont {V.}~\bibnamefont {Moretti}},\ and\ \bibinfo {author} {\bibfnamefont {N.}~\bibnamefont {Pinamonti}},\ }\bibfield  {title} {\bibinfo {title} {Rigorous construction and {{Hadamard}} property of the {{Unruh}} state in {{Schwarzschild}} spacetime},\ }\href {https://doi.org/10.4310/ATMP.2011.v15.n2.a4} {\bibfield  {journal} {\bibinfo  {journal} {Advances in Theoretical and Mathematical Physics}\ }\textbf {\bibinfo {volume} {15}},\ \bibinfo {pages} {355} (\bibinfo {year} {2011})},\ \Eprint {https://arxiv.org/abs/0907.1034} {arXiv:0907.1034 [gr-qc]} \BibitemShut {NoStop}%
\bibitem [{\citenamefont {Prabhu}\ \emph {et~al.}(2022)\citenamefont {Prabhu}, \citenamefont {Satishchandran},\ and\ \citenamefont {Wald}}]{prabhu2022InfraredFiniteScattering}%
  \BibitemOpen
  \bibfield  {author} {\bibinfo {author} {\bibfnamefont {K.}~\bibnamefont {Prabhu}}, \bibinfo {author} {\bibfnamefont {G.}~\bibnamefont {Satishchandran}},\ and\ \bibinfo {author} {\bibfnamefont {R.~M.}\ \bibnamefont {Wald}},\ }\bibfield  {title} {\bibinfo {title} {Infrared finite scattering theory in quantum field theory and quantum gravity},\ }\href {https://doi.org/10.1103/PhysRevD.106.066005} {\bibfield  {journal} {\bibinfo  {journal} {Physical Review D}\ }\textbf {\bibinfo {volume} {106}},\ \bibinfo {eid} {066005} (\bibinfo {year} {2022})},\ \Eprint {https://arxiv.org/abs/2203.14334} {arXiv:2203.14334 [hep-th]} \BibitemShut {NoStop}%
\bibitem [{\citenamefont {Prabhu}\ and\ \citenamefont {Satishchandran}(2024{\natexlab{a}})}]{prabhu2024InfraredFiniteScatteringStatesRepresentations}%
  \BibitemOpen
  \bibfield  {author} {\bibinfo {author} {\bibfnamefont {K.}~\bibnamefont {Prabhu}}\ and\ \bibinfo {author} {\bibfnamefont {G.}~\bibnamefont {Satishchandran}},\ }\bibfield  {title} {\bibinfo {title} {Infrared finite scattering theory: {{Scattering}} states and representations of the {{BMS}} group},\ }\href {https://doi.org/10.1007/JHEP08(2024)055} {\bibfield  {journal} {\bibinfo  {journal} {Journal of High Energy Physics}\ }\textbf {\bibinfo {volume} {2024}},\ \bibinfo {eid} {055} (\bibinfo {year} {2024}{\natexlab{a}})},\ \Eprint {https://arxiv.org/abs/2402.00102} {arXiv:2402.00102 [hep-th]} \BibitemShut {NoStop}%
\bibitem [{\citenamefont {Prabhu}\ and\ \citenamefont {Satishchandran}(2024{\natexlab{b}})}]{prabhu2024InfraredFiniteScatteringAmplitudesSoftTheorems}%
  \BibitemOpen
  \bibfield  {author} {\bibinfo {author} {\bibfnamefont {K.}~\bibnamefont {Prabhu}}\ and\ \bibinfo {author} {\bibfnamefont {G.}~\bibnamefont {Satishchandran}},\ }\bibfield  {title} {\bibinfo {title} {Infrared finite scattering theory: {{Amplitudes}} and soft theorems},\ }\href {https://doi.org/10.1103/PhysRevD.110.085022} {\bibfield  {journal} {\bibinfo  {journal} {Physical Review D}\ }\textbf {\bibinfo {volume} {110}},\ \bibinfo {eid} {085022} (\bibinfo {year} {2024}{\natexlab{b}})},\ \Eprint {https://arxiv.org/abs/2402.18637} {arXiv:2402.18637 [hep-th]} \BibitemShut {NoStop}%
\bibitem [{\citenamefont {Danielson}\ \emph {et~al.}(2022{\natexlab{a}})\citenamefont {Danielson}, \citenamefont {Satishchandran},\ and\ \citenamefont {Wald}}]{danielson2022BlackHolesDecohere}%
  \BibitemOpen
  \bibfield  {author} {\bibinfo {author} {\bibfnamefont {D.~L.}\ \bibnamefont {Danielson}}, \bibinfo {author} {\bibfnamefont {G.}~\bibnamefont {Satishchandran}},\ and\ \bibinfo {author} {\bibfnamefont {R.~M.}\ \bibnamefont {Wald}},\ }\bibfield  {title} {\bibinfo {title} {Black holes decohere quantum superpositions},\ }\href {https://doi.org/10.1142/S0218271822410036} {\bibfield  {journal} {\bibinfo  {journal} {International Journal of Modern Physics D}\ }\textbf {\bibinfo {volume} {31}},\ \bibinfo {eid} {2241003} (\bibinfo {year} {2022}{\natexlab{a}})},\ \Eprint {https://arxiv.org/abs/2205.06279} {arXiv:2205.06279 [hep-th]} \BibitemShut {NoStop}%
\bibitem [{\citenamefont {Danielson}\ \emph {et~al.}(2022{\natexlab{b}})\citenamefont {Danielson}, \citenamefont {Satishchandran},\ and\ \citenamefont {Wald}}]{danielson2022KillingHorizonsDecohere}%
  \BibitemOpen
  \bibfield  {author} {\bibinfo {author} {\bibfnamefont {D.~L.}\ \bibnamefont {Danielson}}, \bibinfo {author} {\bibfnamefont {G.}~\bibnamefont {Satishchandran}},\ and\ \bibinfo {author} {\bibfnamefont {R.~M.}\ \bibnamefont {Wald}},\ }\bibfield  {title} {\bibinfo {title} {Killing horizons decohere quantum superpositions},\ }\href {https://doi.org/10.1103/PhysRevD.108.025007} {\bibfield  {journal} {\bibinfo  {journal} {Physical Review D}\ }\textbf {\bibinfo {volume} {108}},\ \bibinfo {eid} {025007} (\bibinfo {year} {2022}{\natexlab{b}})},\ \Eprint {https://arxiv.org/abs/2301.00026} {arXiv:2301.00026 [hep-th]} \BibitemShut {NoStop}%
\bibitem [{\citenamefont {Danielson}\ \emph {et~al.}(2025)\citenamefont {Danielson}, \citenamefont {Satishchandran},\ and\ \citenamefont {Wald}}]{danielson2025LocalDescriptionDecoherence}%
  \BibitemOpen
  \bibfield  {author} {\bibinfo {author} {\bibfnamefont {D.~L.}\ \bibnamefont {Danielson}}, \bibinfo {author} {\bibfnamefont {G.}~\bibnamefont {Satishchandran}},\ and\ \bibinfo {author} {\bibfnamefont {R.~M.}\ \bibnamefont {Wald}},\ }\bibfield  {title} {\bibinfo {title} {Local description of decoherence of quantum superpositions by black holes and other bodies},\ }\href {https://doi.org/10.1103/PhysRevD.111.025014} {\bibfield  {journal} {\bibinfo  {journal} {Physical Review D}\ }\textbf {\bibinfo {volume} {111}},\ \bibinfo {eid} {025014} (\bibinfo {year} {2025})},\ \Eprint {https://arxiv.org/abs/2407.02567} {arXiv:2407.02567 [hep-th]} \BibitemShut {NoStop}%
\bibitem [{\citenamefont {Strominger}(2001)}]{strominger2001DSCFTCorrespondence}%
  \BibitemOpen
  \bibfield  {author} {\bibinfo {author} {\bibfnamefont {A.}~\bibnamefont {Strominger}},\ }\bibfield  {title} {\bibinfo {title} {The {{dS}}/{{CFT}} correspondence},\ }\href {https://doi.org/10.1088/1126-6708/2001/10/034} {\bibfield  {journal} {\bibinfo  {journal} {Journal of High Energy Physics}\ }\textbf {\bibinfo {volume} {2001}},\ \bibinfo {eid} {034} (\bibinfo {year} {2001})},\ \Eprint {https://arxiv.org/abs/hep-th/0106113} {arXiv:hep-th/0106113} \BibitemShut {NoStop}%
\bibitem [{\citenamefont {Koga}(2001)}]{koga2001AsymptoticSymmetriesKilling}%
  \BibitemOpen
  \bibfield  {author} {\bibinfo {author} {\bibfnamefont {J.-i.}\ \bibnamefont {Koga}},\ }\bibfield  {title} {\bibinfo {title} {Asymptotic symmetries on {{Killing}} horizons},\ }\href {https://doi.org/10.1103/PhysRevD.64.124012} {\bibfield  {journal} {\bibinfo  {journal} {Physical Review D}\ }\textbf {\bibinfo {volume} {64}},\ \bibinfo {eid} {124012} (\bibinfo {year} {2001})},\ \Eprint {https://arxiv.org/abs/gr-qc/0107096} {arXiv:gr-qc/0107096} \BibitemShut {NoStop}%
\bibitem [{\citenamefont {Compère}\ \emph {et~al.}(2019)\citenamefont {Compère}, \citenamefont {Fiorucci},\ and\ \citenamefont {Ruzziconi}}]{compere2019LBMS4GroupdS4}%
  \BibitemOpen
  \bibfield  {author} {\bibinfo {author} {\bibfnamefont {G.}~\bibnamefont {Compère}}, \bibinfo {author} {\bibfnamefont {A.}~\bibnamefont {Fiorucci}},\ and\ \bibinfo {author} {\bibfnamefont {R.}~\bibnamefont {Ruzziconi}},\ }\bibfield  {title} {\bibinfo {title} {The ${\Lambda}\text{-BMS}_4$ group of $\text{dS}_4$ and new boundary conditions for $\text{AdS}_4$},\ }\href {https://doi.org/10.1088/1361-6382/ab3d4b} {\bibfield  {journal} {\bibinfo  {journal} {Classical and Quantum Gravity}\ }\textbf {\bibinfo {volume} {36}},\ \bibinfo {eid} {195017} (\bibinfo {year} {2019})},\ \Eprint {https://arxiv.org/abs/1905.00971} {arXiv:1905.00971 [gr-qc]} \BibitemShut {NoStop}%
\bibitem [{\citenamefont {Ruzziconi}(2020{\natexlab{a}})}]{ruzziconi2020VariousExtensionsBMS}%
  \BibitemOpen
  \bibfield  {author} {\bibinfo {author} {\bibfnamefont {R.}~\bibnamefont {Ruzziconi}},\ }\emph {\bibinfo {title} {On the Various Extensions of the {{BMS}} Group}},\ \href@noop {} {Ph.D. thesis},\ \bibinfo  {school} {Université Libre de Bruxelles}, \bibinfo {address} {Brussels, Belgium} (\bibinfo {year} {2020}{\natexlab{a}}),\ \Eprint {https://arxiv.org/abs/2009.01926} {arXiv:2009.01926 [hep-th]} \BibitemShut {NoStop}%
\bibitem [{\citenamefont {Duval}\ \emph {et~al.}(2014{\natexlab{a}})\citenamefont {Duval}, \citenamefont {Gibbons}, \citenamefont {Horvathy},\ and\ \citenamefont {Zhang}}]{duval2014CarrollNewtonGalilei}%
  \BibitemOpen
  \bibfield  {author} {\bibinfo {author} {\bibfnamefont {C.}~\bibnamefont {Duval}}, \bibinfo {author} {\bibfnamefont {G.~W.}\ \bibnamefont {Gibbons}}, \bibinfo {author} {\bibfnamefont {P.~A.}\ \bibnamefont {Horvathy}},\ and\ \bibinfo {author} {\bibfnamefont {P.~M.}\ \bibnamefont {Zhang}},\ }\bibfield  {title} {\bibinfo {title} {Carroll versus {{Newton}} and {{Galilei}}: Two dual non-{{Einsteinian}} concepts of time},\ }\href {https://doi.org/10.1088/0264-9381/31/8/085016} {\bibfield  {journal} {\bibinfo  {journal} {Classical and Quantum Gravity}\ }\textbf {\bibinfo {volume} {31}},\ \bibinfo {eid} {085016} (\bibinfo {year} {2014}{\natexlab{a}})},\ \Eprint {https://arxiv.org/abs/1402.0657} {arXiv:1402.0657 [gr-qc]} \BibitemShut {NoStop}%
\bibitem [{\citenamefont {Duval}\ \emph {et~al.}(2014{\natexlab{b}})\citenamefont {Duval}, \citenamefont {Gibbons},\ and\ \citenamefont {Horvathy}}]{duval2014ConformalCarrollGroups}%
  \BibitemOpen
  \bibfield  {author} {\bibinfo {author} {\bibfnamefont {C.}~\bibnamefont {Duval}}, \bibinfo {author} {\bibfnamefont {G.~W.}\ \bibnamefont {Gibbons}},\ and\ \bibinfo {author} {\bibfnamefont {P.~A.}\ \bibnamefont {Horvathy}},\ }\bibfield  {title} {\bibinfo {title} {Conformal {{Carroll}} groups and {{BMS}} symmetry},\ }\href {https://doi.org/10.1088/0264-9381/31/9/092001} {\bibfield  {journal} {\bibinfo  {journal} {Classical and Quantum Gravity}\ }\textbf {\bibinfo {volume} {31}},\ \bibinfo {eid} {092001} (\bibinfo {year} {2014}{\natexlab{b}})},\ \Eprint {https://arxiv.org/abs/1402.5894} {arXiv:1402.5894 [gr-qc]} \BibitemShut {NoStop}%
\bibitem [{\citenamefont {Duval}\ \emph {et~al.}(2014{\natexlab{c}})\citenamefont {Duval}, \citenamefont {Gibbons},\ and\ \citenamefont {Horvathy}}]{duval2014ConformalCarrollGroupsa}%
  \BibitemOpen
  \bibfield  {author} {\bibinfo {author} {\bibfnamefont {C.}~\bibnamefont {Duval}}, \bibinfo {author} {\bibfnamefont {G.~W.}\ \bibnamefont {Gibbons}},\ and\ \bibinfo {author} {\bibfnamefont {P.~A.}\ \bibnamefont {Horvathy}},\ }\bibfield  {title} {\bibinfo {title} {Conformal {{Carroll}} groups},\ }\href {https://doi.org/10.1088/1751-8113/47/33/335204} {\bibfield  {journal} {\bibinfo  {journal} {Journal of Physics A: Mathematical and Theoretical}\ }\textbf {\bibinfo {volume} {47}},\ \bibinfo {eid} {335204} (\bibinfo {year} {2014}{\natexlab{c}})},\ \Eprint {https://arxiv.org/abs/1403.4213} {arXiv:1403.4213 [hep-th]} \BibitemShut {NoStop}%
\bibitem [{\citenamefont {Ciambelli}\ \emph {et~al.}(2019)\citenamefont {Ciambelli}, \citenamefont {Leigh}, \citenamefont {Marteau},\ and\ \citenamefont {Petropoulos}}]{ciambelli2019CarrollStructuresNull}%
  \BibitemOpen
  \bibfield  {author} {\bibinfo {author} {\bibfnamefont {L.}~\bibnamefont {Ciambelli}}, \bibinfo {author} {\bibfnamefont {R.~G.}\ \bibnamefont {Leigh}}, \bibinfo {author} {\bibfnamefont {C.}~\bibnamefont {Marteau}},\ and\ \bibinfo {author} {\bibfnamefont {P.~M.}\ \bibnamefont {Petropoulos}},\ }\bibfield  {title} {\bibinfo {title} {Carroll structures, null geometry, and conformal isometries},\ }\href {https://doi.org/10.1103/PhysRevD.100.046010} {\bibfield  {journal} {\bibinfo  {journal} {Physical Review D}\ }\textbf {\bibinfo {volume} {100}},\ \bibinfo {eid} {046010} (\bibinfo {year} {2019})},\ \Eprint {https://arxiv.org/abs/1905.02221} {arXiv:1905.02221 [hep-th]} \BibitemShut {NoStop}%
\bibitem [{\citenamefont {Chandrasekaran}\ \emph {et~al.}(2018)\citenamefont {Chandrasekaran}, \citenamefont {Flanagan},\ and\ \citenamefont {Prabhu}}]{chandrasekaran2018SymmetriesChargesGeneral}%
  \BibitemOpen
  \bibfield  {author} {\bibinfo {author} {\bibfnamefont {V.}~\bibnamefont {Chandrasekaran}}, \bibinfo {author} {\bibfnamefont {E.~E.}\ \bibnamefont {Flanagan}},\ and\ \bibinfo {author} {\bibfnamefont {K.}~\bibnamefont {Prabhu}},\ }\bibfield  {title} {\bibinfo {title} {Symmetries and charges of general relativity at null boundaries},\ }\href {https://doi.org/10.1007/JHEP11(2018)125} {\bibfield  {journal} {\bibinfo  {journal} {Journal of High Energy Physics}\ }\textbf {\bibinfo {volume} {2018}},\ \bibinfo {eid} {125} (\bibinfo {year} {2018})},\ \Eprint {https://arxiv.org/abs/1807.11499} {arXiv:1807.11499 [hep-th]} \BibitemShut {NoStop}%
\bibitem [{\citenamefont {Ashtekar}\ and\ \citenamefont {Romano}(1992)}]{ashtekar1992SpatialInfinityBoundary}%
  \BibitemOpen
  \bibfield  {author} {\bibinfo {author} {\bibfnamefont {A.}~\bibnamefont {Ashtekar}}\ and\ \bibinfo {author} {\bibfnamefont {J.~D.}\ \bibnamefont {Romano}},\ }\bibfield  {title} {\bibinfo {title} {Spatial infinity as a boundary of spacetime},\ }\href {https://doi.org/10.1088/0264-9381/9/4/019} {\bibfield  {journal} {\bibinfo  {journal} {Classical and Quantum Gravity}\ }\textbf {\bibinfo {volume} {9}},\ \bibinfo {pages} {1069} (\bibinfo {year} {1992})}\BibitemShut {NoStop}%
\bibitem [{\citenamefont {Newman}\ and\ \citenamefont {Unti}(1962)}]{newman1962BehaviorAsymptoticallyFlat}%
  \BibitemOpen
  \bibfield  {author} {\bibinfo {author} {\bibfnamefont {E.~T.}\ \bibnamefont {Newman}}\ and\ \bibinfo {author} {\bibfnamefont {T.~W.~J.}\ \bibnamefont {Unti}},\ }\bibfield  {title} {\bibinfo {title} {Behavior of asymptotically flat empty spaces},\ }\href {https://doi.org/10.1063/1.1724303} {\bibfield  {journal} {\bibinfo  {journal} {Journal of Mathematical Physics}\ }\textbf {\bibinfo {volume} {3}},\ \bibinfo {pages} {891} (\bibinfo {year} {1962})}\BibitemShut {NoStop}%
\bibitem [{\citenamefont {Eardley}\ \emph {et~al.}(1973{\natexlab{a}})\citenamefont {Eardley}, \citenamefont {Lee},\ and\ \citenamefont {Lightman}}]{eardley1973GravitationalWaveObservationsPRD}%
  \BibitemOpen
  \bibfield  {author} {\bibinfo {author} {\bibfnamefont {D.~M.}\ \bibnamefont {Eardley}}, \bibinfo {author} {\bibfnamefont {D.~L.}\ \bibnamefont {Lee}},\ and\ \bibinfo {author} {\bibfnamefont {A.~P.}\ \bibnamefont {Lightman}},\ }\bibfield  {title} {\bibinfo {title} {Gravitational-wave observations as a tool for testing relativistic gravity},\ }\href {https://doi.org/10.1103/PhysRevD.8.3308} {\bibfield  {journal} {\bibinfo  {journal} {Physical Review D}\ }\textbf {\bibinfo {volume} {8}},\ \bibinfo {pages} {3308} (\bibinfo {year} {1973}{\natexlab{a}})}\BibitemShut {NoStop}%
\bibitem [{\citenamefont {Eardley}\ \emph {et~al.}(1973{\natexlab{b}})\citenamefont {Eardley}, \citenamefont {Lee}, \citenamefont {Lightman}, \citenamefont {Wagoner},\ and\ \citenamefont {Will}}]{eardley1973GravitationalWaveObservationsPRL}%
  \BibitemOpen
  \bibfield  {author} {\bibinfo {author} {\bibfnamefont {D.~M.}\ \bibnamefont {Eardley}}, \bibinfo {author} {\bibfnamefont {D.~L.}\ \bibnamefont {Lee}}, \bibinfo {author} {\bibfnamefont {A.~P.}\ \bibnamefont {Lightman}}, \bibinfo {author} {\bibfnamefont {R.~V.}\ \bibnamefont {Wagoner}},\ and\ \bibinfo {author} {\bibfnamefont {C.~M.}\ \bibnamefont {Will}},\ }\bibfield  {title} {\bibinfo {title} {Gravitational-wave observations as a tool for testing relativistic gravity},\ }\href {https://doi.org/10.1103/PhysRevLett.30.884} {\bibfield  {journal} {\bibinfo  {journal} {Physical Review Letters}\ }\textbf {\bibinfo {volume} {30}},\ \bibinfo {pages} {884} (\bibinfo {year} {1973}{\natexlab{b}})}\BibitemShut {NoStop}%
\bibitem [{\citenamefont {Will}(2014)}]{will2014ConfrontationGeneralRelativity}%
  \BibitemOpen
  \bibfield  {author} {\bibinfo {author} {\bibfnamefont {C.~M.}\ \bibnamefont {Will}},\ }\bibfield  {title} {\bibinfo {title} {The confrontation between general relativity and experiment},\ }\href {https://doi.org/10.12942/lrr-2014-4} {\bibfield  {journal} {\bibinfo  {journal} {Living Reviews in Relativity}\ }\textbf {\bibinfo {volume} {17}},\ \bibinfo {eid} {4} (\bibinfo {year} {2014})},\ \Eprint {https://arxiv.org/abs/1403.7377} {arXiv:1403.7377 [gr-qc]} \BibitemShut {NoStop}%
\bibitem [{\citenamefont {Wald}(1984)}]{wald1984GeneralRelativity}%
  \BibitemOpen
  \bibfield  {author} {\bibinfo {author} {\bibfnamefont {R.~M.}\ \bibnamefont {Wald}},\ }\href {https://doi.org/10.7208/chicago/9780226870373.001.0001} {\emph {\bibinfo {title} {General Relativity}}}\ (\bibinfo  {publisher} {The University of Chicago Press},\ \bibinfo {address} {Chicago, IL},\ \bibinfo {year} {1984})\BibitemShut {NoStop}%
\bibitem [{\citenamefont {{Lévy-Leblond}}(1965)}]{levy-leblond1965NouvelleLimiteNonrelativiste}%
  \BibitemOpen
  \bibfield  {author} {\bibinfo {author} {\bibfnamefont {J.-M.}\ \bibnamefont {{Lévy-Leblond}}},\ }\bibfield  {title} {\bibinfo {title} {Une nouvelle limite non-relativiste du groupe de {{Poincaré}}},\ }\href {http://www.numdam.org/item/AIHPA_1965__3_1_1_0/} {\bibfield  {journal} {\bibinfo  {journal} {Annales de l'institut Henri Poincaré. Section A, Physique Théorique}\ }\textbf {\bibinfo {volume} {3}},\ \bibinfo {pages} {1} (\bibinfo {year} {1965})}\BibitemShut {NoStop}%
\bibitem [{\citenamefont {Sen~Gupta}(1966)}]{sengupta1966AnalogueGalileiGroup}%
  \BibitemOpen
  \bibfield  {author} {\bibinfo {author} {\bibfnamefont {N.~D.}\ \bibnamefont {Sen~Gupta}},\ }\bibfield  {title} {\bibinfo {title} {On an analogue of the {{Galilei}} group},\ }\href {https://doi.org/10.1007/BF02740871} {\bibfield  {journal} {\bibinfo  {journal} {Il Nuovo Cimento A}\ }\textbf {\bibinfo {volume} {44}},\ \bibinfo {pages} {512} (\bibinfo {year} {1966})}\BibitemShut {NoStop}%
\bibitem [{\citenamefont {Hamilton}(2017)}]{hamilton2017MathematicalGaugeTheory}%
  \BibitemOpen
  \bibfield  {author} {\bibinfo {author} {\bibfnamefont {M.~J.}\ \bibnamefont {Hamilton}},\ }\href {https://doi.org/10.1007/978-3-319-68439-0} {\emph {\bibinfo {title} {Mathematical Gauge Theory}}},\ Universitext\ (\bibinfo  {publisher} {Springer},\ \bibinfo {address} {Cham},\ \bibinfo {year} {2017})\BibitemShut {NoStop}%
\bibitem [{\citenamefont {Kobayashi}\ and\ \citenamefont {Nomizu}(1963)}]{kobayashi1963FoundationsDifferentialGeometry}%
  \BibitemOpen
  \bibfield  {author} {\bibinfo {author} {\bibfnamefont {S.}~\bibnamefont {Kobayashi}}\ and\ \bibinfo {author} {\bibfnamefont {K.}~\bibnamefont {Nomizu}},\ }\href@noop {} {\emph {\bibinfo {title} {Foundations of Differential Geometry}}},\ Vol.~\bibinfo {volume} {I}\ (\bibinfo  {publisher} {Interscience Publishers},\ \bibinfo {address} {New York, NY},\ \bibinfo {year} {1963})\BibitemShut {NoStop}%
\bibitem [{\citenamefont {Kolář}\ \emph {et~al.}(1993)\citenamefont {Kolář}, \citenamefont {Michor},\ and\ \citenamefont {Slovák}}]{kolar1993NaturalOperationsDifferential}%
  \BibitemOpen
  \bibfield  {author} {\bibinfo {author} {\bibfnamefont {I.}~\bibnamefont {Kolář}}, \bibinfo {author} {\bibfnamefont {P.~W.}\ \bibnamefont {Michor}},\ and\ \bibinfo {author} {\bibfnamefont {J.}~\bibnamefont {Slovák}},\ }\href {https://doi.org/10.1007/978-3-662-02950-3} {\emph {\bibinfo {title} {Natural Operations in Differential Geometry}}}\ (\bibinfo  {publisher} {Springer},\ \bibinfo {address} {Berlin},\ \bibinfo {year} {1993})\BibitemShut {NoStop}%
\bibitem [{\citenamefont {Tu}(2017)}]{tu2017DifferentialGeometryConnections}%
  \BibitemOpen
  \bibfield  {author} {\bibinfo {author} {\bibfnamefont {L.~W.}\ \bibnamefont {Tu}},\ }\href {https://doi.org/10.1007/978-3-319-55084-8} {\emph {\bibinfo {title} {Differential Geometry: {{Connections}}, Curvature, and Characteristic Classes}}},\ \bibinfo {series} {Graduate Texts in Mathematics}\ No.\ \bibinfo {number} {275}\ (\bibinfo  {publisher} {Springer},\ \bibinfo {address} {Cham},\ \bibinfo {year} {2017})\BibitemShut {NoStop}%
\bibitem [{\citenamefont {Hawking}\ and\ \citenamefont {Ellis}(1973)}]{hawking1973LargeScaleStructure}%
  \BibitemOpen
  \bibfield  {author} {\bibinfo {author} {\bibfnamefont {S.~W.}\ \bibnamefont {Hawking}}\ and\ \bibinfo {author} {\bibfnamefont {G.~F.~R.}\ \bibnamefont {Ellis}},\ }\href {https://doi.org/10.1017/CBO9780511524646} {\emph {\bibinfo {title} {The Large Scale Structure of Spacetime}}},\ Cambridge Monographs on Mathematical Physics\ (\bibinfo  {publisher} {Cambridge University Press},\ \bibinfo {address} {Cambridge},\ \bibinfo {year} {1973})\BibitemShut {NoStop}%
\bibitem [{\citenamefont {Penrose}(1963)}]{penrose1963AsymptoticPropertiesFields}%
  \BibitemOpen
  \bibfield  {author} {\bibinfo {author} {\bibfnamefont {R.}~\bibnamefont {Penrose}},\ }\bibfield  {title} {\bibinfo {title} {Asymptotic properties of fields and space-times},\ }\href {https://doi.org/10.1103/PhysRevLett.10.66} {\bibfield  {journal} {\bibinfo  {journal} {Physical Review Letters}\ }\textbf {\bibinfo {volume} {10}},\ \bibinfo {pages} {66} (\bibinfo {year} {1963})}\BibitemShut {NoStop}%
\bibitem [{\citenamefont {Penrose}(1965)}]{penrose1965ZeroRestmassFields}%
  \BibitemOpen
  \bibfield  {author} {\bibinfo {author} {\bibfnamefont {R.}~\bibnamefont {Penrose}},\ }\bibfield  {title} {\bibinfo {title} {Zero rest-mass fields including gravitation: Asymptotic behaviour},\ }\href {https://doi.org/10.1098/rspa.1965.0058} {\bibfield  {journal} {\bibinfo  {journal} {Proceedings of the Royal Society of London. Series A. Mathematical and Physical Sciences}\ }\textbf {\bibinfo {volume} {284}},\ \bibinfo {pages} {159} (\bibinfo {year} {1965})}\BibitemShut {NoStop}%
\bibitem [{\citenamefont {Christodoulou}\ and\ \citenamefont {Klainerman}(1993)}]{christodoulou1993GlobalNonlinearStability}%
  \BibitemOpen
  \bibfield  {author} {\bibinfo {author} {\bibfnamefont {D.}~\bibnamefont {Christodoulou}}\ and\ \bibinfo {author} {\bibfnamefont {S.}~\bibnamefont {Klainerman}},\ }\href {https://doi.org/10.1515/9781400863174} {\emph {\bibinfo {title} {The Global Nonlinear Stability of the {{Minkowski}} Space}}},\ \bibinfo {series} {Princeton Mathematical Series}\ No.~\bibinfo {number} {41}\ (\bibinfo  {publisher} {Princeton University Press},\ \bibinfo {address} {Princeton, NJ},\ \bibinfo {year} {1993})\BibitemShut {NoStop}%
\bibitem [{\citenamefont {Bieri}(2007)}]{bieri2007ExtensionStabilityTheorem}%
  \BibitemOpen
  \bibfield  {author} {\bibinfo {author} {\bibfnamefont {L.}~\bibnamefont {Bieri}},\ }\emph {\bibinfo {title} {An Extension of the Stability Theorem of the {{Minkowski}} Space in General Relativity}},\ \href {https://doi.org/10.3929/ETHZ-A-005463083} {Ph.D. thesis},\ \bibinfo  {school} {ETH Zurich}, \bibinfo {address} {Zurich, Switzerland} (\bibinfo {year} {2007})\BibitemShut {NoStop}%
\bibitem [{\citenamefont {Bieri}(2009)}]{bieri2009ExtensionStabilityTheorem}%
  \BibitemOpen
  \bibfield  {author} {\bibinfo {author} {\bibfnamefont {L.}~\bibnamefont {Bieri}},\ }\href@noop {} {\bibinfo {title} {An extension of the stability theorem of the {{Minkowski}} space in general relativity}} (\bibinfo {year} {2009}),\ \Eprint {https://arxiv.org/abs/0904.0620} {arXiv:0904.0620 [gr-qc]} \BibitemShut {NoStop}%
\bibitem [{\citenamefont {Bieri}\ and\ \citenamefont {Zipser}(2009)}]{bieri2009ExtensionsStabilityTheorem}%
  \BibitemOpen
  \bibfield  {author} {\bibinfo {author} {\bibfnamefont {L.}~\bibnamefont {Bieri}}\ and\ \bibinfo {author} {\bibfnamefont {N.}~\bibnamefont {Zipser}},\ }\href {https://doi.org/10.1090/amsip/045} {\emph {\bibinfo {title} {Extensions of the Stability Theorem of the {{Minkowski}} Space in General Relativity}}},\ \bibinfo {series} {{{AMS}}/{{IP}} Studies in Advanced Mathematics}\ No.~\bibinfo {number} {45}\ (\bibinfo  {publisher} {American Mathematical Society},\ \bibinfo {address} {Providence, RI},\ \bibinfo {year} {2009})\BibitemShut {NoStop}%
\bibitem [{\citenamefont {Cachazo}\ and\ \citenamefont {Strominger}(2014)}]{cachazo2014EvidenceNewSoft}%
  \BibitemOpen
  \bibfield  {author} {\bibinfo {author} {\bibfnamefont {F.}~\bibnamefont {Cachazo}}\ and\ \bibinfo {author} {\bibfnamefont {A.}~\bibnamefont {Strominger}},\ }\href@noop {} {\bibinfo {title} {Evidence for a new soft graviton theorem}} (\bibinfo {year} {2014}),\ \Eprint {https://arxiv.org/abs/1404.4091} {arXiv:1404.4091 [hep-th]} \BibitemShut {NoStop}%
\bibitem [{\citenamefont {Pasterski}\ \emph {et~al.}(2016)\citenamefont {Pasterski}, \citenamefont {Strominger},\ and\ \citenamefont {Zhiboedov}}]{pasterski2016NewGravitationalMemories}%
  \BibitemOpen
  \bibfield  {author} {\bibinfo {author} {\bibfnamefont {S.}~\bibnamefont {Pasterski}}, \bibinfo {author} {\bibfnamefont {A.}~\bibnamefont {Strominger}},\ and\ \bibinfo {author} {\bibfnamefont {A.}~\bibnamefont {Zhiboedov}},\ }\bibfield  {title} {\bibinfo {title} {New gravitational memories},\ }\href {https://doi.org/10.1007/JHEP12(2016)053} {\bibfield  {journal} {\bibinfo  {journal} {Journal of High Energy Physics}\ }\textbf {\bibinfo {volume} {2016}},\ \bibinfo {eid} {53} (\bibinfo {year} {2016})},\ \Eprint {https://arxiv.org/abs/1502.06120} {arXiv:1502.06120 [hep-th]} \BibitemShut {NoStop}%
\bibitem [{\citenamefont {Pasterski}(2019)}]{pasterski2019ImplicationsSuperrotations}%
  \BibitemOpen
  \bibfield  {author} {\bibinfo {author} {\bibfnamefont {S.}~\bibnamefont {Pasterski}},\ }\bibfield  {title} {\bibinfo {title} {Implications of superrotations},\ }\href {https://doi.org/10.1016/j.physrep.2019.09.006} {\bibfield  {journal} {\bibinfo  {journal} {Physics Reports}\ }\textbf {\bibinfo {volume} {829}},\ \bibinfo {pages} {1} (\bibinfo {year} {2019})},\ \Eprint {https://arxiv.org/abs/1905.10052} {arXiv:1905.10052 [hep-th]} \BibitemShut {NoStop}%
\bibitem [{\citenamefont {Donnay}\ \emph {et~al.}(2016{\natexlab{a}})\citenamefont {Donnay}, \citenamefont {Giribet}, \citenamefont {González},\ and\ \citenamefont {Pino}}]{donnay2016ExtendedSymmetriesBlack}%
  \BibitemOpen
  \bibfield  {author} {\bibinfo {author} {\bibfnamefont {L.}~\bibnamefont {Donnay}}, \bibinfo {author} {\bibfnamefont {G.}~\bibnamefont {Giribet}}, \bibinfo {author} {\bibfnamefont {H.~A.}\ \bibnamefont {González}},\ and\ \bibinfo {author} {\bibfnamefont {M.}~\bibnamefont {Pino}},\ }\bibfield  {title} {\bibinfo {title} {Extended symmetries at the black hole horizon},\ }\href {https://doi.org/10.1007/JHEP09(2016)100} {\bibfield  {journal} {\bibinfo  {journal} {Journal of High Energy Physics}\ }\textbf {\bibinfo {volume} {2016}},\ \bibinfo {eid} {100} (\bibinfo {year} {2016}{\natexlab{a}})},\ \Eprint {https://arxiv.org/abs/1607.05703} {arXiv:1607.05703 [hep-th]} \BibitemShut {NoStop}%
\bibitem [{\citenamefont {Donnay}\ \emph {et~al.}(2016{\natexlab{b}})\citenamefont {Donnay}, \citenamefont {Giribet}, \citenamefont {Gonzalez},\ and\ \citenamefont {Pino}}]{donnay2016SupertranslationsSuperrotationsBlack}%
  \BibitemOpen
  \bibfield  {author} {\bibinfo {author} {\bibfnamefont {L.}~\bibnamefont {Donnay}}, \bibinfo {author} {\bibfnamefont {G.}~\bibnamefont {Giribet}}, \bibinfo {author} {\bibfnamefont {H.~A.}\ \bibnamefont {Gonzalez}},\ and\ \bibinfo {author} {\bibfnamefont {M.}~\bibnamefont {Pino}},\ }\bibfield  {title} {\bibinfo {title} {Supertranslations and superrotations at the black hole horizon},\ }\href {https://doi.org/10.1103/PhysRevLett.116.091101} {\bibfield  {journal} {\bibinfo  {journal} {Physical Review Letters}\ }\textbf {\bibinfo {volume} {116}},\ \bibinfo {eid} {091101} (\bibinfo {year} {2016}{\natexlab{b}})},\ \Eprint {https://arxiv.org/abs/1511.08687} {arXiv:1511.08687 [hep-th]} \BibitemShut {NoStop}%
\bibitem [{\citenamefont {Grumiller}\ \emph {et~al.}(2020{\natexlab{a}})\citenamefont {Grumiller}, \citenamefont {{Sheikh-Jabbari}},\ and\ \citenamefont {Zwikel}}]{grumiller2020Horizons2020}%
  \BibitemOpen
  \bibfield  {author} {\bibinfo {author} {\bibfnamefont {D.}~\bibnamefont {Grumiller}}, \bibinfo {author} {\bibfnamefont {M.~M.}\ \bibnamefont {{Sheikh-Jabbari}}},\ and\ \bibinfo {author} {\bibfnamefont {C.}~\bibnamefont {Zwikel}},\ }\bibfield  {title} {\bibinfo {title} {Horizons 2020},\ }\href {https://doi.org/10.1142/S0218271820430063} {\bibfield  {journal} {\bibinfo  {journal} {International Journal of Modern Physics D}\ }\textbf {\bibinfo {volume} {29}},\ \bibinfo {eid} {2043006} (\bibinfo {year} {2020}{\natexlab{a}})},\ \Eprint {https://arxiv.org/abs/2005.06936} {arXiv:2005.06936 [hep-th]} \BibitemShut {NoStop}%
\bibitem [{\citenamefont {Grumiller}\ \emph {et~al.}(2020{\natexlab{b}})\citenamefont {Grumiller}, \citenamefont {Pérez}, \citenamefont {{Sheikh-Jabbari}}, \citenamefont {Troncoso},\ and\ \citenamefont {Zwikel}}]{grumiller2020SpacetimeStructureGeneric}%
  \BibitemOpen
  \bibfield  {author} {\bibinfo {author} {\bibfnamefont {D.}~\bibnamefont {Grumiller}}, \bibinfo {author} {\bibfnamefont {A.}~\bibnamefont {Pérez}}, \bibinfo {author} {\bibfnamefont {M.~M.}\ \bibnamefont {{Sheikh-Jabbari}}}, \bibinfo {author} {\bibfnamefont {R.}~\bibnamefont {Troncoso}},\ and\ \bibinfo {author} {\bibfnamefont {C.}~\bibnamefont {Zwikel}},\ }\bibfield  {title} {\bibinfo {title} {Spacetime structure near generic horizons and soft hair},\ }\href {https://doi.org/10.1103/PhysRevLett.124.041601} {\bibfield  {journal} {\bibinfo  {journal} {Physical Review Letters}\ }\textbf {\bibinfo {volume} {124}},\ \bibinfo {eid} {041601} (\bibinfo {year} {2020}{\natexlab{b}})},\ \Eprint {https://arxiv.org/abs/1908.09833} {arXiv:1908.09833 [hep-th]} \BibitemShut {NoStop}%
\bibitem [{\citenamefont {Adami}\ \emph {et~al.}(2021)\citenamefont {Adami}, \citenamefont {Grumiller}, \citenamefont {{Sheikh-Jabbari}}, \citenamefont {Taghiloo}, \citenamefont {Yavartanoo},\ and\ \citenamefont {Zwikel}}]{adami2021NullBoundaryPhase}%
  \BibitemOpen
  \bibfield  {author} {\bibinfo {author} {\bibfnamefont {H.}~\bibnamefont {Adami}}, \bibinfo {author} {\bibfnamefont {D.}~\bibnamefont {Grumiller}}, \bibinfo {author} {\bibfnamefont {M.~M.}\ \bibnamefont {{Sheikh-Jabbari}}}, \bibinfo {author} {\bibfnamefont {V.}~\bibnamefont {Taghiloo}}, \bibinfo {author} {\bibfnamefont {H.}~\bibnamefont {Yavartanoo}},\ and\ \bibinfo {author} {\bibfnamefont {C.}~\bibnamefont {Zwikel}},\ }\bibfield  {title} {\bibinfo {title} {Null boundary phase space: Slicings, news and memory},\ }\href {https://doi.org/10.1007/JHEP11(2021)155} {\bibfield  {journal} {\bibinfo  {journal} {Journal of High Energy Physics}\ }\textbf {\bibinfo {volume} {2021}},\ \bibinfo {eid} {155} (\bibinfo {year} {2021})},\ \Eprint {https://arxiv.org/abs/2110.04218} {arXiv:2110.04218 [hep-th]} \BibitemShut {NoStop}%
\bibitem [{\citenamefont {Bondi}(1960)}]{bondi1960GravitationalWavesGeneral}%
  \BibitemOpen
  \bibfield  {author} {\bibinfo {author} {\bibfnamefont {H.}~\bibnamefont {Bondi}},\ }\bibfield  {title} {\bibinfo {title} {Gravitational waves in general relativity},\ }\href {https://doi.org/10.1038/186535a0} {\bibfield  {journal} {\bibinfo  {journal} {Nature}\ }\textbf {\bibinfo {volume} {186}},\ \bibinfo {pages} {535} (\bibinfo {year} {1960})}\BibitemShut {NoStop}%
\bibitem [{\citenamefont {Campiglia}\ and\ \citenamefont {Laddha}(2014)}]{campiglia2014AsymptoticSymmetriesSubleading}%
  \BibitemOpen
  \bibfield  {author} {\bibinfo {author} {\bibfnamefont {M.}~\bibnamefont {Campiglia}}\ and\ \bibinfo {author} {\bibfnamefont {A.}~\bibnamefont {Laddha}},\ }\bibfield  {title} {\bibinfo {title} {Asymptotic symmetries and subleading soft graviton theorem},\ }\href {https://doi.org/10.1103/PhysRevD.90.124028} {\bibfield  {journal} {\bibinfo  {journal} {Physical Review D}\ }\textbf {\bibinfo {volume} {90}},\ \bibinfo {eid} {124028} (\bibinfo {year} {2014})},\ \Eprint {https://arxiv.org/abs/1408.2228} {arXiv:1408.2228 [hep-th]} \BibitemShut {NoStop}%
\bibitem [{\citenamefont {Wald}\ and\ \citenamefont {Zoupas}(2000)}]{wald2000GeneralDefinitionConserved}%
  \BibitemOpen
  \bibfield  {author} {\bibinfo {author} {\bibfnamefont {R.~M.}\ \bibnamefont {Wald}}\ and\ \bibinfo {author} {\bibfnamefont {A.}~\bibnamefont {Zoupas}},\ }\bibfield  {title} {\bibinfo {title} {A general definition of ``conserved quantities'' in general relativity and other theories of gravity},\ }\href {https://doi.org/10.1103/PhysRevD.61.084027} {\bibfield  {journal} {\bibinfo  {journal} {Physical Review D}\ }\textbf {\bibinfo {volume} {61}},\ \bibinfo {eid} {084027} (\bibinfo {year} {2000})},\ \Eprint {https://arxiv.org/abs/gr-qc/9911095} {arXiv:gr-qc/9911095} \BibitemShut {NoStop}%
\bibitem [{\citenamefont {Gieres}(2023)}]{gieres2023CovariantCanonicalFormulations}%
  \BibitemOpen
  \bibfield  {author} {\bibinfo {author} {\bibfnamefont {F.}~\bibnamefont {Gieres}},\ }\bibfield  {title} {\bibinfo {title} {Covariant canonical formulations of classical field theories},\ }\href {https://doi.org/10.21468/SciPostPhysLectNotes.77} {\bibfield  {journal} {\bibinfo  {journal} {SciPost Physics Lecture Notes}\ ,\ \bibinfo {eid} {77}} (\bibinfo {year} {2023})},\ \Eprint {https://arxiv.org/abs/2109.07330} {arXiv:2109.07330 [hep-th]} \BibitemShut {NoStop}%
\bibitem [{\citenamefont {Ruzziconi}(2020{\natexlab{b}})}]{ruzziconi2020AsymptoticSymmetriesGauge}%
  \BibitemOpen
  \bibfield  {author} {\bibinfo {author} {\bibfnamefont {R.}~\bibnamefont {Ruzziconi}},\ }\bibfield  {title} {\bibinfo {title} {Asymptotic symmetries in the gauge fixing approach and the {{BMS}} group},\ }in\ \href {https://doi.org/10.22323/1.384.0003} {\emph {\bibinfo {booktitle} {Proceedings of {{XV Modave Summer School}} in {{Mathematical Physics}} — {{PoS}}({{Modave2019}})}}}\ (\bibinfo  {publisher} {Sissa Medialab},\ \bibinfo {address} {Modave, Belgium},\ \bibinfo {year} {2020})\ \Eprint {https://arxiv.org/abs/1910.08367} {arXiv:1910.08367 [hep-th]} \BibitemShut {NoStop}%
\bibitem [{\citenamefont {Carter}(1966)}]{carter1966CompleteAnalyticExtension}%
  \BibitemOpen
  \bibfield  {author} {\bibinfo {author} {\bibfnamefont {B.}~\bibnamefont {Carter}},\ }\bibfield  {title} {\bibinfo {title} {Complete analytic extension of the symmetry axis of {{Kerr}}'s solution of {{Einstein}}'s equations},\ }\href {https://doi.org/10.1103/PhysRev.141.1242} {\bibfield  {journal} {\bibinfo  {journal} {Physical Review}\ }\textbf {\bibinfo {volume} {141}},\ \bibinfo {pages} {1242} (\bibinfo {year} {1966})}\BibitemShut {NoStop}%
\bibitem [{\citenamefont {Carter}(1969)}]{carter1969KillingHorizonsOrthogonally}%
  \BibitemOpen
  \bibfield  {author} {\bibinfo {author} {\bibfnamefont {B.}~\bibnamefont {Carter}},\ }\bibfield  {title} {\bibinfo {title} {Killing horizons and orthogonally transitive groups in space-time},\ }\href {https://doi.org/10.1063/1.1664763} {\bibfield  {journal} {\bibinfo  {journal} {Journal of Mathematical Physics}\ }\textbf {\bibinfo {volume} {10}},\ \bibinfo {pages} {70} (\bibinfo {year} {1969})}\BibitemShut {NoStop}%
\bibitem [{\citenamefont {Dyer}\ and\ \citenamefont {Honig}(1979)}]{dyer1979ConformalKillingHorizons}%
  \BibitemOpen
  \bibfield  {author} {\bibinfo {author} {\bibfnamefont {C.~C.}\ \bibnamefont {Dyer}}\ and\ \bibinfo {author} {\bibfnamefont {E.}~\bibnamefont {Honig}},\ }\bibfield  {title} {\bibinfo {title} {Conformal {{Killing}} horizons},\ }\href {https://doi.org/10.1063/1.524078} {\bibfield  {journal} {\bibinfo  {journal} {Journal of Mathematical Physics}\ }\textbf {\bibinfo {volume} {20}},\ \bibinfo {pages} {409} (\bibinfo {year} {1979})}\BibitemShut {NoStop}%
\bibitem [{\citenamefont {Sultana}\ and\ \citenamefont {Dyer}(2004)}]{sultana2004ConformalKillingHorizons}%
  \BibitemOpen
  \bibfield  {author} {\bibinfo {author} {\bibfnamefont {J.}~\bibnamefont {Sultana}}\ and\ \bibinfo {author} {\bibfnamefont {C.~C.}\ \bibnamefont {Dyer}},\ }\bibfield  {title} {\bibinfo {title} {Conformal {{Killing}} horizons},\ }\href {https://doi.org/10.1063/1.1814417} {\bibfield  {journal} {\bibinfo  {journal} {Journal of Mathematical Physics}\ }\textbf {\bibinfo {volume} {45}},\ \bibinfo {pages} {4764} (\bibinfo {year} {2004})}\BibitemShut {NoStop}%
\bibitem [{\citenamefont {Nielsen}\ and\ \citenamefont {Shoom}(2018)}]{nielsen2018ConformalKillingHorizons}%
  \BibitemOpen
  \bibfield  {author} {\bibinfo {author} {\bibfnamefont {A.~B.}\ \bibnamefont {Nielsen}}\ and\ \bibinfo {author} {\bibfnamefont {A.~A.}\ \bibnamefont {Shoom}},\ }\bibfield  {title} {\bibinfo {title} {Conformal {{Killing}} horizons and their thermodynamics},\ }\href {https://doi.org/10.1088/1361-6382/aab505} {\bibfield  {journal} {\bibinfo  {journal} {Classical and Quantum Gravity}\ }\textbf {\bibinfo {volume} {35}},\ \bibinfo {eid} {105008} (\bibinfo {year} {2018})},\ \Eprint {https://arxiv.org/abs/1708.08015} {arXiv:1708.08015 [gr-qc]} \BibitemShut {NoStop}%
\bibitem [{\citenamefont {Freidel}\ \emph {et~al.}(2021)\citenamefont {Freidel}, \citenamefont {Oliveri}, \citenamefont {Pranzetti},\ and\ \citenamefont {Speziale}}]{freidel2021WeylBMSGroup}%
  \BibitemOpen
  \bibfield  {author} {\bibinfo {author} {\bibfnamefont {L.}~\bibnamefont {Freidel}}, \bibinfo {author} {\bibfnamefont {R.}~\bibnamefont {Oliveri}}, \bibinfo {author} {\bibfnamefont {D.}~\bibnamefont {Pranzetti}},\ and\ \bibinfo {author} {\bibfnamefont {S.}~\bibnamefont {Speziale}},\ }\bibfield  {title} {\bibinfo {title} {The {{Weyl}} {{BMS}} group and {{Einstein}}'s equations},\ }\href {https://doi.org/10.1007/JHEP07(2021)170} {\bibfield  {journal} {\bibinfo  {journal} {Journal of High Energy Physics}\ }\textbf {\bibinfo {volume} {2021}},\ \bibinfo {eid} {170} (\bibinfo {year} {2021})},\ \Eprint {https://arxiv.org/abs/2104.05793} {arXiv:2104.05793 [hep-th]} \BibitemShut {NoStop}%
\bibitem [{\citenamefont {Haco}\ \emph {et~al.}(2017)\citenamefont {Haco}, \citenamefont {Hawking}, \citenamefont {Perry},\ and\ \citenamefont {Bourjaily}}]{haco2017ConformalBMSGroup}%
  \BibitemOpen
  \bibfield  {author} {\bibinfo {author} {\bibfnamefont {S.~J.}\ \bibnamefont {Haco}}, \bibinfo {author} {\bibfnamefont {S.~W.}\ \bibnamefont {Hawking}}, \bibinfo {author} {\bibfnamefont {M.~J.}\ \bibnamefont {Perry}},\ and\ \bibinfo {author} {\bibfnamefont {J.~L.}\ \bibnamefont {Bourjaily}},\ }\bibfield  {title} {\bibinfo {title} {The conformal {{BMS}} group},\ }\href {https://doi.org/10.1007/JHEP11(2017)012} {\bibfield  {journal} {\bibinfo  {journal} {Journal of High Energy Physics}\ }\textbf {\bibinfo {volume} {2017}},\ \bibinfo {eid} {12} (\bibinfo {year} {2017})},\ \Eprint {https://arxiv.org/abs/1701.08110} {arXiv:1701.08110 [hep-th]} \BibitemShut {NoStop}%
\bibitem [{\citenamefont {Prinz}\ and\ \citenamefont {Schmeding}(2022{\natexlab{a}})}]{prinz2022LieTheoryAsymptoticNU}%
  \BibitemOpen
  \bibfield  {author} {\bibinfo {author} {\bibfnamefont {D.}~\bibnamefont {Prinz}}\ and\ \bibinfo {author} {\bibfnamefont {A.}~\bibnamefont {Schmeding}},\ }\bibfield  {title} {\bibinfo {title} {Lie theory for asymptotic symmetries in general relativity: {{The}} {{NU}} group},\ }\href {https://doi.org/10.1088/1361-6382/ac776c} {\bibfield  {journal} {\bibinfo  {journal} {Classical and Quantum Gravity}\ }\textbf {\bibinfo {volume} {39}},\ \bibinfo {eid} {155005} (\bibinfo {year} {2022}{\natexlab{a}})},\ \Eprint {https://arxiv.org/abs/2109.11476} {arXiv:2109.11476 [gr-qc]} \BibitemShut {NoStop}%
\bibitem [{\citenamefont {Prinz}\ and\ \citenamefont {Schmeding}(2022{\natexlab{b}})}]{prinz2022LieTheoryAsymptoticBMS}%
  \BibitemOpen
  \bibfield  {author} {\bibinfo {author} {\bibfnamefont {D.}~\bibnamefont {Prinz}}\ and\ \bibinfo {author} {\bibfnamefont {A.}~\bibnamefont {Schmeding}},\ }\bibfield  {title} {\bibinfo {title} {Lie theory for asymptotic symmetries in general relativity: {{The}} {{BMS}} group},\ }\href {https://doi.org/10.1088/1361-6382/ac4ae2} {\bibfield  {journal} {\bibinfo  {journal} {Classical and Quantum Gravity}\ }\textbf {\bibinfo {volume} {39}},\ \bibinfo {eid} {065004} (\bibinfo {year} {2022}{\natexlab{b}})},\ \Eprint {https://arxiv.org/abs/2106.12513} {arXiv:2106.12513 [gr-qc]} \BibitemShut {NoStop}%
\bibitem [{\citenamefont {Brans}\ and\ \citenamefont {Dicke}(1961)}]{brans1961MachsPrincipleRelativistic}%
  \BibitemOpen
  \bibfield  {author} {\bibinfo {author} {\bibfnamefont {C.}~\bibnamefont {Brans}}\ and\ \bibinfo {author} {\bibfnamefont {R.~H.}\ \bibnamefont {Dicke}},\ }\bibfield  {title} {\bibinfo {title} {Mach's principle and a relativistic theory of gravitation},\ }\href {https://doi.org/10.1103/PhysRev.124.925} {\bibfield  {journal} {\bibinfo  {journal} {Physical Review}\ }\textbf {\bibinfo {volume} {124}},\ \bibinfo {pages} {925} (\bibinfo {year} {1961})}\BibitemShut {NoStop}%
\bibitem [{\citenamefont {Hou}\ and\ \citenamefont {Zhu}(2021)}]{hou2021GravitationalMemoryEffects}%
  \BibitemOpen
  \bibfield  {author} {\bibinfo {author} {\bibfnamefont {S.}~\bibnamefont {Hou}}\ and\ \bibinfo {author} {\bibfnamefont {Z.-H.}\ \bibnamefont {Zhu}},\ }\bibfield  {title} {\bibinfo {title} {Gravitational memory effects and {{Bondi}}-{{Metzner}}-{{Sachs}} symmetries in scalar-tensor theories},\ }\href {https://doi.org/10.1007/JHEP01(2021)083} {\bibfield  {journal} {\bibinfo  {journal} {Journal of High Energy Physics}\ }\textbf {\bibinfo {volume} {2021}},\ \bibinfo {eid} {83} (\bibinfo {year} {2021})},\ \Eprint {https://arxiv.org/abs/2005.01310} {arXiv:2005.01310 [gr-qc]} \BibitemShut {NoStop}%
\bibitem [{\citenamefont {Seraj}(2021)}]{seraj2021GravitationalBreathingMemory}%
  \BibitemOpen
  \bibfield  {author} {\bibinfo {author} {\bibfnamefont {A.}~\bibnamefont {Seraj}},\ }\bibfield  {title} {\bibinfo {title} {Gravitational breathing memory and dual symmetries},\ }\href {https://doi.org/10.1007/JHEP05(2021)283} {\bibfield  {journal} {\bibinfo  {journal} {Journal of High Energy Physics}\ }\textbf {\bibinfo {volume} {2021}},\ \bibinfo {eid} {283} (\bibinfo {year} {2021})},\ \Eprint {https://arxiv.org/abs/2103.12185} {arXiv:2103.12185 [hep-th]} \BibitemShut {NoStop}%
\bibitem [{\citenamefont {Campiglia}\ \emph {et~al.}(2018)\citenamefont {Campiglia}, \citenamefont {Coito},\ and\ \citenamefont {Mizera}}]{campiglia2018CanScalarsHave}%
  \BibitemOpen
  \bibfield  {author} {\bibinfo {author} {\bibfnamefont {M.}~\bibnamefont {Campiglia}}, \bibinfo {author} {\bibfnamefont {L.}~\bibnamefont {Coito}},\ and\ \bibinfo {author} {\bibfnamefont {S.}~\bibnamefont {Mizera}},\ }\bibfield  {title} {\bibinfo {title} {Can scalars have asymptotic symmetries?},\ }\href {https://doi.org/10.1103/PhysRevD.97.046002} {\bibfield  {journal} {\bibinfo  {journal} {Physical Review D}\ }\textbf {\bibinfo {volume} {97}},\ \bibinfo {eid} {046002} (\bibinfo {year} {2018})},\ \Eprint {https://arxiv.org/abs/1703.07885} {arXiv:1703.07885 [hep-th]} \BibitemShut {NoStop}%
\bibitem [{\citenamefont {Campiglia}\ \emph {et~al.}(2019)\citenamefont {Campiglia}, \citenamefont {Freidel}, \citenamefont {Hopfmüller},\ and\ \citenamefont {Soni}}]{campiglia2019ScalarAsymptoticCharges}%
  \BibitemOpen
  \bibfield  {author} {\bibinfo {author} {\bibfnamefont {M.}~\bibnamefont {Campiglia}}, \bibinfo {author} {\bibfnamefont {L.}~\bibnamefont {Freidel}}, \bibinfo {author} {\bibfnamefont {F.}~\bibnamefont {Hopfmüller}},\ and\ \bibinfo {author} {\bibfnamefont {R.~M.}\ \bibnamefont {Soni}},\ }\bibfield  {title} {\bibinfo {title} {Scalar asymptotic charges and dual large gauge transformations},\ }\href {https://doi.org/10.1007/JHEP04(2019)003} {\bibfield  {journal} {\bibinfo  {journal} {Journal of High Energy Physics}\ }\textbf {\bibinfo {volume} {2019}},\ \bibinfo {eid} {3} (\bibinfo {year} {2019})},\ \Eprint {https://arxiv.org/abs/1810.04213} {arXiv:1810.04213 [hep-th]} \BibitemShut {NoStop}%
\bibitem [{\citenamefont {Nagy}\ and\ \citenamefont {Peraza}(2023)}]{nagy2023RadiativePhaseSpace}%
  \BibitemOpen
  \bibfield  {author} {\bibinfo {author} {\bibfnamefont {S.}~\bibnamefont {Nagy}}\ and\ \bibinfo {author} {\bibfnamefont {J.}~\bibnamefont {Peraza}},\ }\bibfield  {title} {\bibinfo {title} {Radiative phase space extensions at all orders in $r$ for self-dual {{Yang-Mills}} and {{Gravity}}},\ }\href {https://doi.org/10.1007/JHEP02(2023)202} {\bibfield  {journal} {\bibinfo  {journal} {Journal of High Energy Physics}\ }\textbf {\bibinfo {volume} {2023}},\ \bibinfo {eid} {202} (\bibinfo {year} {2023})},\ \Eprint {https://arxiv.org/abs/2211.12991} {arXiv:2211.12991 [hep-th]} \BibitemShut {NoStop}%
\bibitem [{\citenamefont {Stückelberg}(1938)}]{stuckelberg1938WechselwirkungskrafteElektrodynamikUnd}%
  \BibitemOpen
  \bibfield  {author} {\bibinfo {author} {\bibfnamefont {E.~C.~G.}\ \bibnamefont {Stückelberg}},\ }\bibfield  {title} {\bibinfo {title} {Die {{Wechselwirkungskräfte}} in der {{Elektrodynamik}} und in der {{Feldtheorie}} der {{Kernkräfte}}. {{Teil I}}},\ }\href {https://doi.org/10.5169/SEALS-110852} {\bibfield  {journal} {\bibinfo  {journal} {Helvetica Physica Acta}\ }\textbf {\bibinfo {volume} {11}},\ \bibinfo {pages} {225} (\bibinfo {year} {1938})}\BibitemShut {NoStop}%
\bibitem [{\citenamefont {Hinterbichler}(2012)}]{hinterbichler2012TheoreticalAspectsMassive}%
  \BibitemOpen
  \bibfield  {author} {\bibinfo {author} {\bibfnamefont {K.}~\bibnamefont {Hinterbichler}},\ }\bibfield  {title} {\bibinfo {title} {Theoretical aspects of massive gravity},\ }\href {https://doi.org/10.1103/RevModPhys.84.671} {\bibfield  {journal} {\bibinfo  {journal} {Reviews of Modern Physics}\ }\textbf {\bibinfo {volume} {84}},\ \bibinfo {pages} {671} (\bibinfo {year} {2012})},\ \Eprint {https://arxiv.org/abs/1105.3735} {arXiv:1105.3735 [hep-th]} \BibitemShut {NoStop}%
\bibitem [{\citenamefont {Boyer}(1969)}]{boyer1969GeodesicKillingOrbits}%
  \BibitemOpen
  \bibfield  {author} {\bibinfo {author} {\bibfnamefont {R.~H.}\ \bibnamefont {Boyer}},\ }\bibfield  {title} {\bibinfo {title} {Geodesic {{Killing}} orbits and bifurcate {{Killing}} horizons},\ }\href {https://doi.org/10.1098/rspa.1969.0116} {\bibfield  {journal} {\bibinfo  {journal} {Proceedings of the Royal Society of London A: Mathematical and Physical Sciences}\ }\textbf {\bibinfo {volume} {311}},\ \bibinfo {pages} {245} (\bibinfo {year} {1969})}\BibitemShut {NoStop}%
\bibitem [{\citenamefont {Kay}\ and\ \citenamefont {Wald}(1991)}]{kay1991TheoremsUniquenessThermal}%
  \BibitemOpen
  \bibfield  {author} {\bibinfo {author} {\bibfnamefont {B.~S.}\ \bibnamefont {Kay}}\ and\ \bibinfo {author} {\bibfnamefont {R.~M.}\ \bibnamefont {Wald}},\ }\bibfield  {title} {\bibinfo {title} {Theorems on the uniqueness and thermal properties of stationary, nonsingular, quasifree states on spacetimes with a bifurcate killing horizon},\ }\href {https://doi.org/10.1016/0370-1573(91)90015-E} {\bibfield  {journal} {\bibinfo  {journal} {Physics Reports}\ }\textbf {\bibinfo {volume} {207}},\ \bibinfo {pages} {49} (\bibinfo {year} {1991})}\BibitemShut {NoStop}%
\bibitem [{\citenamefont {Wald}(1994)}]{wald1994QuantumFieldTheory}%
  \BibitemOpen
  \bibfield  {author} {\bibinfo {author} {\bibfnamefont {R.~M.}\ \bibnamefont {Wald}},\ }\href {https://www.bibliovault.org/BV.landing.epl?ISBN=9780226870274} {\emph {\bibinfo {title} {Quantum Field Theory in Curved Spacetime and Black Hole Thermodynamics}}},\ Chicago Lectures in Physics\ (\bibinfo  {publisher} {The University of Chicago Press},\ \bibinfo {address} {Chicago, IL},\ \bibinfo {year} {1994})\BibitemShut {NoStop}%
\bibitem [{\citenamefont {Bousso}(1999)}]{bousso1999HolographyGeneralSpacetimes}%
  \BibitemOpen
  \bibfield  {author} {\bibinfo {author} {\bibfnamefont {R.}~\bibnamefont {Bousso}},\ }\bibfield  {title} {\bibinfo {title} {Holography in general space-times},\ }\href {https://doi.org/10.1088/1126-6708/1999/06/028} {\bibfield  {journal} {\bibinfo  {journal} {Journal of High Energy Physics}\ }\textbf {\bibinfo {volume} {1999}},\ \bibinfo {eid} {028} (\bibinfo {year} {1999})},\ \Eprint {https://arxiv.org/abs/hep-th/9906022} {arXiv:hep-th/9906022} \BibitemShut {NoStop}%
\bibitem [{\citenamefont {Bousso}(2002)}]{bousso2002HolographicPrinciple}%
  \BibitemOpen
  \bibfield  {author} {\bibinfo {author} {\bibfnamefont {R.}~\bibnamefont {Bousso}},\ }\bibfield  {title} {\bibinfo {title} {The holographic principle},\ }\href {https://doi.org/10.1103/RevModPhys.74.825} {\bibfield  {journal} {\bibinfo  {journal} {Reviews of Modern Physics}\ }\textbf {\bibinfo {volume} {74}},\ \bibinfo {pages} {825} (\bibinfo {year} {2002})},\ \Eprint {https://arxiv.org/abs/hep-th/0203101} {arXiv:hep-th/0203101} \BibitemShut {NoStop}%
\bibitem [{\citenamefont {Bunch}\ and\ \citenamefont {Davies}(1978)}]{bunch1978QuantumFieldTheory}%
  \BibitemOpen
  \bibfield  {author} {\bibinfo {author} {\bibfnamefont {T.~S.}\ \bibnamefont {Bunch}}\ and\ \bibinfo {author} {\bibfnamefont {P.~C.~W.}\ \bibnamefont {Davies}},\ }\bibfield  {title} {\bibinfo {title} {Quantum field theory in de {{Sitter}} space: Renormalization by point-splitting},\ }\href {https://doi.org/10.1098/rspa.1978.0060} {\bibfield  {journal} {\bibinfo  {journal} {Proceedings of the Royal Society of London. A. Mathematical and Physical Sciences}\ }\textbf {\bibinfo {volume} {360}},\ \bibinfo {pages} {117} (\bibinfo {year} {1978})}\BibitemShut {NoStop}%
\bibitem [{\citenamefont {Chernikov}\ and\ \citenamefont {Tagirov}(1968)}]{chernikov1968QuantumTheoryScalar}%
  \BibitemOpen
  \bibfield  {author} {\bibinfo {author} {\bibfnamefont {N.~A.}\ \bibnamefont {Chernikov}}\ and\ \bibinfo {author} {\bibfnamefont {E.~A.}\ \bibnamefont {Tagirov}},\ }\bibfield  {title} {\bibinfo {title} {Quantum theory of scalar field in de {{Sitter}} space-time},\ }\href {https://www.numdam.org/item/AIHPA_1968__9_2_109_0/} {\bibfield  {journal} {\bibinfo  {journal} {Annales de l'Institut Henri Poincaré. Section A, Physique Théorique}\ }\textbf {\bibinfo {volume} {9}},\ \bibinfo {pages} {109} (\bibinfo {year} {1968})}\BibitemShut {NoStop}%
\bibitem [{\citenamefont {Hartle}\ and\ \citenamefont {Hawking}(1976)}]{hartle1976PathintegralDerivationBlackhole}%
  \BibitemOpen
  \bibfield  {author} {\bibinfo {author} {\bibfnamefont {J.~B.}\ \bibnamefont {Hartle}}\ and\ \bibinfo {author} {\bibfnamefont {S.~W.}\ \bibnamefont {Hawking}},\ }\bibfield  {title} {\bibinfo {title} {Path-integral derivation of black-hole radiance},\ }\href {https://doi.org/10.1103/PhysRevD.13.2188} {\bibfield  {journal} {\bibinfo  {journal} {Physical Review D}\ }\textbf {\bibinfo {volume} {13}},\ \bibinfo {pages} {2188} (\bibinfo {year} {1976})}\BibitemShut {NoStop}%
\bibitem [{\citenamefont {{Wolfram Research}}(2024)}]{wolframresearch2024Mathematica140}%
  \BibitemOpen
  \bibfield  {author} {\bibinfo {author} {\bibnamefont {{Wolfram Research}}},\ }\href {https://www.wolfram.com/mathematica} {\bibinfo {title} {Mathematica 14.0}},\ \bibinfo {howpublished} {Wolfram Research} (\bibinfo {year} {2024})\BibitemShut {NoStop}%
\bibitem [{\citenamefont {Shoshany}(2021)}]{shoshany2021OGReObjectOrientedGeneral}%
  \BibitemOpen
  \bibfield  {author} {\bibinfo {author} {\bibfnamefont {B.}~\bibnamefont {Shoshany}},\ }\bibfield  {title} {\bibinfo {title} {{{OGRe}}: {{An}} object-oriented general relativity package for {{Mathematica}}},\ }\href {https://doi.org/10.21105/joss.03416} {\bibfield  {journal} {\bibinfo  {journal} {Journal of Open Source Software}\ }\textbf {\bibinfo {volume} {6}},\ \bibinfo {eid} {3416} (\bibinfo {year} {2021})},\ \Eprint {https://arxiv.org/abs/2109.04193} {arXiv:2109.04193 [cs.MS]} \BibitemShut {NoStop}%
\bibitem [{\citenamefont {Ellis}(2017)}]{ellis2017TikZFeynmanFeynmanDiagrams}%
  \BibitemOpen
  \bibfield  {author} {\bibinfo {author} {\bibfnamefont {J.~P.}\ \bibnamefont {Ellis}},\ }\bibfield  {title} {\bibinfo {title} {Ti{\emph{k}}{{Z-Feynman}}: {{Feynman}} diagrams with {{Ti}}{\emph{k}}{{Z}}},\ }\href {https://doi.org/10.1016/j.cpc.2016.08.019} {\bibfield  {journal} {\bibinfo  {journal} {Computer Physics Communications}\ }\textbf {\bibinfo {volume} {210}},\ \bibinfo {pages} {103} (\bibinfo {year} {2017})},\ \Eprint {https://arxiv.org/abs/1601.05437} {arXiv:1601.05437 [hep-ph]} \BibitemShut {NoStop}%
\end{thebibliography}%
